\documentclass[aps,twocolumn,noshowkeys]{revtex4-2}
\usepackage{amsmath}
\usepackage{graphicx}
\usepackage{bm}
\usepackage{hyperref}
\usepackage{xcolor}

\hypersetup{colorlinks=true, citecolor=blue, urlcolor=blue, linkcolor=blue}

\usepackage{multirow}

\usepackage[T2A]{fontenc}
\usepackage[utf8]{inputenc}

\newcommand{\eEDM}{{\em e}EDM}

\newcommand{\de}{d_\mathrm{e}}

\newcommand{\WaxeN}{W_{\textrm{ax}}^{(eN)}}

\newcommand{\Waxee}{W_{\textrm{ax}}^{(ee)}}

\newcommand{\Ti}{$\mathcal{T}$}
\newcommand{\Par}{$\mathcal{P}$}
\newcommand{\CP}{$\mathcal{CP}$}
\newcommand{\Ci}{$\mathcal{C}$}

\usepackage{color}

\begin{document}
\title{Updated constraints on \Ti,\Par-violating axionlike-particle-mediated electron-electron and electron-nucleus interactions from HfF$^+$ experiment}

\date{14.04.2023}

\begin{abstract}
Recently, upper bounds on the static time-reversal (\Ti) and spatial parity (\Par)-violating electron electric dipole moment (\eEDM) and dimensionless constant characterizing the strength of the \Ti,\Par-violating scalar-pseudoscalar nucleus-electron interaction have been updated in the JILA experiment using the HfF$^+$ cations [T.S. Roussy et al, \href{https://doi.org/10.48550/arXiv.2212.11841}{
arxiv:2212.11841 (2022)}]. We considered other two sources of the \Ti,\Par\,-violation in HfF$^+$ -- axionlike-particle-mediated (ALP) scalar-pseudoscalar electron-electron and nucleus-electron interactions. To estimate the magnitude of effects, induced by such interactions in HfF$^+$ we have developed and applied a method which implies direct use of the \textit{ab initio} relativistic coupled cluster theory to calculate molecular parameters that characterize the interactions. Using these parameters, we showed that an order of magnitude updated laboratory constraints on the ALP-mediated electron-electron and nucleus-electron interactions can be derived from the experimental data on \Ti,\Par-violating effects in HfF$^+$ for a wide range of ALP masses.
\end{abstract}

\author{Sergey D. Prosnyak$^{1,2}$, Daniel E. Maison$^{1}$ and Leonid V. Skripnikov$^{1,2}$} 
 \homepage{http://www.qchem.pnpi.spb.ru}
 \email{skripnikov\_lv@pnpi.nrcki.ru}
\affiliation{$^1$Petersburg Nuclear Physics Institute named by B.P.\ Konstantinov of National Research Center ``Kurchatov Institute'' (NRC ``Kurchatov Institute'' - PNPI), 1 Orlova roscha mcr., Gatchina, 188300 Leningrad region, Russia}
\affiliation{$^2$Saint Petersburg State University, 7/9
Universitetskaya Naberezhnaya, St. Petersburg, 199034 Russia}

\maketitle
\section{Introduction}

Experiments with paramagnetic heavy-atom molecules aiming to search for the effects of violation of the time-reversal (\Ti) and spatial parity (\Par) symmetries of fundamental interactions are considered to be very sensitive to new physics beyond the Standard Model (SM) \cite{Safronova:18,eEDM_snowmass:2022}. One of the most discussed sources of the \Ti,\Par-violation is the permanent electron electric dipole moment (\eEDM\ or $\de$). 
Indeed, the \eEDM\ should be directed along its spin $\mathbf{s}$: $\mathbf{d}_e=d_e \mathbf{s}/s$~\cite{Safronova:18,eEDM_snowmass:2022,Safronova2023}. The interaction of such electric dipole moment with an external electric field $\mathbf{E}$ is $H=-\mathbf{d}_e \cdot \mathbf{E} \propto d_e \mathbf{s} \cdot \mathbf{E}$.
Spin $\mathbf{s}$ is \Ti-odd, while $\mathbf{E}$ is \Ti-even. Thus $H$ is \Ti-odd. At the same time $\mathbf{s}$ is \Par-even, but $\mathbf{E}$ is \Par-odd. Thus, $H$ is simultaneously \Ti- and \Par- odd.
The expected value of the \eEDM\ within the extensions to the SM is many orders of magnitude bigger than estimates within the SM~\cite{eEDMLimit:2022,Yamaguchi:2021,Engel:2013,Chubukov:2016,Commins:98,Chupp:2019}. Recently, an updated upper bound on \eEDM\ has been established by the JILA group using the hafnium monofluoride molecular cations $^{180}$Hf$^{19}$F$^+$ trapped by the rotating electric field:  $|\de|<4.1\times 10^{-30}$ $\ensuremath{e \cdot {\rm cm}}$ (90\% confidence)~\cite{newlimit1}. The experiment was carried out on the first excited metastable electronic state $^3\Delta_1$ of HfF$^+$.
Molecules in such states have very close-lying states of opposite parity, so-called $\Omega$-doublet energy level structure. Therefore \Ti,\Par-odd effects are strongly enhanced~\cite{Labzowsky:78,Sushkov:78} in these molecules. In addition, the use of the $^3\Delta_1$ electronic state leads to the suppression of systematic errors \cite{ComminsFest,Kawall:04b,Vutha:2010,caldwell2022systematic}.
The obtained constraint on \eEDM\ using HfF$^+$ is better by the factor of 2.4 than the previous most stringent one obtained using the ThO molecular beam~\cite{ACME:18}.
Experiments to search for \eEDM\ are also sensitive to the \Ti,\Par-violating scalar-pseudoscalar nucleus-electron interaction~\cite{Gorshkov:79,Safronova:18,Jung:13,Pospelov:14,Skripnikov:2020b}. Indeed, the same experiment on HfF$^+$~\cite{newlimit1} provided also an updated constraint on the dimensionless constant characterizing the strength of this interaction.  At present, several more experiments to search for the \eEDM\ using other molecular systems are under preparation. Among them are ThF$^+$\cite{ThFp:2016}, YbOH~\cite{kozyryev2017precision, isaev2017laser}, BaF~\cite{BaF:2018}, YbF~\cite{YbF:2020}, LuOH$^+$~\cite{Maison:2022b} and some others.

The search for the symmetry-violating effects is of crucial importance for understanding the fundamental laws of Nature as several phenomena have not been explained  yet within the Standard model. 
About sixty years ago the violation of the combined \CP\ symmetry  (\Ci\ is the charge conjugation) was discovered in the kaon decay~\cite{Christenson:64}. It showed that weak interactions do not conserve \CP. However, up to now it is not clear why the \CP\ symmetry seems to be 
conserved in the quantum chromodynamics (QCD) sector of the SM. In principle, there is no reason for such conservation as there are natural terms in the Lagrangian of QCD that can violate the \CP-symmetry. In general, such mechanism can lead to the neutron electric dipole moment value that is about 10$^8$ times larger than the present constraint~\cite{abel2020_neutronEDM}. Therefore, this ``fine tuning'' problem is called the strong \CP\ problem.
Note, that according to the \Ci\Par\Ti\ theorem, the \CP\ violation leads to violation of the \Ti symmetry. The \CP-nonconservation is one of the necessary conditions for the baryon asymmetry in our Universe, i.e. an imbalance between the amount of matter and anti-matter~\cite{Sakharov1967,MatterAntimatter:2003,Safronova:18}. However, the very weak level of \CP\ symmetry violation in the SM is not enough to explain the imbalance. Another unsolved problem is the unknown nature of the dark matter which makes up about 25\% of the universe. There are numerous efforts to search for the dark matter particles~\cite{Drukier1986, Ahmed2011, Aprile2017,Aalseth2011, adhikari2019experiment, kane2008dark}. Among them, the pseudoscalar spin-$0$ axion or axionlike particle (ALP) are very popular candidates~\cite{preskill1983cosmology, abbott1983cosmological,dine1983not}.
The axion~\cite{weinberg1978new,wilczek1978problem} is considered as a quasi-Nambu-Goldstone boson due to the spontaneous breaking of the Peccei-Quinn symmetry $U_{\textrm{PQ}}(1)$~\cite{peccei1977cp}. This symmetry violation is considered as a solution to the strong \CP-problem in QCD~\cite{Kim:2010}. Therefore, the search for the axion is so important for modern physics. 

The interaction of axions and axionlike-particles with other particles of the SM can be \CP-preserving and \CP-violating. In the present paper we are interested in \CP-violating interactions with electrons and nucleons. Due to its high importance there are numerous astrophysical and laboratory experiments and research programs that are aimed to search for ALPs and study their properties. In the last decade such affords have accelerated. The present status and constraints on axion and ALP properties can be found e.g. in Refs. \cite{Hare:2020,QCDaxion:2020,Kim:2010,youdin1996limits, ni1999search, Duffy:2006, Zavattini:2006, hammond2007new, hoedl2011improved, barth2013cast, pugnat2014search, flambaum2018resonant,Roussy:2021,aybas2021,adams2023axion,sym13112150,Budker:2023a,sym14102165,sym12010025}.

According to Refs.~\cite{Graham:2011,graham2013new}, the interaction of the cosmic axionlike particles with fermions can induce \textit{oscillating} \Ti,\Par-violating molecular and atomic EDMs and one can try to observe this effect.
The oscillation frequency is determined by the ALP mass. The oscillation amplitude is proportional to the square root of the ALP density \cite{Graham:2011,graham2013new}. One can estimate this density assuming that the dark matter consists mostly of
ALPs. Later, there was a proposal to search for the oscillating EDM within the solid-state nuclear magnetic resonance technique~\cite{budker2014proposal}. Recently, the results of such experiment were reported~\cite{aybas2021} and expressed~\cite{Skripnikov:16a,aybas2021} in terms of the oscillating nuclear Schiff moment. The latter was interpreted in terms of axion to the gluon field coupling constant~\cite{aybas2021}. Significant restrictions on the ALP-gluon coupling constants were also deduced from the analysis of the previous experimental data on the HfF$^+$ molecular cation in terms of the oscillating dipole moment~\cite{Roussy:2021}.

In Ref.~\cite{Stadnik:2018} the effect of the exchange of a virtual ALP between electron and electron as well as between electron and nucleus was studied. Such effects can induce the \textit{static} permanent \Ti,\Par-violating atomic and molecular EDMs. Corresponding calculations for various atoms and some atomic-based estimates for molecules were carried out in Ref.~\cite{Stadnik:2018}. 
In Refs.~\cite{Maison:2021,maison2021axion,Maison:2022} the ALP-exchange effect was studied theoretically for the Fr atom and YbOH molecule. In the latter case for the electron-electron ALP-mediated interaction we considered only the limit of light ALP.
The present paper aims to interpret the recent experimental data on the static \Ti,\Par-violating molecular EDM of HfF$^+$~\cite{newlimit1} in terms of the virtual ALP exchange between electrons and between electron and nucleus. For this, we developed an approach that allows us to \textit{directly} calculate the contribution of such effects in \textit{molecules} for a very wide range of ALP masses including limiting cases of very light and very heavy ALPs.

\section{Theory}
There are numerous possible couplings of axions with other particles.
Let us consider the Lagrangian of the interaction of ALPs with fermions, which can be written as follows~\cite{moody1984new}:
\begin{equation}\label{lagrangian}
    \mathcal{L}_{\textrm{int}} = a \sum_\psi \bar{\psi} \left(g_\psi^s + ig_\psi^p \gamma_5 \right) \psi \ .
\end{equation}
In this expression $a$ and $\psi$ are the axion and fermion fields, respectively; $\bar{\psi} = \psi^{+} \gamma_0$; $g_\psi^s$ ($g_\psi^p$) denotes the scalar (pseudoscalar) axion to fermion coupling constant; $\gamma_0$ and $\gamma_5$ are Dirac matrices, defined as in Ref. \cite{Khriplovich:91}. The summation in the Lagrangian (\ref{lagrangian}) is over all types of fermions in the system. 
The corresponding \Ti,\Par-violating interaction between two electrons, which is mediated by the axion with mass $m_a$ can be described by the Yukawa-type two-electron interaction  \cite{moody1984new,Stadnik:2018}:
\begin{equation} 
\label{ee_potential}
    V_{ee}(\boldsymbol{r}_1, \boldsymbol{r}_2) = 
    +i\frac{g_e^s g_e^p}{4\pi} \frac{e^{-m_a |\boldsymbol{r}_1 - \boldsymbol{r}_2|}}{|\boldsymbol{r}_1 - \boldsymbol{r}_2|} \gamma_0 \gamma_5 \ ,
\end{equation}
where $\boldsymbol{r}_1$ and $\boldsymbol{r}_2$ are positions of electrons, $\gamma_0$ and $\gamma_5$ matrices refer to the second electron, $g_e^s$ ($g_e^p$) is the scalar (pseudoscalar) coupling constant of the electron-axion interaction. Within the QCD axion models  $g_e^s$ and $m_a$ are related~\cite{moody1984new}. However, in many studies, one considers instead a more general case in which these parameters are considered independent, so the arbitrary mass particle is implied. In the present paper we follow this approach and do not distinguish the axion and ALP.

The ALP-mediated \Ti,\Par-violating interaction between electron and nucleon can be expressed as follows~\cite{moody1984new,Stadnik:2018}:
\begin{equation}
\label{eN_potential}
    V_{eN} (\boldsymbol{r}) = +i \frac{g_N^sg_e^p}{4 \pi}
    \frac{e^{-m_a |\boldsymbol{r} - \boldsymbol{R}|}}{|\boldsymbol{r} - \boldsymbol{R}|} \gamma_0 \gamma_5,
\end{equation}
where $\boldsymbol{R}$ is the position of the nucleon and $\boldsymbol{r}$ is the position of the electron; $g_N^s$ is the scalar ALP-nucleon coupling constant; $N$ in the nucleon (proton or neutron);  $\gamma$ matrices refer to the electron.

The considered \Ti,\Par-violating electron-electron interaction inside a molecule can be characterized by the following parameter~\cite{Maison:2021,Dmitriev:92}:
\begin{equation}
    \Waxee(m_a) = \frac{1}{\Omega} \frac{1}{g_e^s g_e^p}
    \langle \Psi | 
    \mathop{{\sum}'}_{i,j=1}^{N_e} V_{ee}(\boldsymbol{r}_i, \boldsymbol{r}_j) | \Psi\rangle,
\label{WaxeeME}    
\end{equation}
where $\Psi$ is the molecular electronic wave function, $\Omega$ is the projection of the total electronic angular momentum on the molecular axis and $N_e$ is the number of electrons. For the considered case of the first excited electronic state of the HfF$^+$ cation $\Omega=1$. The prime index in the sum means the terms with $i = j$ should be omitted.

The ALP-mediated electron-nucleus interaction can be characterized by the molecular parameter $\WaxeN$:
\begin{equation}
    \WaxeN(m_a) = 
    \frac{1}{\Omega}
    \langle \Psi | \sum\limits_{i=1}^{N_\textrm{e}} \sum\limits_{N} 
    \frac{1}{g_N^s g_e^p}
    V_{\textrm{eN}}(\boldsymbol{r}_i)
    | \Psi \rangle,
\label{WaxeNME}        
\end{equation}
where index $i$ runs over all electrons; the inner sum runs over all the nucleons of the Hf nucleus. Note, that $\Waxee(m_a)$ and $\WaxeN$ are functions of the ALP mass.

The characteristic \Ti,\Par-violating energy shift of the electronic level induced by the interaction (\ref{ee_potential}) can be expressed as:
\begin{equation}
\label{Etp_ee}
    \delta E = g_e^s g_e^p \Omega \Waxee(m_a),
\end{equation}
while the \Ti,\Par-violating energy shift induced by the interaction (\ref{eN_potential}) can be expressed as:
\begin{equation}
\label{Etp_eN}
    \delta E = \bar{g}_{N}^s g_e^p  \Omega \WaxeN(m_a).
\end{equation}
Here $\bar{g}_{N}^s$ is the ALP-nucleon coupling constant averaged over the Hf nucleus nucleons: $\bar{g}_{N}^s = (Zg_p^s+N_n g_n^s)/A$; $g_p^s$ and $g_n^s$ are scalar ALP-proton and ALP-neutron coupling constants, $Z$ is the charge of the Hf nucleus, $N_n$ is its neutron number and $A=Z+N_n$: $Z=72$, $N_n=108$, $A=180$.

\section{Implementation and computational details}

One can see from Eqns.~(\ref{Etp_ee}) and (\ref{Etp_eN}), that molecular parameters (\ref{WaxeeME}) and (\ref{WaxeNME}) are required to extract the values of the products of the unknown coupling constants. However, they cannot be measured in practice. The situation is analogues to the problem of extracting the value of the \eEDM\ where one should know the value of the internal effective electric field acting on the electron inside a molecule.
To solve such problems of the experiment interpretation, molecular electronic structure calculation is required.
In the present study we used the Dirac-Coulomb (DC) Hamiltonian to describe HfF$^+$ electronic structure:
\begin{widetext}
\begin{eqnarray}
 H_{DC} &=& \Lambda_+ \left[ \sum\limits_j\, \left[ \, \boldsymbol{\alpha}_j \cdot {\bf{p}}_j + \beta_j  
+V_{\rm nuc}(j) \right]
+ \sum\limits_{j<k}\,
V_{C}(r_{jk}) \right] \Lambda_+.
 \label{EQ:HAMILTONIAN}
\end{eqnarray}
\end{widetext}
Here $\alpha$, $\beta$ are Dirac matrices, $\bf{p}$ in the electron momentum, $V_{\rm nuc}$ is the nuclear subsystem potential, $V_{C}$ is the Coulomb electron-electron interaction operator, $\Lambda_+$ are projectors on the positive-energy states of the Dirac picture and the summation index runs over all electrons.
%
To calculate the electronic wave function we used the method of the relativistic coupled cluster (CC)~\cite{Visscher:96a,Bartlett:2007}. In this approach one uses the following exponential ansatz for the many-electron wave function $\Psi$:
\begin{equation} 
\label{expAnsatz}
    \Psi = e^{\hat T} \Phi_0.
\end{equation}
Here $\Phi_0$ is the reference Slater determinant providing zero-order approximation to the many-electron molecular wave function.
In the present work $\Phi_0$ was obtained within the Dirac-Hartree-Fock method.
Wave function $\Psi$ takes into account effects of electron correlation.
Cluster operator $\hat{T}$ can be written as follows:
\begin{equation}
\label{CCexpansion}
    \hat T = \hat{T}_1 + \hat{T}_2 + \hat{T}_3 + \dots,
\end{equation}
where $\hat{T}_1$, $\hat{T}_2$,... are the excitation operators of different orders:
$$
    \hat{T}_1 = \sum\limits_{\substack{i \in occ \\ b \in virt}} t_{i}^{b} a_{b}^\dagger a_{i};
    \quad
    \hat{T}_2 =\frac{1}{2!} \sum\limits_{\substack{i_1<i_2 \in occ \\ b_1<b_2 \in virt}} t_{i_1 i_2}^{b_1 b_2} 
    a_{b_1}^\dagger a_{b_2}^\dagger a_{i_2} a_{i_1}.
$$
Quantities $t_{\dots}^{\dots}$ are called cluster amplitudes. They can be determined by solving coupled cluster equations~\cite{Visscher:96a,Bartlett:2007}. $a_b^\dagger$ and $a_i$ are creation and annihilation operators of one-electron states $b$ and $i$. Indexes $i_1,i_2...$ refer to the occupied states, i.e. those included in the reference determinant $\Phi_0$, while $b_1,b_2,...$ refer to the unoccupied (virtual) orbitals.  
In the present paper we considered the following levels of theory: (i) the coupled cluster method with single and double cluster amplitudes, CCSD,  (ii) the coupled cluster method with single, double and triple cluster amplitudes, CCSDT and (iii) the coupled cluster method with single, double and noniterative triple cluster amplitudes, CCSD(T). The latter is usually quite a good approximation to the full iterative CCSDT model, but requires much less computational resources. The simplification is due to the use of the perturbation theory to estimate the contribution of triple cluster amplitudes. As the CCSD(T) method has a very good ratio between accuracy and numerical complexity, it is called ``the golden standard'' of quantum chemistry.
One can see from Eq.~(\ref{CCexpansion}) that even if the cluster operator is truncated after $\hat{T}_2$ term, the higher-order (e.g. quadruple) excitations will be effectively taken into account due to the exponential ansatz for the many-electron wave function (\ref{expAnsatz}).
One of the most important features of the CC approach is its size-extensivity property~\cite{Bartlett:2007}. It means that correlation energy properly scales with the number of electrons. This feature is very important for the description of many-electron systems and calculation of symmetry-violating effects~\cite{Skripnikov:15a,Skripnikov:16b,Skripnikov:17a}.

To calculate molecular parameters of interest within the CC theory, we used the finite-field approach. For this, we added required operator (\ref{ee_potential}) or (\ref{eN_potential}) to the electronic Hamiltonian with a small coefficient (``weak field'') and calculated the derivative of the total electronic energy with respect to this coefficient within the standard numerical technique. In this way we obtained required matrix elements (\ref{WaxeeME}) and (\ref{WaxeNME}). This approach is well-known for calculating such properties as the molecule-frame dipole moment, but it is also widely used to calculate many others. Usually, one studies one-electron properties as we did it here for (\ref{eN_potential}). But in the present study we also applied the finite-field approach to calculate the two-electron characteristic (\ref{WaxeNME}).
Technically, for this one should calculate corresponding two-electron integrals of the operation of interest over molecular bispinors. Let us describe this problem in some detail.

The only attempt to \textit{directly} calculate the effect (\ref{WaxeeME}), (\ref{Etp_ee}) in molecules was undertaken in Ref.~\cite{maison2021axion}. However, it was done only for the case when the exponent in Eq.~(\ref{ee_potential}) can be approximated by two terms (the low-mass ALP case). Calculation of the matrix elements of the operator~(\ref{ee_potential}) without such an approximation over molecular bispinors can be reduced to calculation of ``primitive'' integrals $\langle ab|\frac{e^{-m_a r}}{r}|cd\rangle$ over primitive Gaussian-type basis functions $a,b,c,d$ of the form $x^n y^m z^k e^{-\beta r^2}$ after taking into account the structure of $\gamma$-matrices.
Here $n,m,k$ are nonnegative integers and their sum defines the angular momentum of a given basis function, while $\beta >0$ is the exponent parameter.
In the case of a diatomic molecule, basis functions $a,b,c,d$ can be centered on different nuclei, which complicates such a calculation (see Ref.~\cite{Maison:2022} for one-center problem with Gaussian-type functions). In the corresponding problem of calculating Coulomb interaction integrals $\langle ab|\frac{1}{r}|cd \rangle$, the computational algorithm usually involves the evaluation of the Boys function~\cite{boys1950electronic, helgaker2013molecular}:
\begin{equation}
    F_m(T) = \int_{0}^{1} dt t^{2m} e^{-Tt^2}.
\end{equation}
Calculation of integrals $\langle ab|\frac{e^{-m_a r}}{r}|cd\rangle$ is similar, but the Boys function must be replaced by a more complicated special function~\cite{Tenno04}:
\begin{equation}
    G_m(T, U) = \int_{0}^{1} dt t^{2m} e^{-Tt^2 + U(1-\frac{1}{t^2})}.
\end{equation}
This allows one to use a large number of different generalizations of classical algorithms of electron repulsion integrals evaluation such as 
McMurchie–Davidson~\cite{McMurchie1978}, Obara–Saika~\cite{Obara_Saika_1986}, PRISM of Gill et al.~\cite{Gill1991}, Pople–Hehre~\cite{Pople1978}, Head-Gordon–Pople~\cite{Head_Gordon_Pople_1988}, and others~\cite{Hamilton1991, Tenno1993, Yanai2000, Nakai2004}. In the present paper we applied the algorithm for the Yukawa potential integrals evaluation implemented in the LIBINT computational library~\cite{valeev2020libint}.
The approach is analogous to the so-called Rys quadrature method of Dupuis et al.~\cite{Dupuis1976, King1976, Rys1983} for the electron repulsion integrals, later elaborated by Lindh et al.~\cite{Lindh1991}, which is known to be well suited for evaluation of integrals over orbitals with high angular momenta compared to other algorithms.
However, the code implemented in the LIBINT library  was designed for the problems of explicitly-correlated electronic structure methods~\cite{Tenno04,ten2007new,shiozaki2009evaluation,Valeev2020,Kallay2021,Kallay:2023}.
It appeared that the implemented range for variables $T$ and $U$~\cite{valeev2020libint} was not enough for our purposes. Therefore, we made  certain modifications to carry out calculations of the present interest.
As mentioned above, in the original version of the LIBINT library, a tailored Gaussian quadrature was used as the main method for calculating integrals~\cite{shiozaki2009evaluation}. This algorithm uses the precalculated values of the Chebyshev expansion coefficients of the positions and weights of the grid points, and is thus limited to the case of $0 \leq T \leq T_{max}$ and $U_{min} \leq U \leq 10^3$, where $T_{max}=1024$ and $U_{min}=10^{-7}$.
If it turns out that $T>T_{max}$ or $U<U_{min}$, the ``scheme 1'' from Ref.~\cite{ten2007new} is used, which implies calculating 
\begin{equation}
    G_{-1} = \frac{e^{-T}}{4} \sqrt{\frac{\pi}{U}}\left[ e^{k^2}\textnormal{erfc}(k) + e^{\lambda^2}\textnormal{erfc}(\lambda) \right],
\end{equation}
\begin{equation}
    G_{0} = \frac{e^{-T}}{4} \sqrt{\frac{\pi}{T}}\left[ e^{k^2}\textnormal{erfc}(k) - e^{\lambda^2}\textnormal{erfc}(\lambda) \right],
    \label{G0}
\end{equation}
where $\textnormal{erfc}$ is the complementary error function,
\begin{equation}
    k = -\sqrt{T} + \sqrt{U},
\end{equation}
\begin{equation}
    \lambda = \sqrt{T} + \sqrt{U},
\end{equation}
and obtaining all the remaining $G_m$ values with the help of the upward recurrence relations 
\begin{equation}
    G_m = \frac{1}{2T}[(2m-1)G_{m-1} + 2UG_{m-2} - e^{-T}].
    \label{URR}
\end{equation}

To consider the case of $T=0$ we added a function that implements the upward recurrence relations at $T=0$ following Ref.~\cite{ten2007new}. The first element of the relations is defined as follows:
\begin{equation}
    G_0 = 1 - e^U\sqrt{\pi U}\textnormal{erfc}(\sqrt{U}).
\label{G0T0}    
\end{equation}
The remaining elements were found within the recurrence relation
\begin{equation}
    G_m(0, U) = \frac{1}{2m+1}[1 - 2UG_{m-1}(0, U)].
\label{G0T0m}        
\end{equation}

In addition, numerical instabilities in the ``scheme 1'' were found to occur at small $T$, since in Eq.~(\ref{URR}) $T$ appears in the denominator. To solve this problem, we implemented ``scheme 3'' from Ref.~\cite{ten2007new} for $T<0.1$ at $U<U_{min}$. 
According to this scheme, $G_m(T, U)$ can be found using the equation: 
\begin{equation}
    G_m(T, U) = \sum_{k=0}^{\infty} \frac{(-T)^k}{k!}
    G_{m+k}(0, U),
\end{equation}
where in practice we considered $k$ up to $k=8$; $G_m(0, U)$ were calculated according to Eqs.~(\ref{G0T0}) and (\ref{G0T0m}).

In Ref.~\cite{maison2021axion} we considered 
only
the case with ALP mass about 1 meV. 
For some axion models this is sufficient~\cite{Peccei2008,adams2023axion,Kim:2010}, but in general all ALP masses are of interest~\cite{HeavyAxion:2021,Giannotti:2010}.
Using the procedure described above we
substantially increased the range of masses, for which the molecular effect can be calculated.
Molecular parameter (\ref{WaxeeME}) (as well as (\ref{WaxeNME})) under consideration can be nonzero only for electronic states with unpaired electrons. In the considered $^3\Delta_1$ state of HfF$^+$ there are two such electrons and both of them are localized mainly on Hf. They approximately correspond to $5d$ and $6s$ states of the Hf$^{++}$ ion. Therefore, one can expect only a small contribution to the parameter (\ref{WaxeeME}) from the two-center integrals in case of high ALP mass due to the short range of this interaction (\ref{ee_potential}). Indeed, according to our direct calculation for $m_a=10^4$ eV the contribution of the two-center integrals is negligible. This leads to a significant simplification of the computational procedure. Therefore, we used this approach for calculating parameters (\ref{WaxeeME}) for $m_a \ge 10^5$~eV.
%
Finally, it is interesting to consider the high-mass ALP limit. In this case one can use the relation \cite{Maison:2022}:
\begin{equation} 
    \frac{e^{-m_a r}}{4 \pi r}  \approx  \frac{1}{m_a^2}  \delta(\boldsymbol{r}),  
\label{DeltSimpl}    
\end{equation}
which is valid for very high masses.
One can see from this Eq., that the ALP mass can be factorized out and the molecular effects (\ref{Etp_ee}) and (\ref{Etp_eN}) can be characterized by only corresponding ALP-mass-independent molecular parameters~\cite{Stadnik:2018,Maison:2022} (see below). As a convenient numerical implementation of the Dirac delta function we used the Gaussian geminal operator $c \cdot e^{-\alpha r^2}$, where $c$ is the normalization constant and in practice $\alpha$ should be much larger than any exponential parameter $\beta$ in primitive Gaussian basis functions. Using such a substitution we could employ the existing implementation of corresponding two-electron integrals in the LIBINT library.

Finally, after all primitive integrals $\langle ab|\frac{e^{-m_a r}}{r}|cd\rangle$ were calculated we performed 4-index transformation \cite{abe2004four} from the basis of primitive Gaussian functions to the basis of complex molecular bispinors taking into account the structure of $\gamma$ matrices.

The values of the $\WaxeN(m_a)$ molecular constants for different ALP masses were calculated using the relativistic CCSD(T) method within the Dirac-Coulomb Hamiltonian. In these calculations, 70 electrons of HfF$^+$ were included in the correlation treatment ($1s2s2p$ electrons of Hf were excluded), and the cutoff for virtual orbital energies was set to 1000~$E_h$. We used the Dyall's uncontracted AE4Z~\cite{Dyall:04,Dyall:12} basis set for Hf and AE2Z~\cite{Dyall:04,Dyall:2010,Dyall:2016} basis set for F; below we call this combination of basis sets as LBas.
In addition, we also calculated the following corrections to the obtained $\WaxeN$ values:
(i) Correction on higher-order electron correlation effects. It was calculated as the difference between the $\WaxeN$ values obtained at the coupled cluster with single double and iterative triple amplitudes, CCSDT~\cite{Oleynichenko:20}, level, and at the CCSD(T) one. Here 42 outer-core and valence electrons of HfF$^+$ were correlated ($1s..4p$ electrons of Hf and $1s$ electrons of F were excluded) and the basis set was reduced to the SBas set. The latter corresponds to the AE2Z~\cite{Dyall:04,Dyall:2010,Dyall:2016} on both atoms.
(ii) The influence of the inner $1s2s2p$ electrons of Hf (which were excluded in the leading calculation) was calculated at the CCSD(T) level using the SBas basis set. In this  calculation, the cutoff for virtual orbital energies was increased up to 10000 $E_h$~\cite{Skripnikov:17a, Skripnikov:15a}.
(iii) Finally, we estimated the contribution of the Gaunt electron-electron interaction effect as the difference between Dirac-Coulomb-Gaunt and Dirac-Coulomb results obtained within the CCSD(T) approach using the SBas basis set.
Calculations of $\Waxee(m_a)$ constant were much more challenging due to the two-electron nature of the considered operator (\ref{ee_potential}). Therefore, these calculations were performed at the CCSD(T) level, using the SBas basis set, and correlating all the electrons. Note, that for the case of YbOH molecule such level of theory was found to give quite accurate results~~\cite{maison2021axion}.
In all electronic structure calculations the experimental distance~\cite{Cossel:12} between Hf and F was used.

In all calculation we used the Gaussian nuclear charge distribution model which is well-suited for molecular problems~\cite{Visscher:1997}. Similar to Ref.~\cite{Stadnik:2018}, we considered the electron-nucleus interaction in the form~(\ref{eN_potential}), i.e. we do not consider finite-nuclear size effects in the operator~(\ref{eN_potential}). The related problems were studied e.g. in Refs.~\cite{sahoo2017improved,gharibnejad2015dark}.

Dirac-Hartree-Fock calculations, required to obtain the reference Slater determinant in Eq.~(\ref{expAnsatz}) were performed within the relativistic electronic structure package DIRAC ~\cite{DIRAC19,Saue:2020}. Correlation calculations were performed using DIRAC~\cite{DIRAC19,Saue:2020} and MRCC~\cite{MRCC2020,Kallay:1,Kallay:2} codes. Matrix elements~(\ref{WaxeNME}) were calculated using the code developed in work~\cite{Maison:2021}. To calculate matrix elements~(\ref{WaxeeME}) we used the code developed in the present work and in the LIBINT computational library~\cite{valeev2020libint}. The code to calculate the Gaunt interelectron interaction matrix elements over molecular bispinors was developed in works~\cite{Maison:2019,Maison:20a}.

\section{Results and discussion}

\begin{table*}
    \caption{The calculated values of $\WaxeN(m_a)$ molecular parameters 
    at various levels of electronic structure theory. The  ``Final'' column is the sum of CCSD(T) results and corrections described in the main text. The last column provides limits on $|\bar{g}_N^s g_e^p|$ derived from the experimental data~\cite{newlimit1} corresponding to ALP masses given in the first column.}
    \centering
    \begin{tabular*}{0.62\textwidth}{lccccr}
    \hline
    \hline
\multirow{2}{*}{$m_a$, eV} & \multicolumn{4}{c}{$\WaxeN(m_a)$, $m_e c / \hbar$} & \multirow{2}{*}{\begin{tabular}[c]{@{}l@{}}$|\bar{g}_N^s g_e^p|$ \\ limit, $\hbar c$\end{tabular}} \\ \cline{2-5}
                           & DHF   & CCSD  & CCSD(T)  & Final  &       \\
    \hline
    1 & $+1.20 \cdot 10^{-5}$& $+1.72 \cdot 10^{-5}$& $+1.68\cdot 10^{-5}$ & $+1.67\cdot10^{-5}$ & $1.11\cdot 10^{-20}$
    \\
    10 & $+1.20 \cdot 10^{-5}$& $+1.72 \cdot 10^{-5}$& $+1.68\cdot 10^{-5}$ & $+1.67\cdot10^{-5}$ & $1.11\cdot 10^{-20}$
    \\
    10$^2$ & $+1.19 \cdot 10^{-5}$& $+1.73 \cdot 10^{-5}$& $+1.68\cdot 10^{-5}$ & $+1.66\cdot10^{-5}$ & $1.11\cdot 10^{-20}$
    \\
    10$^3$ & $+1.14 \cdot 10^{-5}$& $+1.60 \cdot 10^{-5}$& $+1.56\cdot 10^{-5}$ & $+1.54\cdot10^{-5}$ & $1.19\cdot 10^{-20}$
    \\
    10$^4$ &$+2.87\cdot 10^{-6}$ & $+3.59\cdot 10^{-6}$& $+3.55\cdot 10^{-6}$ & $+3.30\cdot10^{-6}$ & $5.25\cdot 10^{-20}$
    \\
    10$^5$ & $-8.36\cdot 10^{-6}$ & $-1.14\cdot 10^{-5}$ & $-1.12\cdot 10^{-5}$ & $-1.15\cdot10^{-5}$ & $1.66\cdot 10^{-20}$
    \\
    10$^6$ & $-3.88\cdot 10^{-6}$ & $-6.43\cdot 10^{-6}$ & $-6.27\cdot 10^{-6}$ & $-6.41\cdot10^{-6}$ & $2.97\cdot 10^{-20}$
    \\
    10$^7$ & $-1.72 \cdot 10^{-7}$ & $-2.86 \cdot 10^{-7}$ & $-2.79 \cdot 10^{-7}$ & $-2.85\cdot10^{-7}$ & $6.67\cdot 10^{-19}$
    \\
    10$^8$ & $-3.19 \cdot 10^{-9}$ & $-5.33 \cdot 10^{-9}$ & $-5.19 \cdot 10^{-9}$ & $-5.30\cdot10^{-9}$ & $3.59\cdot 10^{-17}$
    \\
    10$^9$ & $-3.53 \cdot 10^{-11}$ & $-5.90 \cdot 10^{-11}$ & $-5.74 \cdot 10^{-11}$ & $-5.85\cdot10^{-11}$ & $3.24\cdot 10^{-15}$
    \\
    10$^{10}$ & $-3.54 \cdot 10^{-13}$ & $-5.91 \cdot 10^{-13}$ & $-5.75 \cdot 10^{-13}$ & $-5.87\cdot10^{-13}$ &  $3.24\cdot 10^{-13}$\\
    \hline
    \end{tabular*}
    \label{tab1}
\end{table*}

The calculated values of $\WaxeN$ constants for a wide range ($1\div 10^{10}$~eV) of ALP masses are given in Table~\ref{tab1} at different levels of theory: Dirac-Hartree-Fock (DHF), coupled cluster with single and double amplitudes (CCSD) and at the CCSD(T) level. The ``Final'' column of Table~\ref{tab1} gives the values that take into account corrections described in the previous section. It can be seen that electron correlation effects (i.e. effects beyond the DHF level of theory) are very important for description of $\WaxeN$ constants and their role increases with the transition from small masses to large ones. At the same time good convergence with respect to the correlation effects treatment was achieved.
Higher-order correlation effects beyond the CCSD(T) model contribute within 1-2\% for all ALP masses. According to our calculations, the contribution of corrections to the CCSD(T) result described above does not exceed 5\% in total for every mass considered.

Similar to the YbOH molecule case, the function $\WaxeN(m_a)$ changes its sign in the region between $m_a=10^4$~eV and $m_a=10^5$~eV.
Interestingly, the values of $\WaxeN(m_a)$ for the considered HfF$^+$ case are approximately twice smaller than $\WaxeN(m_a)$ for the YbOH case for all ALP masses. This fact can be related to the different values of $\Omega$ in Eq.~(\ref{WaxeNME}): $\Omega=1$ for HfF$^+$, while for YbOH $\Omega=1/2$. However, the energy shifts (\ref{Etp_eN}) of HfF$^+$ and YbOH are comparable.

The calculated values of $\Waxee(m_a)$ constants for a variety of ALP masses values ($1\div 10^{10}$~eV) are given in Table~\ref{tabWee}. It can be seen that electron correlation effects are more substantial for the case of high-mass ALPs than for the light ALPs.
One can note that for ALP masses below $m_a=10^2$~eV the values of $\Waxee(m_a)$ are almost equal. 
In this case $\Waxee$ is approximately twice smaller than $\Waxee$ for the YbOH case~\cite{maison2021axion}. Again it can be attributed to the different $\Omega$ values for YbOH and HfF$^+$. One can see a certain change of $\Waxee(m_a)$ near the point $m_a=10^3$~eV and the drop by the factor of 3 for $m_a=10^4$~eV and finally the change of sign for higher-mass ALPs as in the case of $\WaxeN$.

In Ref.~\cite{newlimit1} the constraint on the \Ti,\Par-violating energy shift was interpreted in terms of the electron electric dipole moment $\de$ according to the relation:
\begin{equation}
\label{Eedm}
    \delta E = \de \Omega W_d,
\end{equation}
where $E_{\rm eff}=W_d|\Omega|\approx 23$ GV/cm~\cite{Skripnikov:17c,Fleig:17,Petrov:07a,Petrov:09b}.
Using the relations (\ref{Etp_ee}) and (\ref{Etp_eN}) one can interpret the obtained constraint on the energy~\cite{newlimit1}:
\begin{equation}
\delta E \approx 23~\mu{\rm Hz}     
\label{newlim}
\end{equation}
in terms of limits on the product of ALP coupling constants. Derived limits for corresponding values of ALP masses are given in the last columns of Tables \ref{tab1} and \ref{tabWee}.
Let us also consider the following two limiting cases.

\textit{Low-mass limit.}
As it was noted above, it can be seen from Table \ref{tab1} that for the light ALPs ($m_a < 1\ \textrm{keV}$), the $\WaxeN(m_a)$ constant is almost independent of the $m_a$ value. Thus, the same value of $\WaxeN$ can be used for a very wide range of ALP masses. This can be explained by the fact that for light ALPs, the characteristic radius of the Yukawa-type interaction $R_{\textrm{Yu}} = 1/m_a (\textrm{relativistic units}) = \hbar/m_a c$ is significantly larger than the molecule size~\cite{Stadnik:2018}. For example, the characteristic molecular distance of $1$ Bohr corresponds to $m_a\approx 4$~keV. Thus, for such ALPs one can reinterpret the experimental data~\cite{newlimit1} in terms of the limit on $\bar{g}_{N}^s g_e^p$: $|\bar{g}_{N}^s g_e^p|/(\hbar c) \lesssim 1.1 \times 10^{-20}$.
This constraint is an order of magnitude better than the previous restriction for $|\bar{g}_{N}^s g_e^p|$, which was deduced from the ThO experiment~\cite{ACME:18, Stadnik:2018}. 

For the case of light ALPs one can also reinterpret experiment~\cite{newlimit1} in terms of the constraint on the 
$g_e^s g_e^p$ product: $|g_e^s g_e^p|/(\hbar c) \lesssim 2.2 \times 10^{-20}$.
This constraint is also an order of magnitude better than that deduced~\cite{Stadnik:2018} from the ThO experiment~\cite{ACME:18}.

\begin{table*}
    \caption{The calculated values of $\Waxee(m_a)$ molecular parameters at various levels of theory. 
    The last column provides limits on $|g_e^s g_e^p|$ product derived from the experimental data~\cite{newlimit1} corresponding to ALP masses given in the first column.}
    \centering
    \begin{tabular*}{0.55\textwidth}{lcccr}
    \hline
    \hline
\multirow{2}{*}{$m_a$, eV} & \multicolumn{3}{c}{$\Waxee(m_a)$, $m_ec/\hbar$} & \multirow{2}{*}{\begin{tabular}[c]{@{}l@{}}$|g_e^s g_e^p|$ \\ limit, $\hbar c$\end{tabular}} \\ \cline{2-4}
                           & DHF   & CCSD  & CCSD(T)  (Final)  &       \\                                                                                                  
    \hline
    1  & $+6.35 \cdot 10^{-6}$ &  $+8.83 \cdot 10^{-6}$ &     $+8.63\cdot 10^{-6}$ & $2.16\cdot 10^{-20}$ 
    \\
    10 & $+6.35 \cdot 10^{-6}$ &  $+8.83 \cdot 10^{-6}$ &     $+8.63\cdot 10^{-6}$ & $2.16\cdot 10^{-20}$
    \\
    $10^2$ & $+6.34 \cdot 10^{-6}$ &  $+8.81 \cdot 10^{-6}$ & $+8.61 \cdot 10^{-6}$ & $2.16\cdot 10^{-20}$
    \\
    $10^3$ & $+5.67 \cdot 10^{-6}$ &  $+7.81 \cdot 10^{-6}$ & $+7.64 \cdot 10^{-6}$  & $2.44\cdot 10^{-20}$
    \\
    $10^4$ & $+1.98 \cdot 10^{-6}$ &  $+2.49\cdot 10^{-6}$ & $+2.46 \cdot 10^{-6}$  & $7.57\cdot 10^{-20}$
    \\
    $10^5$ & $+7.73 \cdot 10^{-8}$ &  $+1.64\cdot 10^{-7}$ & $+1.59 \cdot 10^{-7}$  & $1.17\cdot 10^{-18}$
    \\
    $10^6$ & $-4.01 \cdot 10^{-9}$ &  $-5.77\cdot 10^{-9}$ & $-5.67 \cdot 10^{-9}$  & $3.28\cdot 10^{-17}$
    \\
    $10^7$ & $-6.83 \cdot 10^{-11}$ &  $-1.11\cdot 10^{-10}$ & $-1.08 \cdot 10^{-10}$  & $1.72\cdot 10^{-15}$
    \\
    $10^8$ & $-6.90\cdot 10^{-13}$ &  $-1.12\cdot 10^{-12}$ & $-1.09 \cdot 10^{-12}$  & $1.70\cdot 10^{-13}$
    \\
    $10^9$ & $-6.94 \cdot 10^{-15}$ &  $-1.12\cdot 10^{-14}$ & $-1.09 \cdot 10^{-14}$  & $1.69\cdot 10^{-11}$
    \\
    $10^{10}$ & $-6.97 \cdot 10^{-17}$& $-1.12\cdot 10^{-16}$& $-1.10 \cdot 10^{-16}$ & $1.67\cdot 10^{-9}$\\
    \hline
    \end{tabular*}
    \label{tabWee}
\end{table*}

\textit{High-mass limit.}
For the case of ALPs with high masses ($m_a\ge~1~\textrm{GeV}$) there is an approximate relation~\cite{Stadnik:2018,Maison:2021}: $\WaxeN(m_a) \simeq \widetilde{W}^{eN} m_a^{-2}$. The $\widetilde{W}^{eN}$ parameter does not depend on $m_a$. Therefore, in this case the energy shift (\ref{Etp_eN}) can be parameterized as follows~\cite{maison2021axion}:
\begin{equation} 
    \delta E \approx  \frac{\bar{g}_N^s g_e^p}{m_a^2} \Omega\widetilde{W}^{eN},
\end{equation}
where
\begin{equation}
\widetilde{W}^{eN} = \lim_{m_a \rightarrow +\infty} m_a^2  \WaxeN(m_a).
\end{equation}
So we have from Table \ref{tab1}:
$$
|\widetilde{W}^{eN}| \approx 5.87 \times 10^{-11} \ \textrm{GeV}^{2}\times \frac{m_e c }{\hbar}.    
$$
Thus, from the experimental constraint on the energy shift (\ref{newlim})~\cite{newlimit1} one can derive: $|\bar{g}_N^s g_e^p|/(\hbar c m_a^{2})~\lesssim 3.2 \cdot 10^{-15}\  \textrm{GeV}^{-2}$.
This constraint is slightly better than the constraint~\cite{Stadnik:2018}, which follows from the ThO experiment~\cite{ACME:18}. 

Similar dependence on ALP masses also holds for the electron-electron molecular parameter~(\ref{WaxeeME}) due to the relation~(\ref{DeltSimpl}):
\begin{equation}
\widetilde{W}^{ee} = \lim_{m_a \rightarrow +\infty} m_a^2  \Waxee(m_a).
\end{equation}
According to the present calculation 
$$
|\widetilde{W}^{ee}| \approx 1.1 \times 10^{-14} \ \textrm{GeV}^{2}\times \frac{m_e c }{\hbar}. 
$$
This corresponds to the following limit:
 $|g_e^s g_e^p|/(\hbar c m_a^{2}) \lesssim 1.7 \cdot 10^{-11}\  \textrm{GeV}^{-2}$.
 This is about three times better than the corresponding best constraint, which follows from the ThO experiment~\cite{Stadnik:2018,ACME:18}.
For convenience, Table \ref{TConstraints} summarizes obtained constraints for the cases of light and high-mass ALPs limits described above.

The combination of the experimental data obtained using the HfF$^+$ cation~\cite{newlimit1} and molecular parameters calculated in the present work allows one to derive updated laboratory constraints on the products of ALP coupling constants for ALP masses $m_a \ge 10^{-2}$~eV (see e.g. Fig. 2 of Ref.~\cite{Stadnik:2018} for a compilation of different constraints).
In Ref.~\cite{Hare:2020} it was reviewed that it is possible to obtain more restrictive indirect  bounds for the product of $g_{N}^s g_e^p$ by combining independent constraints on $g_{N}^s$ and $g_e^p$ from completely different sources, e.g. laboratory experiments and astrophysical stellar energy-loss bounds. However, as it was stressed~\cite{Hare:2020} there may be some mechanisms that can spoil the astrophysical bounds and therefore purely laboratory experiments of Earth are highly needed.

\begin{table}[!h]
\caption{Summary of the derived constraints on the combinations of ALP coupling constants for limiting cases of light- and high-mass ALPs.}
\centering
\begin{tabular}{ll}
\hline
\hline
Limit                                                        & Value                 \\
\hline
$|\bar{g}_{N}^s g_e^p|/(\hbar c)$, $m_a \ll 1\ \textrm{keV}$     & $1.1 \times 10^{-20}$ \\
$|g_e^s g_e^p|/(\hbar c)$, $m_a \ll 1\ \textrm{keV}$            & $2.2 \times 10^{-20}$ \\
$|\bar{g}_N^s g_e^p|/(\hbar c m_a^{2})$, $m_a \ge 1\ \textrm{GeV}$  & $3.2 \cdot 10^{-15}~\textrm{GeV}^{-2}$ \\
$|g_e^s g_e^p|/(\hbar c m_a^{2})$, $m_a \ge 1\ \textrm{GeV}$        & $1.7 \cdot 10^{-11}~\textrm{GeV}^{-2}$ \\
\hline
\hline
\end{tabular}
\label{TConstraints}
\end{table}

\section{Conclusions}
The effects of the \Ti,\Par-violating axionlike-particles-mediated scalar-pseudoscalar electron-electron and nucleus-electron interactions were studied for the HfF$^+$ cation in the metastable $^3\Delta_1$ electronic state. To solve this problem for the nucleus-electron interaction, we used the code developed previously~\cite{maison2021axion}, while to describe the electron-electron interaction we used the code developed in the present work. Our approach was applied to calculate molecular constants that characterize the interactions. They link the experimental constraint on the \Ti,\Par-violating energy shift with the products of the ALP-electron scalar and  ALP-electron pseudoscalar or ALP-nucleus scalar and ALP-electron pseudoscalar coupling constants. The obtained values of the constants were used to reinterpret the JILA experiment on the HfF$^+$ molecular cation~\cite{newlimit1} aimed to search for the electron electric dipole moment in terms of the products of ALP coupling constants. The updated constraints for a wide range of ALP masses were deduced from the experimental data~\cite{newlimit1}.

The developed procedure can be used for the interpretation of further experiments with paramagnetic molecules and atoms aimed to search for the \Ti,\Par-violating effects in terms of the scalar-pseudoscalar ALP-mediated interactions with arbitrary mass ALPs.

\begin{acknowledgments}
Electronic structure calculations were carried out using computing resources of the federal collective usage center Complex for Simulation and Data Processing for Mega-science Facilities at National Research Centre ``Kurchatov Institute'', http://ckp.nrcki.ru/, and partly using the computing resources of the quantum chemistry laboratory.

Molecular coupled cluster electronic structure calculations were supported by the Russian Science Foundation Grant No. 19-72-10019 (https://rscf.ru/project/22-72-41010/). Calculations of the $\Waxee(m_a)$ matrix elements and development of corresponding code were supported by the Foundation for the Advancement of Theoretical Physics and Mathematics ``BASIS'' Grant according to Projects No. 21-1-2-47-1 and 20-1-5-76-1.
\end{acknowledgments}


\begin{thebibliography}{124}%
\makeatletter
\providecommand \@ifxundefined [1]{%
 \@ifx{#1\undefined}
}%
\providecommand \@ifnum [1]{%
 \ifnum #1\expandafter \@firstoftwo
 \else \expandafter \@secondoftwo
 \fi
}%
\providecommand \@ifx [1]{%
 \ifx #1\expandafter \@firstoftwo
 \else \expandafter \@secondoftwo
 \fi
}%
\providecommand \natexlab [1]{#1}%
\providecommand \enquote  [1]{``#1''}%
\providecommand \bibnamefont  [1]{#1}%
\providecommand \bibfnamefont [1]{#1}%
\providecommand \citenamefont [1]{#1}%
\providecommand \href@noop [0]{\@secondoftwo}%
\providecommand \href [0]{\begingroup \@sanitize@url \@href}%
\providecommand \@href[1]{\@@startlink{#1}\@@href}%
\providecommand \@@href[1]{\endgroup#1\@@endlink}%
\providecommand \@sanitize@url [0]{\catcode `\\12\catcode `\$12\catcode
  `\&12\catcode `\#12\catcode `\^12\catcode `\_12\catcode `\%12\relax}%
\providecommand \@@startlink[1]{}%
\providecommand \@@endlink[0]{}%
\providecommand \url  [0]{\begingroup\@sanitize@url \@url }%
\providecommand \@url [1]{\endgroup\@href {#1}{\urlprefix }}%
\providecommand \urlprefix  [0]{URL }%
\providecommand \Eprint [0]{\href }%
\providecommand \doibase [0]{https://doi.org/}%
\providecommand \selectlanguage [0]{\@gobble}%
\providecommand \bibinfo  [0]{\@secondoftwo}%
\providecommand \bibfield  [0]{\@secondoftwo}%
\providecommand \translation [1]{[#1]}%
\providecommand \BibitemOpen [0]{}%
\providecommand \bibitemStop [0]{}%
\providecommand \bibitemNoStop [0]{.\EOS\space}%
\providecommand \EOS [0]{\spacefactor3000\relax}%
\providecommand \BibitemShut  [1]{\csname bibitem#1\endcsname}%
\let\auto@bib@innerbib\@empty
\bibitem [{\citenamefont {Safronova}\ \emph {et~al.}(2018)\citenamefont
  {Safronova}, \citenamefont {Budker}, \citenamefont {DeMille}, \citenamefont
  {Kimball}, \citenamefont {Derevianko},\ and\ \citenamefont
  {Clark}}]{Safronova:18}%
  \BibitemOpen
  \bibfield  {author} {\bibinfo {author} {\bibfnamefont {M.~S.}\ \bibnamefont
  {Safronova}}, \bibinfo {author} {\bibfnamefont {D.}~\bibnamefont {Budker}},
  \bibinfo {author} {\bibfnamefont {D.}~\bibnamefont {DeMille}}, \bibinfo
  {author} {\bibfnamefont {D.~F.~J.}\ \bibnamefont {Kimball}}, \bibinfo
  {author} {\bibfnamefont {A.}~\bibnamefont {Derevianko}},\ and\ \bibinfo
  {author} {\bibfnamefont {C.~W.}\ \bibnamefont {Clark}},\ }\bibfield  {title}
  {\bibinfo {title} {Search for new physics with atoms and molecules},\ }\href
  {https://doi.org/10.1103/RevModPhys.90.025008} {\bibfield  {journal}
  {\bibinfo  {journal} {Rev.\ Mod.\ Phys.}\ }\textbf {\bibinfo {volume} {90}},\
  \bibinfo {pages} {025008} (\bibinfo {year} {2018})}\BibitemShut {NoStop}%
\bibitem [{\citenamefont {Alarcon}\ \emph {et~al.}()\citenamefont {Alarcon},
  \citenamefont {Alexander}, \citenamefont {Anastassopoulos}, \citenamefont
  {Aoki}, \citenamefont {Baartman}, \citenamefont {Baeßler}, \citenamefont
  {Bartoszek}, \citenamefont {Beck}, \citenamefont {Bedeschi}, \citenamefont
  {Berger}, \citenamefont {Berz}, \citenamefont {Bethlem}, \citenamefont
  {Bhattacharya}, \citenamefont {Blaskiewicz}, \citenamefont {Blum},
  \citenamefont {Bowcock}, \citenamefont {Borschevsky}, \citenamefont {Brown},
  \citenamefont {Budker}, \citenamefont {Burdin}, \citenamefont {Casey},
  \citenamefont {Casse}, \citenamefont {Cantatore}, \citenamefont {Cheng},
  \citenamefont {Chupp}, \citenamefont {Cianciolo}, \citenamefont {Cirigliano},
  \citenamefont {Clayton}, \citenamefont {Crawford}, \citenamefont {Das},
  \citenamefont {Davoudiasl}, \citenamefont {de~Vries}, \citenamefont
  {DeMille}, \citenamefont {Denisov}, \citenamefont {Diwan}, \citenamefont
  {Doyle}, \citenamefont {Engel}, \citenamefont {Fanourakis}, \citenamefont
  {Fatemi}, \citenamefont {Filippone}, \citenamefont {Flambaum}, \citenamefont
  {Fleig}, \citenamefont {Fomin}, \citenamefont {Fischer}, \citenamefont
  {Gabrielse}, \citenamefont {Ruiz}, \citenamefont {Gardikiotis}, \citenamefont
  {Gatti}, \citenamefont {Geraci}, \citenamefont {Gooding}, \citenamefont
  {Golub}, \citenamefont {Graham}, \citenamefont {Gray}, \citenamefont
  {Griffith}, \citenamefont {Haciomeroglu}, \citenamefont {Gwinner},
  \citenamefont {Hoekstra}, \citenamefont {Hoffstaetter}, \citenamefont
  {Huang}, \citenamefont {Hutzler}, \citenamefont {Incagli}, \citenamefont
  {Ito}, \citenamefont {Izubuchi}, \citenamefont {Jayich}, \citenamefont
  {Jeong}, \citenamefont {Kaplan}, \citenamefont {Karuza}, \citenamefont
  {Kawall}, \citenamefont {Kim}, \citenamefont {Koop}, \citenamefont {Korsch},
  \citenamefont {Korobkina}, \citenamefont {Lebedev}, \citenamefont {Lee},
  \citenamefont {Lee}, \citenamefont {Lehnert}, \citenamefont {Leung},
  \citenamefont {Liu}, \citenamefont {Long}, \citenamefont {Lusiani},
  \citenamefont {Marciano}, \citenamefont {Maroudas}, \citenamefont
  {Matlashov}, \citenamefont {Matsumoto}, \citenamefont {Mawhorter},
  \citenamefont {Meot}, \citenamefont {Mereghetti}, \citenamefont {Miller},
  \citenamefont {Morse}, \citenamefont {Mott}, \citenamefont {Omarov},
  \citenamefont {Orozco}, \citenamefont {O'Shaughnessy}, \citenamefont {Ozben},
  \citenamefont {Park}, \citenamefont {Pattie}, \citenamefont {Petrov},
  \citenamefont {Piacentino}, \citenamefont {Plaster}, \citenamefont
  {Podobedov}, \citenamefont {Poelker}, \citenamefont {Pocanic}, \citenamefont
  {Prasannaa}, \citenamefont {Price}, \citenamefont {Ramsey-Musolf},
  \citenamefont {Raparia}, \citenamefont {Rajendran}, \citenamefont {Reece},
  \citenamefont {Reid}, \citenamefont {Rescia}, \citenamefont {Ritz},
  \citenamefont {Roberts}, \citenamefont {Safronova}, \citenamefont {Sakemi},
  \citenamefont {Schmidt-Wellenburg}, \citenamefont {Shindler}, \citenamefont
  {Semertzidis}, \citenamefont {Silenko}, \citenamefont {Singh}, \citenamefont
  {Skripnikov}, \citenamefont {Soni}, \citenamefont {Stephenson}, \citenamefont
  {Suleiman}, \citenamefont {Sunaga}, \citenamefont {Syphers}, \citenamefont
  {Syritsyn}, \citenamefont {Tarbutt}, \citenamefont {Thoerngren},
  \citenamefont {Timmermans}, \citenamefont {Tishchenko}, \citenamefont
  {Titov}, \citenamefont {Tsoupas}, \citenamefont {Tzamarias}, \citenamefont
  {Variola}, \citenamefont {Venanzoni}, \citenamefont {Vilella}, \citenamefont
  {Vossebeld}, \citenamefont {Winter}, \citenamefont {Won}, \citenamefont
  {Zelenski}, \citenamefont {Zelevinsky}, \citenamefont {Zhou},\ and\
  \citenamefont {Zioutas}}]{eEDM_snowmass:2022}%
  \BibitemOpen
  \bibfield  {author} {\bibinfo {author} {\bibfnamefont {R.}~\bibnamefont
  {Alarcon}}, \bibinfo {author} {\bibfnamefont {J.}~\bibnamefont {Alexander}},
  \bibinfo {author} {\bibfnamefont {V.}~\bibnamefont {Anastassopoulos}},
  \bibinfo {author} {\bibfnamefont {T.}~\bibnamefont {Aoki}}, \bibinfo {author}
  {\bibfnamefont {R.}~\bibnamefont {Baartman}}, \bibinfo {author}
  {\bibfnamefont {S.}~\bibnamefont {Baeßler}}, \bibinfo {author}
  {\bibfnamefont {L.}~\bibnamefont {Bartoszek}}, \bibinfo {author}
  {\bibfnamefont {D.~H.}\ \bibnamefont {Beck}}, \bibinfo {author}
  {\bibfnamefont {F.}~\bibnamefont {Bedeschi}}, \bibinfo {author}
  {\bibfnamefont {R.}~\bibnamefont {Berger}}, \bibinfo {author} {\bibfnamefont
  {M.}~\bibnamefont {Berz}}, \bibinfo {author} {\bibfnamefont {H.~L.}\
  \bibnamefont {Bethlem}}, \bibinfo {author} {\bibfnamefont {T.}~\bibnamefont
  {Bhattacharya}}, \bibinfo {author} {\bibfnamefont {M.}~\bibnamefont
  {Blaskiewicz}}, \bibinfo {author} {\bibfnamefont {T.}~\bibnamefont {Blum}},
  \bibinfo {author} {\bibfnamefont {T.}~\bibnamefont {Bowcock}}, \bibinfo
  {author} {\bibfnamefont {A.}~\bibnamefont {Borschevsky}}, \bibinfo {author}
  {\bibfnamefont {K.}~\bibnamefont {Brown}}, \bibinfo {author} {\bibfnamefont
  {D.}~\bibnamefont {Budker}}, \bibinfo {author} {\bibfnamefont
  {S.}~\bibnamefont {Burdin}}, \bibinfo {author} {\bibfnamefont {B.~C.}\
  \bibnamefont {Casey}}, \bibinfo {author} {\bibfnamefont {G.}~\bibnamefont
  {Casse}}, \bibinfo {author} {\bibfnamefont {G.}~\bibnamefont {Cantatore}},
  \bibinfo {author} {\bibfnamefont {L.}~\bibnamefont {Cheng}}, \bibinfo
  {author} {\bibfnamefont {T.}~\bibnamefont {Chupp}}, \bibinfo {author}
  {\bibfnamefont {V.}~\bibnamefont {Cianciolo}}, \bibinfo {author}
  {\bibfnamefont {V.}~\bibnamefont {Cirigliano}}, \bibinfo {author}
  {\bibfnamefont {S.~M.}\ \bibnamefont {Clayton}}, \bibinfo {author}
  {\bibfnamefont {C.}~\bibnamefont {Crawford}}, \bibinfo {author}
  {\bibfnamefont {B.~P.}\ \bibnamefont {Das}}, \bibinfo {author} {\bibfnamefont
  {H.}~\bibnamefont {Davoudiasl}}, \bibinfo {author} {\bibfnamefont
  {J.}~\bibnamefont {de~Vries}}, \bibinfo {author} {\bibfnamefont
  {D.}~\bibnamefont {DeMille}}, \bibinfo {author} {\bibfnamefont
  {D.}~\bibnamefont {Denisov}}, \bibinfo {author} {\bibfnamefont {M.~V.}\
  \bibnamefont {Diwan}}, \bibinfo {author} {\bibfnamefont {J.~M.}\ \bibnamefont
  {Doyle}}, \bibinfo {author} {\bibfnamefont {J.}~\bibnamefont {Engel}},
  \bibinfo {author} {\bibfnamefont {G.}~\bibnamefont {Fanourakis}}, \bibinfo
  {author} {\bibfnamefont {R.}~\bibnamefont {Fatemi}}, \bibinfo {author}
  {\bibfnamefont {B.~W.}\ \bibnamefont {Filippone}}, \bibinfo {author}
  {\bibfnamefont {V.~V.}\ \bibnamefont {Flambaum}}, \bibinfo {author}
  {\bibfnamefont {T.}~\bibnamefont {Fleig}}, \bibinfo {author} {\bibfnamefont
  {N.}~\bibnamefont {Fomin}}, \bibinfo {author} {\bibfnamefont
  {W.}~\bibnamefont {Fischer}}, \bibinfo {author} {\bibfnamefont
  {G.}~\bibnamefont {Gabrielse}}, \bibinfo {author} {\bibfnamefont {R.~F.~G.}\
  \bibnamefont {Ruiz}}, \bibinfo {author} {\bibfnamefont {A.}~\bibnamefont
  {Gardikiotis}}, \bibinfo {author} {\bibfnamefont {C.}~\bibnamefont {Gatti}},
  \bibinfo {author} {\bibfnamefont {A.}~\bibnamefont {Geraci}}, \bibinfo
  {author} {\bibfnamefont {J.}~\bibnamefont {Gooding}}, \bibinfo {author}
  {\bibfnamefont {B.}~\bibnamefont {Golub}}, \bibinfo {author} {\bibfnamefont
  {P.}~\bibnamefont {Graham}}, \bibinfo {author} {\bibfnamefont
  {F.}~\bibnamefont {Gray}}, \bibinfo {author} {\bibfnamefont {W.~C.}\
  \bibnamefont {Griffith}}, \bibinfo {author} {\bibfnamefont {S.}~\bibnamefont
  {Haciomeroglu}}, \bibinfo {author} {\bibfnamefont {G.}~\bibnamefont
  {Gwinner}}, \bibinfo {author} {\bibfnamefont {S.}~\bibnamefont {Hoekstra}},
  \bibinfo {author} {\bibfnamefont {G.~H.}\ \bibnamefont {Hoffstaetter}},
  \bibinfo {author} {\bibfnamefont {H.}~\bibnamefont {Huang}}, \bibinfo
  {author} {\bibfnamefont {N.~R.}\ \bibnamefont {Hutzler}}, \bibinfo {author}
  {\bibfnamefont {M.}~\bibnamefont {Incagli}}, \bibinfo {author} {\bibfnamefont
  {T.~M.}\ \bibnamefont {Ito}}, \bibinfo {author} {\bibfnamefont
  {T.}~\bibnamefont {Izubuchi}}, \bibinfo {author} {\bibfnamefont {A.~M.}\
  \bibnamefont {Jayich}}, \bibinfo {author} {\bibfnamefont {H.}~\bibnamefont
  {Jeong}}, \bibinfo {author} {\bibfnamefont {D.}~\bibnamefont {Kaplan}},
  \bibinfo {author} {\bibfnamefont {M.}~\bibnamefont {Karuza}}, \bibinfo
  {author} {\bibfnamefont {D.}~\bibnamefont {Kawall}}, \bibinfo {author}
  {\bibfnamefont {O.}~\bibnamefont {Kim}}, \bibinfo {author} {\bibfnamefont
  {I.}~\bibnamefont {Koop}}, \bibinfo {author} {\bibfnamefont {W.}~\bibnamefont
  {Korsch}}, \bibinfo {author} {\bibfnamefont {E.}~\bibnamefont {Korobkina}},
  \bibinfo {author} {\bibfnamefont {V.}~\bibnamefont {Lebedev}}, \bibinfo
  {author} {\bibfnamefont {J.}~\bibnamefont {Lee}}, \bibinfo {author}
  {\bibfnamefont {S.}~\bibnamefont {Lee}}, \bibinfo {author} {\bibfnamefont
  {R.}~\bibnamefont {Lehnert}}, \bibinfo {author} {\bibfnamefont {K.~K.~H.}\
  \bibnamefont {Leung}}, \bibinfo {author} {\bibfnamefont {C.-Y.}\ \bibnamefont
  {Liu}}, \bibinfo {author} {\bibfnamefont {J.}~\bibnamefont {Long}}, \bibinfo
  {author} {\bibfnamefont {A.}~\bibnamefont {Lusiani}}, \bibinfo {author}
  {\bibfnamefont {W.~J.}\ \bibnamefont {Marciano}}, \bibinfo {author}
  {\bibfnamefont {M.}~\bibnamefont {Maroudas}}, \bibinfo {author}
  {\bibfnamefont {A.}~\bibnamefont {Matlashov}}, \bibinfo {author}
  {\bibfnamefont {N.}~\bibnamefont {Matsumoto}}, \bibinfo {author}
  {\bibfnamefont {R.}~\bibnamefont {Mawhorter}}, \bibinfo {author}
  {\bibfnamefont {F.}~\bibnamefont {Meot}}, \bibinfo {author} {\bibfnamefont
  {E.}~\bibnamefont {Mereghetti}}, \bibinfo {author} {\bibfnamefont {J.~P.}\
  \bibnamefont {Miller}}, \bibinfo {author} {\bibfnamefont {W.~M.}\
  \bibnamefont {Morse}}, \bibinfo {author} {\bibfnamefont {J.}~\bibnamefont
  {Mott}}, \bibinfo {author} {\bibfnamefont {Z.}~\bibnamefont {Omarov}},
  \bibinfo {author} {\bibfnamefont {L.~A.}\ \bibnamefont {Orozco}}, \bibinfo
  {author} {\bibfnamefont {C.~M.}\ \bibnamefont {O'Shaughnessy}}, \bibinfo
  {author} {\bibfnamefont {C.}~\bibnamefont {Ozben}}, \bibinfo {author}
  {\bibfnamefont {S.}~\bibnamefont {Park}}, \bibinfo {author} {\bibfnamefont
  {R.~W.}\ \bibnamefont {Pattie}}, \bibinfo {author} {\bibfnamefont {A.~N.}\
  \bibnamefont {Petrov}}, \bibinfo {author} {\bibfnamefont {G.~M.}\
  \bibnamefont {Piacentino}}, \bibinfo {author} {\bibfnamefont {B.~R.}\
  \bibnamefont {Plaster}}, \bibinfo {author} {\bibfnamefont {B.}~\bibnamefont
  {Podobedov}}, \bibinfo {author} {\bibfnamefont {M.}~\bibnamefont {Poelker}},
  \bibinfo {author} {\bibfnamefont {D.}~\bibnamefont {Pocanic}}, \bibinfo
  {author} {\bibfnamefont {V.~S.}\ \bibnamefont {Prasannaa}}, \bibinfo {author}
  {\bibfnamefont {J.}~\bibnamefont {Price}}, \bibinfo {author} {\bibfnamefont
  {M.~J.}\ \bibnamefont {Ramsey-Musolf}}, \bibinfo {author} {\bibfnamefont
  {D.}~\bibnamefont {Raparia}}, \bibinfo {author} {\bibfnamefont
  {S.}~\bibnamefont {Rajendran}}, \bibinfo {author} {\bibfnamefont
  {M.}~\bibnamefont {Reece}}, \bibinfo {author} {\bibfnamefont
  {A.}~\bibnamefont {Reid}}, \bibinfo {author} {\bibfnamefont {S.}~\bibnamefont
  {Rescia}}, \bibinfo {author} {\bibfnamefont {A.}~\bibnamefont {Ritz}},
  \bibinfo {author} {\bibfnamefont {B.~L.}\ \bibnamefont {Roberts}}, \bibinfo
  {author} {\bibfnamefont {M.~S.}\ \bibnamefont {Safronova}}, \bibinfo {author}
  {\bibfnamefont {Y.}~\bibnamefont {Sakemi}}, \bibinfo {author} {\bibfnamefont
  {P.}~\bibnamefont {Schmidt-Wellenburg}}, \bibinfo {author} {\bibfnamefont
  {A.}~\bibnamefont {Shindler}}, \bibinfo {author} {\bibfnamefont {Y.~K.}\
  \bibnamefont {Semertzidis}}, \bibinfo {author} {\bibfnamefont
  {A.}~\bibnamefont {Silenko}}, \bibinfo {author} {\bibfnamefont {J.~T.}\
  \bibnamefont {Singh}}, \bibinfo {author} {\bibfnamefont {L.~V.}\ \bibnamefont
  {Skripnikov}}, \bibinfo {author} {\bibfnamefont {A.}~\bibnamefont {Soni}},
  \bibinfo {author} {\bibfnamefont {E.}~\bibnamefont {Stephenson}}, \bibinfo
  {author} {\bibfnamefont {R.}~\bibnamefont {Suleiman}}, \bibinfo {author}
  {\bibfnamefont {A.}~\bibnamefont {Sunaga}}, \bibinfo {author} {\bibfnamefont
  {M.}~\bibnamefont {Syphers}}, \bibinfo {author} {\bibfnamefont
  {S.}~\bibnamefont {Syritsyn}}, \bibinfo {author} {\bibfnamefont {M.~R.}\
  \bibnamefont {Tarbutt}}, \bibinfo {author} {\bibfnamefont {P.}~\bibnamefont
  {Thoerngren}}, \bibinfo {author} {\bibfnamefont {R.~G.~E.}\ \bibnamefont
  {Timmermans}}, \bibinfo {author} {\bibfnamefont {V.}~\bibnamefont
  {Tishchenko}}, \bibinfo {author} {\bibfnamefont {A.~V.}\ \bibnamefont
  {Titov}}, \bibinfo {author} {\bibfnamefont {N.}~\bibnamefont {Tsoupas}},
  \bibinfo {author} {\bibfnamefont {S.}~\bibnamefont {Tzamarias}}, \bibinfo
  {author} {\bibfnamefont {A.}~\bibnamefont {Variola}}, \bibinfo {author}
  {\bibfnamefont {G.}~\bibnamefont {Venanzoni}}, \bibinfo {author}
  {\bibfnamefont {E.}~\bibnamefont {Vilella}}, \bibinfo {author} {\bibfnamefont
  {J.}~\bibnamefont {Vossebeld}}, \bibinfo {author} {\bibfnamefont
  {P.}~\bibnamefont {Winter}}, \bibinfo {author} {\bibfnamefont
  {E.}~\bibnamefont {Won}}, \bibinfo {author} {\bibfnamefont {A.}~\bibnamefont
  {Zelenski}}, \bibinfo {author} {\bibfnamefont {T.}~\bibnamefont
  {Zelevinsky}}, \bibinfo {author} {\bibfnamefont {Y.}~\bibnamefont {Zhou}},\
  and\ \bibinfo {author} {\bibfnamefont {K.}~\bibnamefont {Zioutas}},\
  }\bibfield  {title} {\bibinfo {title} {Electric dipole moments and the search
  for new physics},\ }\bibinfo {note} {arXiv:2203.08103
  [hep-ph](2022)}\BibitemShut {NoStop}%
\bibitem [{\citenamefont {Safronova}(2023)}]{Safronova2023}%
  \BibitemOpen
  \bibfield  {author} {\bibinfo {author} {\bibfnamefont {M.~S.}\ \bibnamefont
  {Safronova}},\ }\bibinfo {title} {Searches for new physics},\ in\ \href
  {https://doi.org/10.1007/978-3-030-73893-8_32} {\emph {\bibinfo {booktitle}
  {Springer Handbook of Atomic, Molecular, and Optical Physics}}},\ \bibinfo
  {editor} {edited by\ \bibinfo {editor} {\bibfnamefont {G.~W.~F.}\
  \bibnamefont {Drake}}}\ (\bibinfo  {publisher} {Springer International
  Publishing},\ \bibinfo {address} {Cham},\ \bibinfo {year} {2023})\ pp.\
  \bibinfo {pages} {471--484}\BibitemShut {NoStop}%
\bibitem [{\citenamefont {Ema}\ \emph {et~al.}(2022)\citenamefont {Ema},
  \citenamefont {Gao},\ and\ \citenamefont {Pospelov}}]{eEDMLimit:2022}%
  \BibitemOpen
  \bibfield  {author} {\bibinfo {author} {\bibfnamefont {Y.}~\bibnamefont
  {Ema}}, \bibinfo {author} {\bibfnamefont {T.}~\bibnamefont {Gao}},\ and\
  \bibinfo {author} {\bibfnamefont {M.}~\bibnamefont {Pospelov}},\ }\bibfield
  {title} {\bibinfo {title} {Standard model prediction for paramagnetic
  electric dipole moments},\ }\href
  {https://doi.org/10.1103/PhysRevLett.129.231801} {\bibfield  {journal}
  {\bibinfo  {journal} {Phys. Rev. Lett.}\ }\textbf {\bibinfo {volume} {129}},\
  \bibinfo {pages} {231801} (\bibinfo {year} {2022})}\BibitemShut {NoStop}%
\bibitem [{\citenamefont {Yamaguchi}\ and\ \citenamefont
  {Yamanaka}(2021)}]{Yamaguchi:2021}%
  \BibitemOpen
  \bibfield  {author} {\bibinfo {author} {\bibfnamefont {Y.}~\bibnamefont
  {Yamaguchi}}\ and\ \bibinfo {author} {\bibfnamefont {N.}~\bibnamefont
  {Yamanaka}},\ }\bibfield  {title} {\bibinfo {title} {Quark level and hadronic
  contributions to the electric dipole moment of charged leptons in the
  standard model},\ }\href {https://doi.org/10.1103/PhysRevD.103.013001}
  {\bibfield  {journal} {\bibinfo  {journal} {Phys. Rev. D}\ }\textbf {\bibinfo
  {volume} {103}},\ \bibinfo {pages} {013001} (\bibinfo {year}
  {2021})}\BibitemShut {NoStop}%
\bibitem [{\citenamefont {Engel}\ \emph {et~al.}(2013)\citenamefont {Engel},
  \citenamefont {Ramsey-Musolf},\ and\ \citenamefont {van Kolck}}]{Engel:2013}%
  \BibitemOpen
  \bibfield  {author} {\bibinfo {author} {\bibfnamefont {J.}~\bibnamefont
  {Engel}}, \bibinfo {author} {\bibfnamefont {M.~J.}\ \bibnamefont
  {Ramsey-Musolf}},\ and\ \bibinfo {author} {\bibfnamefont {U.}~\bibnamefont
  {van Kolck}},\ }\bibfield  {title} {\bibinfo {title} {Electric dipole moments
  of nucleons, nuclei, and atoms: The standard model and beyond},\ }\href
  {https://doi.org/https://doi.org/10.1016/j.ppnp.2013.03.003} {\bibfield
  {journal} {\bibinfo  {journal} {Progr. Part. Nuc. Phys.}\ }\textbf {\bibinfo
  {volume} {71}},\ \bibinfo {pages} {21 } (\bibinfo {year} {2013})}\BibitemShut
  {NoStop}%
\bibitem [{\citenamefont {Chubukov}\ and\ \citenamefont
  {Labzowsky}(2016)}]{Chubukov:2016}%
  \BibitemOpen
  \bibfield  {author} {\bibinfo {author} {\bibfnamefont {D.~V.}\ \bibnamefont
  {Chubukov}}\ and\ \bibinfo {author} {\bibfnamefont {L.~N.}\ \bibnamefont
  {Labzowsky}},\ }\bibfield  {title} {\bibinfo {title}
  {$\mathcal{P},\mathcal{T}$-odd electron-nucleus interaction in atomic systems
  as an exchange by higgs bosons},\ }\href
  {https://doi.org/10.1103/PhysRevA.93.062503} {\bibfield  {journal} {\bibinfo
  {journal} {Phys. Rev. A}\ }\textbf {\bibinfo {volume} {93}},\ \bibinfo
  {pages} {062503} (\bibinfo {year} {2016})}\BibitemShut {NoStop}%
\bibitem [{\citenamefont {Commins}(1998)}]{Commins:98}%
  \BibitemOpen
  \bibfield  {author} {\bibinfo {author} {\bibfnamefont {E.~D.}\ \bibnamefont
  {Commins}},\ }\bibfield  {title} {\bibinfo {title} {Electric dipole moments
  of leptons},\ }\href {https://doi.org/10.1016/S1049-250X(08)60110-X}
  {\bibfield  {journal} {\bibinfo  {journal} {Adv.\ At.\ Mol.\ Opt.\ Phys.}\
  }\textbf {\bibinfo {volume} {40}},\ \bibinfo {pages} {1} (\bibinfo {year}
  {1998})}\BibitemShut {NoStop}%
\bibitem [{\citenamefont {Chupp}\ \emph {et~al.}(2019)\citenamefont {Chupp},
  \citenamefont {Fierlinger}, \citenamefont {Ramsey-Musolf},\ and\
  \citenamefont {Singh}}]{Chupp:2019}%
  \BibitemOpen
  \bibfield  {author} {\bibinfo {author} {\bibfnamefont {T.~E.}\ \bibnamefont
  {Chupp}}, \bibinfo {author} {\bibfnamefont {P.}~\bibnamefont {Fierlinger}},
  \bibinfo {author} {\bibfnamefont {M.~J.}\ \bibnamefont {Ramsey-Musolf}},\
  and\ \bibinfo {author} {\bibfnamefont {J.~T.}\ \bibnamefont {Singh}},\
  }\bibfield  {title} {\bibinfo {title} {Electric dipole moments of atoms,
  molecules, nuclei, and particles},\ }\href
  {https://doi.org/10.1103/RevModPhys.91.015001} {\bibfield  {journal}
  {\bibinfo  {journal} {Rev. Mod. Phys.}\ }\textbf {\bibinfo {volume} {91}},\
  \bibinfo {pages} {015001} (\bibinfo {year} {2019})}\BibitemShut {NoStop}%
\bibitem [{\citenamefont {Roussy}\ \emph {et~al.}(2022)\citenamefont {Roussy},
  \citenamefont {Caldwell}, \citenamefont {Wright}, \citenamefont {Cairncross},
  \citenamefont {Shagam}, \citenamefont {Ng}, \citenamefont {Schlossberger},
  \citenamefont {Park}, \citenamefont {Wang}, \citenamefont {Ye},\ and\
  \citenamefont {Cornell}}]{newlimit1}%
  \BibitemOpen
  \bibfield  {author} {\bibinfo {author} {\bibfnamefont {T.~S.}\ \bibnamefont
  {Roussy}}, \bibinfo {author} {\bibfnamefont {L.}~\bibnamefont {Caldwell}},
  \bibinfo {author} {\bibfnamefont {T.}~\bibnamefont {Wright}}, \bibinfo
  {author} {\bibfnamefont {W.~B.}\ \bibnamefont {Cairncross}}, \bibinfo
  {author} {\bibfnamefont {Y.}~\bibnamefont {Shagam}}, \bibinfo {author}
  {\bibfnamefont {K.~B.}\ \bibnamefont {Ng}}, \bibinfo {author} {\bibfnamefont
  {N.}~\bibnamefont {Schlossberger}}, \bibinfo {author} {\bibfnamefont {S.~Y.}\
  \bibnamefont {Park}}, \bibinfo {author} {\bibfnamefont {A.}~\bibnamefont
  {Wang}}, \bibinfo {author} {\bibfnamefont {J.}~\bibnamefont {Ye}},\ and\
  \bibinfo {author} {\bibfnamefont {E.~A.}\ \bibnamefont {Cornell}},\
  }\href@noop {} {\bibinfo {title} {A new bound on the electron's electric
  dipole moment}} (\bibinfo {year} {2022}),\ \Eprint
  {https://arxiv.org/abs/2212.11841} {arXiv:2212.11841 [physics.atom-ph]}
  \BibitemShut {NoStop}%
\bibitem [{\citenamefont {Labzowsky}(1978)}]{Labzowsky:78}%
  \BibitemOpen
  \bibfield  {author} {\bibinfo {author} {\bibfnamefont {L.~N.}\ \bibnamefont
  {Labzowsky}},\ }\bibfield  {title} {\bibinfo {title} {{$\Lambda$} doubling
  and parity nonconservation effects in the spectra of diatomic molecules},\
  }\href@noop {} {\bibfield  {journal} {\bibinfo  {journal} {Sov.\ Phys.\ --\
  JETP}\ }\textbf {\bibinfo {volume} {48}},\ \bibinfo {pages} {434} (\bibinfo
  {year} {1978})}\BibitemShut {NoStop}%
\bibitem [{\citenamefont {Sushkov}\ and\ \citenamefont
  {Flambaum}(1978)}]{Sushkov:78}%
  \BibitemOpen
  \bibfield  {author} {\bibinfo {author} {\bibfnamefont {O.~P.}\ \bibnamefont
  {Sushkov}}\ and\ \bibinfo {author} {\bibfnamefont {V.~V.}\ \bibnamefont
  {Flambaum}},\ }\bibfield  {title} {\bibinfo {title} {Parity breaking effects
  in diatomic molecules},\ }\href@noop {} {\bibfield  {journal} {\bibinfo
  {journal} {Sov.\ Phys.\ --\ JETP}\ }\textbf {\bibinfo {volume} {48}},\
  \bibinfo {pages} {608} (\bibinfo {year} {1978})}\BibitemShut {NoStop}%
\bibitem [{\citenamefont {DeMille}\ \emph {et~al.}(2001)\citenamefont
  {DeMille}, \citenamefont {Bay}, \citenamefont {Bickman}, \citenamefont
  {Kawall}, \citenamefont {Hunter}, \citenamefont {Krause}, \citenamefont
  {Maxwell},\ and\ \citenamefont {Ulmer}}]{ComminsFest}%
  \BibitemOpen
  \bibfield  {author} {\bibinfo {author} {\bibfnamefont {D.}~\bibnamefont
  {DeMille}}, \bibinfo {author} {\bibfnamefont {F.}~\bibnamefont {Bay}},
  \bibinfo {author} {\bibfnamefont {S.}~\bibnamefont {Bickman}}, \bibinfo
  {author} {\bibfnamefont {D.}~\bibnamefont {Kawall}}, \bibinfo {author}
  {\bibfnamefont {L.}~\bibnamefont {Hunter}}, \bibinfo {author} {\bibfnamefont
  {D.}~\bibnamefont {Krause}}, \bibinfo {author} {\bibfnamefont
  {S.}~\bibnamefont {Maxwell}},\ and\ \bibinfo {author} {\bibfnamefont
  {K.}~\bibnamefont {Ulmer}},\ }\href@noop {} {\emph {\bibinfo {title} {Art and
  Symmetry in Experimental Physics: Festschrift for Eugene D. Commins}}}\
  (\bibinfo  {publisher} {AIP Conf. Proc. {\bf 596}, ed. D.Budker,
  P.H.Buck\-sbaum, and S.J.Freedman, Melville, NY, p. 72},\ \bibinfo {year}
  {2001})\BibitemShut {NoStop}%
\bibitem [{\citenamefont {Kawall}\ \emph {et~al.}(2004)\citenamefont {Kawall},
  \citenamefont {Bay}, \citenamefont {Bickman}, \citenamefont {Jiang},\ and\
  \citenamefont {DeMille}}]{Kawall:04b}%
  \BibitemOpen
  \bibfield  {author} {\bibinfo {author} {\bibfnamefont {D.}~\bibnamefont
  {Kawall}}, \bibinfo {author} {\bibfnamefont {F.}~\bibnamefont {Bay}},
  \bibinfo {author} {\bibfnamefont {S.}~\bibnamefont {Bickman}}, \bibinfo
  {author} {\bibfnamefont {Y.}~\bibnamefont {Jiang}},\ and\ \bibinfo {author}
  {\bibfnamefont {D.}~\bibnamefont {DeMille}},\ }\bibfield  {title} {\bibinfo
  {title} {Precision zeeman-stark spectroscopy of the metastable
  {$a(1)[^3\Sigma^+]$} state of {PbO}},\ }\href
  {https://doi.org/10.1103/PhysRevLett.92.133007} {\bibfield  {journal}
  {\bibinfo  {journal} {Phys.\ Rev.\ Lett.}\ }\textbf {\bibinfo {volume}
  {92}},\ \bibinfo {pages} {133007} (\bibinfo {year} {2004})}\BibitemShut
  {NoStop}%
\bibitem [{\citenamefont {Vutha}\ \emph {et~al.}(2010)\citenamefont {Vutha},
  \citenamefont {Campbell}, \citenamefont {Gurevich}, \citenamefont {Hutzler},
  \citenamefont {Parsons}, \citenamefont {Patterson}, \citenamefont {Petrik},
  \citenamefont {Spaun}, \citenamefont {Doyle}, \citenamefont {Gabrielse},\
  and\ \citenamefont {DeMille}}]{Vutha:2010}%
  \BibitemOpen
  \bibfield  {author} {\bibinfo {author} {\bibfnamefont {A.~C.}\ \bibnamefont
  {Vutha}}, \bibinfo {author} {\bibfnamefont {W.~C.}\ \bibnamefont {Campbell}},
  \bibinfo {author} {\bibfnamefont {Y.~V.}\ \bibnamefont {Gurevich}}, \bibinfo
  {author} {\bibfnamefont {N.~R.}\ \bibnamefont {Hutzler}}, \bibinfo {author}
  {\bibfnamefont {M.}~\bibnamefont {Parsons}}, \bibinfo {author} {\bibfnamefont
  {D.}~\bibnamefont {Patterson}}, \bibinfo {author} {\bibfnamefont
  {E.}~\bibnamefont {Petrik}}, \bibinfo {author} {\bibfnamefont
  {B.}~\bibnamefont {Spaun}}, \bibinfo {author} {\bibfnamefont {J.~M.}\
  \bibnamefont {Doyle}}, \bibinfo {author} {\bibfnamefont {G.}~\bibnamefont
  {Gabrielse}},\ and\ \bibinfo {author} {\bibfnamefont {D.}~\bibnamefont
  {DeMille}},\ }\bibfield  {title} {\bibinfo {title} {Search for the electric
  dipole moment of the electron with thorium monoxide},\ }\href
  {https://doi.org/10.1088/0953-4075/43/7/074007} {\bibfield  {journal}
  {\bibinfo  {journal} {J.\ Phys.\ B}\ }\textbf {\bibinfo {volume} {43}},\
  \bibinfo {pages} {074007} (\bibinfo {year} {2010})}\BibitemShut {NoStop}%
\bibitem [{\citenamefont {Caldwell}\ \emph {et~al.}(2022)\citenamefont
  {Caldwell}, \citenamefont {Roussy}, \citenamefont {Wright}, \citenamefont
  {Cairncross}, \citenamefont {Shagam}, \citenamefont {Ng}, \citenamefont
  {Schlossberger}, \citenamefont {Park}, \citenamefont {Wang}, \citenamefont
  {Ye},\ and\ \citenamefont {Cornell}}]{caldwell2022systematic}%
  \BibitemOpen
  \bibfield  {author} {\bibinfo {author} {\bibfnamefont {L.}~\bibnamefont
  {Caldwell}}, \bibinfo {author} {\bibfnamefont {T.~S.}\ \bibnamefont
  {Roussy}}, \bibinfo {author} {\bibfnamefont {T.}~\bibnamefont {Wright}},
  \bibinfo {author} {\bibfnamefont {W.~B.}\ \bibnamefont {Cairncross}},
  \bibinfo {author} {\bibfnamefont {Y.}~\bibnamefont {Shagam}}, \bibinfo
  {author} {\bibfnamefont {K.~B.}\ \bibnamefont {Ng}}, \bibinfo {author}
  {\bibfnamefont {N.}~\bibnamefont {Schlossberger}}, \bibinfo {author}
  {\bibfnamefont {S.~Y.}\ \bibnamefont {Park}}, \bibinfo {author}
  {\bibfnamefont {A.}~\bibnamefont {Wang}}, \bibinfo {author} {\bibfnamefont
  {J.}~\bibnamefont {Ye}},\ and\ \bibinfo {author} {\bibfnamefont {E.~A.}\
  \bibnamefont {Cornell}},\ }\href@noop {} {\bibinfo {title} {Systematic and
  statistical uncertainty evaluation of the {HfF$^+$} electron electric dipole
  moment experiment}} (\bibinfo {year} {2022}),\ \Eprint
  {https://arxiv.org/abs/2212.11837} {arXiv:2212.11837 [physics.atom-ph]}
  \BibitemShut {NoStop}%
\bibitem [{\citenamefont {Andreev}\ \emph {et~al.}(2018)\citenamefont
  {Andreev}, \citenamefont {Ang}, \citenamefont {DeMille}, \citenamefont
  {Doyle}, \citenamefont {Gabrielse}, \citenamefont {Haefner}, \citenamefont
  {Hutzler}, \citenamefont {Lasner}, \citenamefont {Meisenhelder},
  \citenamefont {O'Leary} \emph {et~al.}}]{ACME:18}%
  \BibitemOpen
  \bibfield  {author} {\bibinfo {author} {\bibfnamefont {V.}~\bibnamefont
  {Andreev}}, \bibinfo {author} {\bibfnamefont {D.}~\bibnamefont {Ang}},
  \bibinfo {author} {\bibfnamefont {D.}~\bibnamefont {DeMille}}, \bibinfo
  {author} {\bibfnamefont {J.}~\bibnamefont {Doyle}}, \bibinfo {author}
  {\bibfnamefont {G.}~\bibnamefont {Gabrielse}}, \bibinfo {author}
  {\bibfnamefont {J.}~\bibnamefont {Haefner}}, \bibinfo {author} {\bibfnamefont
  {N.}~\bibnamefont {Hutzler}}, \bibinfo {author} {\bibfnamefont
  {Z.}~\bibnamefont {Lasner}}, \bibinfo {author} {\bibfnamefont
  {C.}~\bibnamefont {Meisenhelder}}, \bibinfo {author} {\bibfnamefont
  {B.}~\bibnamefont {O'Leary}}, \emph {et~al.},\ }\bibfield  {title} {\bibinfo
  {title} {Improved limit on the electric dipole moment of the electron},\
  }\href {https://doi.org/10.1038/s41586-018-0599-8} {\bibfield  {journal}
  {\bibinfo  {journal} {Nature}\ }\textbf {\bibinfo {volume} {562}},\ \bibinfo
  {pages} {355} (\bibinfo {year} {2018})}\BibitemShut {NoStop}%
\bibitem [{\citenamefont {Gorshkow}\ \emph {et~al.}(1979)\citenamefont
  {Gorshkow}, \citenamefont {Labzovsky},\ and\ \citenamefont
  {Moskalyov}}]{Gorshkov:79}%
  \BibitemOpen
  \bibfield  {author} {\bibinfo {author} {\bibfnamefont {V.~G.}\ \bibnamefont
  {Gorshkow}}, \bibinfo {author} {\bibfnamefont {L.~N.}\ \bibnamefont
  {Labzovsky}},\ and\ \bibinfo {author} {\bibfnamefont {A.~N.}\ \bibnamefont
  {Moskalyov}},\ }\bibfield  {title} {\bibinfo {title} {Space and time parity
  nonconservation effects in the spectra of diatomic molecules},\ }\href@noop
  {} {\bibfield  {journal} {\bibinfo  {journal} {Sov.\ Phys.\ --\ JETP}\
  }\textbf {\bibinfo {volume} {49}},\ \bibinfo {pages} {209} (\bibinfo {year}
  {1979})}\BibitemShut {NoStop}%
\bibitem [{\citenamefont {Jung}(2013)}]{Jung:13}%
  \BibitemOpen
  \bibfield  {author} {\bibinfo {author} {\bibfnamefont {M.}~\bibnamefont
  {Jung}},\ }\bibfield  {title} {\bibinfo {title} {A robust limit for the
  electric dipole moment of the electron},\ }\href
  {https://doi.org/10.1007/JHEP05(2013)168} {\bibfield  {journal} {\bibinfo
  {journal} {J. High Energy Phys.}\ }\textbf {\bibinfo {volume} {2013}}\bibinfo
   {number} { (5)},\ \bibinfo {pages} {168}}\BibitemShut {NoStop}%
\bibitem [{\citenamefont {Pospelov}\ and\ \citenamefont
  {Ritz}(2014)}]{Pospelov:14}%
  \BibitemOpen
\bibfield  {number} {  }\bibfield  {author} {\bibinfo {author} {\bibfnamefont
  {M.}~\bibnamefont {Pospelov}}\ and\ \bibinfo {author} {\bibfnamefont
  {A.}~\bibnamefont {Ritz}},\ }\bibfield  {title} {\bibinfo {title} {Ckm
  benchmarks for electron electric dipole moment experiments},\ }\href
  {https://doi.org/10.1103/PhysRevD.89.056006} {\bibfield  {journal} {\bibinfo
  {journal} {Phys.\ Rev.\ D}\ }\textbf {\bibinfo {volume} {89}},\ \bibinfo
  {pages} {056006} (\bibinfo {year} {2014})}\BibitemShut {NoStop}%
\bibitem [{\citenamefont {Fleig}\ and\ \citenamefont
  {Skripnikov}(2020)}]{Skripnikov:2020b}%
  \BibitemOpen
  \bibfield  {author} {\bibinfo {author} {\bibfnamefont {T.}~\bibnamefont
  {Fleig}}\ and\ \bibinfo {author} {\bibfnamefont {L.~V.}\ \bibnamefont
  {Skripnikov}},\ }\bibfield  {title} {\bibinfo {title} {{P,T}-violating and
  magnetic hyperfine interactions in atomic thallium},\ }\href
  {https://doi.org/10.3390/sym12040498} {\bibfield  {journal} {\bibinfo
  {journal} {Symmetry}\ }\textbf {\bibinfo {volume} {12}},\ \bibinfo {pages}
  {498} (\bibinfo {year} {2020})}\BibitemShut {NoStop}%
\bibitem [{\citenamefont {Gresh}\ \emph {et~al.}(2016)\citenamefont {Gresh},
  \citenamefont {Cossel}, \citenamefont {Zhou}, \citenamefont {Ye},\ and\
  \citenamefont {Cornell}}]{ThFp:2016}%
  \BibitemOpen
  \bibfield  {author} {\bibinfo {author} {\bibfnamefont {D.~N.}\ \bibnamefont
  {Gresh}}, \bibinfo {author} {\bibfnamefont {K.~C.}\ \bibnamefont {Cossel}},
  \bibinfo {author} {\bibfnamefont {Y.}~\bibnamefont {Zhou}}, \bibinfo {author}
  {\bibfnamefont {J.}~\bibnamefont {Ye}},\ and\ \bibinfo {author}
  {\bibfnamefont {E.~A.}\ \bibnamefont {Cornell}},\ }\bibfield  {title}
  {\bibinfo {title} {Broadband velocity modulation spectroscopy of {ThF$^+$}
  for use in a measurement of the electron electric dipole moment},\ }\href
  {https://doi.org/https://doi.org/10.1016/j.jms.2015.11.001} {\bibfield
  {journal} {\bibinfo  {journal} {J. Mol. Spectrosc.}\ }\textbf {\bibinfo
  {volume} {319}},\ \bibinfo {pages} {1} (\bibinfo {year} {2016})}\BibitemShut
  {NoStop}%
\bibitem [{\citenamefont {Kozyryev}\ and\ \citenamefont
  {Hutzler}(2017)}]{kozyryev2017precision}%
  \BibitemOpen
  \bibfield  {author} {\bibinfo {author} {\bibfnamefont {I.}~\bibnamefont
  {Kozyryev}}\ and\ \bibinfo {author} {\bibfnamefont {N.~R.}\ \bibnamefont
  {Hutzler}},\ }\bibfield  {title} {\bibinfo {title} {Precision measurement of
  time-reversal symmetry violation with laser-cooled polyatomic molecules},\
  }\href {https://doi.org/10.1103/PhysRevLett.119.133002} {\bibfield  {journal}
  {\bibinfo  {journal} {Phys.\ Rev.\ Lett.}\ }\textbf {\bibinfo {volume}
  {119}},\ \bibinfo {pages} {133002} (\bibinfo {year} {2017})}\BibitemShut
  {NoStop}%
\bibitem [{\citenamefont {Isaev}\ \emph {et~al.}(2017)\citenamefont {Isaev},
  \citenamefont {Zaitsevskii},\ and\ \citenamefont {Eliav}}]{isaev2017laser}%
  \BibitemOpen
  \bibfield  {author} {\bibinfo {author} {\bibfnamefont {T.~A.}\ \bibnamefont
  {Isaev}}, \bibinfo {author} {\bibfnamefont {A.~V.}\ \bibnamefont
  {Zaitsevskii}},\ and\ \bibinfo {author} {\bibfnamefont {E.}~\bibnamefont
  {Eliav}},\ }\bibfield  {title} {\bibinfo {title} {Laser-coolable polyatomic
  molecules with heavy nuclei},\ }\href
  {https://doi.org/10.1088/1361-6455/aa8f34} {\bibfield  {journal} {\bibinfo
  {journal} {J. Phys. B: At. Mol. Opt. Phys.}\ }\textbf {\bibinfo {volume}
  {50}},\ \bibinfo {pages} {225101} (\bibinfo {year} {2017})}\BibitemShut
  {NoStop}%
\bibitem [{\citenamefont {Aggarwal}\ \emph {et~al.}(2018)\citenamefont
  {Aggarwal}, \citenamefont {Bethlem}, \citenamefont {Borschevsky},
  \citenamefont {Denis}, \citenamefont {Esajas}, \citenamefont {Haase},
  \citenamefont {Hao}, \citenamefont {Hoekstra}, \citenamefont {Jungmann},
  \citenamefont {Meijknecht} \emph {et~al.}}]{BaF:2018}%
  \BibitemOpen
  \bibfield  {author} {\bibinfo {author} {\bibfnamefont {P.}~\bibnamefont
  {Aggarwal}}, \bibinfo {author} {\bibfnamefont {H.~L.}\ \bibnamefont
  {Bethlem}}, \bibinfo {author} {\bibfnamefont {A.}~\bibnamefont
  {Borschevsky}}, \bibinfo {author} {\bibfnamefont {M.}~\bibnamefont {Denis}},
  \bibinfo {author} {\bibfnamefont {K.}~\bibnamefont {Esajas}}, \bibinfo
  {author} {\bibfnamefont {P.~A.~B.}\ \bibnamefont {Haase}}, \bibinfo {author}
  {\bibfnamefont {Y.}~\bibnamefont {Hao}}, \bibinfo {author} {\bibfnamefont
  {S.}~\bibnamefont {Hoekstra}}, \bibinfo {author} {\bibfnamefont
  {K.}~\bibnamefont {Jungmann}}, \bibinfo {author} {\bibfnamefont {T.~B.}\
  \bibnamefont {Meijknecht}}, \emph {et~al.},\ }\bibfield  {title} {\bibinfo
  {title} {Measuring the electric dipole moment of the electron in baf},\
  }\href {https://doi.org/10.1140/epjd/e2018-90192-9} {\bibfield  {journal}
  {\bibinfo  {journal} {Eur. Phys. J. D}\ }\textbf {\bibinfo {volume} {72}},\
  \bibinfo {pages} {197} (\bibinfo {year} {2018})}\BibitemShut {NoStop}%
\bibitem [{\citenamefont {Fitch}\ \emph {et~al.}(2020)\citenamefont {Fitch},
  \citenamefont {Lim}, \citenamefont {Hinds}, \citenamefont {Sauer},\ and\
  \citenamefont {Tarbutt}}]{YbF:2020}%
  \BibitemOpen
  \bibfield  {author} {\bibinfo {author} {\bibfnamefont {N.~J.}\ \bibnamefont
  {Fitch}}, \bibinfo {author} {\bibfnamefont {J.}~\bibnamefont {Lim}}, \bibinfo
  {author} {\bibfnamefont {E.~A.}\ \bibnamefont {Hinds}}, \bibinfo {author}
  {\bibfnamefont {B.~E.}\ \bibnamefont {Sauer}},\ and\ \bibinfo {author}
  {\bibfnamefont {M.~R.}\ \bibnamefont {Tarbutt}},\ }\bibfield  {title}
  {\bibinfo {title} {Methods for measuring the electron’s electric dipole
  moment using ultracold ybf molecules},\ }\href
  {https://doi.org/10.1088/2058-9565/abc931} {\bibfield  {journal} {\bibinfo
  {journal} {Quantum Sci. Technol.}\ }\textbf {\bibinfo {volume} {6}},\
  \bibinfo {pages} {014006} (\bibinfo {year} {2020})}\BibitemShut {NoStop}%
\bibitem [{\citenamefont {Maison}\ \emph {et~al.}(2022)\citenamefont {Maison},
  \citenamefont {Skripnikov}, \citenamefont {Penyazkov}, \citenamefont {Grau},\
  and\ \citenamefont {Petrov}}]{Maison:2022b}%
  \BibitemOpen
  \bibfield  {author} {\bibinfo {author} {\bibfnamefont {D.~E.}\ \bibnamefont
  {Maison}}, \bibinfo {author} {\bibfnamefont {L.~V.}\ \bibnamefont
  {Skripnikov}}, \bibinfo {author} {\bibfnamefont {G.}~\bibnamefont
  {Penyazkov}}, \bibinfo {author} {\bibfnamefont {M.}~\bibnamefont {Grau}},\
  and\ \bibinfo {author} {\bibfnamefont {A.~N.}\ \bibnamefont {Petrov}},\
  }\bibfield  {title} {\bibinfo {title} {$\mathcal{T},\mathcal{P}$-odd effects
  in the ${\mathrm{luoh}}^{+}$ cation},\ }\href
  {https://doi.org/10.1103/PhysRevA.106.062827} {\bibfield  {journal} {\bibinfo
   {journal} {Phys. Rev. A}\ }\textbf {\bibinfo {volume} {106}},\ \bibinfo
  {pages} {062827} (\bibinfo {year} {2022})}\BibitemShut {NoStop}%
\bibitem [{\citenamefont {Christenson}\ \emph {et~al.}(1964)\citenamefont
  {Christenson}, \citenamefont {Cronin}, \citenamefont {Fitch},\ and\
  \citenamefont {Turlay}}]{Christenson:64}%
  \BibitemOpen
  \bibfield  {author} {\bibinfo {author} {\bibfnamefont {J.~H.}\ \bibnamefont
  {Christenson}}, \bibinfo {author} {\bibfnamefont {J.~W.}\ \bibnamefont
  {Cronin}}, \bibinfo {author} {\bibfnamefont {V.~L.}\ \bibnamefont {Fitch}},\
  and\ \bibinfo {author} {\bibfnamefont {R.}~\bibnamefont {Turlay}},\
  }\bibfield  {title} {\bibinfo {title} {Evidence for the {$2\pi$} decay of the
  {$K_2^0$} meson},\ }\href {https://doi.org/10.1103/PhysRevLett.13.138}
  {\bibfield  {journal} {\bibinfo  {journal} {Phys.\ Rev.\ Lett.}\ }\textbf
  {\bibinfo {volume} {13}},\ \bibinfo {pages} {138} (\bibinfo {year}
  {1964})}\BibitemShut {NoStop}%
\bibitem [{\citenamefont {Abel}\ \emph {et~al.}(2020)\citenamefont {Abel},
  \citenamefont {Afach}, \citenamefont {Ayres}, \citenamefont {Baker},
  \citenamefont {Ban}, \citenamefont {Bison}, \citenamefont {Bodek},
  \citenamefont {Bondar}, \citenamefont {Burghoff}, \citenamefont {Chanel},
  \citenamefont {Chowdhuri}, \citenamefont {Chiu}, \citenamefont {Clement},
  \citenamefont {Crawford}, \citenamefont {Daum}, \citenamefont {Emmenegger},
  \citenamefont {Ferraris-Bouchez}, \citenamefont {Fertl}, \citenamefont
  {Flaux}, \citenamefont {Franke}, \citenamefont {Fratangelo}, \citenamefont
  {Geltenbort}, \citenamefont {Green}, \citenamefont {Griffith}, \citenamefont
  {van~der Grinten}, \citenamefont {Gruji\ifmmode~\acute{c}\else \'{c}\fi{}},
  \citenamefont {Harris}, \citenamefont {Hayen}, \citenamefont {Heil},
  \citenamefont {Henneck}, \citenamefont {H\'elaine}, \citenamefont {Hild},
  \citenamefont {Hodge}, \citenamefont {Horras}, \citenamefont {Iaydjiev},
  \citenamefont {Ivanov}, \citenamefont {Kasprzak}, \citenamefont {Kermaidic},
  \citenamefont {Kirch}, \citenamefont {Knecht}, \citenamefont {Knowles},
  \citenamefont {Koch}, \citenamefont {Koss}, \citenamefont {Komposch},
  \citenamefont {Kozela}, \citenamefont {Kraft}, \citenamefont {Krempel},
  \citenamefont {Ku\ifmmode~\acute{z}\else \'{z}\fi{}niak}, \citenamefont
  {Lauss}, \citenamefont {Lefort}, \citenamefont {Lemi\`ere}, \citenamefont
  {Leredde}, \citenamefont {Mohanmurthy}, \citenamefont {Mtchedlishvili},
  \citenamefont {Musgrave}, \citenamefont {Naviliat-Cuncic}, \citenamefont
  {Pais}, \citenamefont {Piegsa}, \citenamefont {Pierre}, \citenamefont
  {Pignol}, \citenamefont {Plonka-Spehr}, \citenamefont {Prashanth},
  \citenamefont {Qu\'em\'ener}, \citenamefont {Rawlik}, \citenamefont
  {Rebreyend}, \citenamefont {Rien\"acker}, \citenamefont {Ries}, \citenamefont
  {Roccia}, \citenamefont {Rogel}, \citenamefont {Rozpedzik}, \citenamefont
  {Schnabel}, \citenamefont {Schmidt-Wellenburg}, \citenamefont {Severijns},
  \citenamefont {Shiers}, \citenamefont {Tavakoli~Dinani}, \citenamefont
  {Thorne}, \citenamefont {Virot}, \citenamefont {Voigt}, \citenamefont {Weis},
  \citenamefont {Wursten}, \citenamefont {Wyszynski}, \citenamefont {Zejma},
  \citenamefont {Zenner},\ and\ \citenamefont
  {Zsigmond}}]{abel2020_neutronEDM}%
  \BibitemOpen
  \bibfield  {author} {\bibinfo {author} {\bibfnamefont {C.}~\bibnamefont
  {Abel}}, \bibinfo {author} {\bibfnamefont {S.}~\bibnamefont {Afach}},
  \bibinfo {author} {\bibfnamefont {N.~J.}\ \bibnamefont {Ayres}}, \bibinfo
  {author} {\bibfnamefont {C.~A.}\ \bibnamefont {Baker}}, \bibinfo {author}
  {\bibfnamefont {G.}~\bibnamefont {Ban}}, \bibinfo {author} {\bibfnamefont
  {G.}~\bibnamefont {Bison}}, \bibinfo {author} {\bibfnamefont
  {K.}~\bibnamefont {Bodek}}, \bibinfo {author} {\bibfnamefont
  {V.}~\bibnamefont {Bondar}}, \bibinfo {author} {\bibfnamefont
  {M.}~\bibnamefont {Burghoff}}, \bibinfo {author} {\bibfnamefont
  {E.}~\bibnamefont {Chanel}}, \bibinfo {author} {\bibfnamefont
  {Z.}~\bibnamefont {Chowdhuri}}, \bibinfo {author} {\bibfnamefont {P.-J.}\
  \bibnamefont {Chiu}}, \bibinfo {author} {\bibfnamefont {B.}~\bibnamefont
  {Clement}}, \bibinfo {author} {\bibfnamefont {C.~B.}\ \bibnamefont
  {Crawford}}, \bibinfo {author} {\bibfnamefont {M.}~\bibnamefont {Daum}},
  \bibinfo {author} {\bibfnamefont {S.}~\bibnamefont {Emmenegger}}, \bibinfo
  {author} {\bibfnamefont {L.}~\bibnamefont {Ferraris-Bouchez}}, \bibinfo
  {author} {\bibfnamefont {M.}~\bibnamefont {Fertl}}, \bibinfo {author}
  {\bibfnamefont {P.}~\bibnamefont {Flaux}}, \bibinfo {author} {\bibfnamefont
  {B.}~\bibnamefont {Franke}}, \bibinfo {author} {\bibfnamefont
  {A.}~\bibnamefont {Fratangelo}}, \bibinfo {author} {\bibfnamefont
  {P.}~\bibnamefont {Geltenbort}}, \bibinfo {author} {\bibfnamefont
  {K.}~\bibnamefont {Green}}, \bibinfo {author} {\bibfnamefont {W.~C.}\
  \bibnamefont {Griffith}}, \bibinfo {author} {\bibfnamefont {M.}~\bibnamefont
  {van~der Grinten}}, \bibinfo {author} {\bibfnamefont {Z.~D.}\ \bibnamefont
  {Gruji\ifmmode~\acute{c}\else \'{c}\fi{}}}, \bibinfo {author} {\bibfnamefont
  {P.~G.}\ \bibnamefont {Harris}}, \bibinfo {author} {\bibfnamefont
  {L.}~\bibnamefont {Hayen}}, \bibinfo {author} {\bibfnamefont
  {W.}~\bibnamefont {Heil}}, \bibinfo {author} {\bibfnamefont {R.}~\bibnamefont
  {Henneck}}, \bibinfo {author} {\bibfnamefont {V.}~\bibnamefont {H\'elaine}},
  \bibinfo {author} {\bibfnamefont {N.}~\bibnamefont {Hild}}, \bibinfo {author}
  {\bibfnamefont {Z.}~\bibnamefont {Hodge}}, \bibinfo {author} {\bibfnamefont
  {M.}~\bibnamefont {Horras}}, \bibinfo {author} {\bibfnamefont
  {P.}~\bibnamefont {Iaydjiev}}, \bibinfo {author} {\bibfnamefont {S.~N.}\
  \bibnamefont {Ivanov}}, \bibinfo {author} {\bibfnamefont {M.}~\bibnamefont
  {Kasprzak}}, \bibinfo {author} {\bibfnamefont {Y.}~\bibnamefont {Kermaidic}},
  \bibinfo {author} {\bibfnamefont {K.}~\bibnamefont {Kirch}}, \bibinfo
  {author} {\bibfnamefont {A.}~\bibnamefont {Knecht}}, \bibinfo {author}
  {\bibfnamefont {P.}~\bibnamefont {Knowles}}, \bibinfo {author} {\bibfnamefont
  {H.-C.}\ \bibnamefont {Koch}}, \bibinfo {author} {\bibfnamefont {P.~A.}\
  \bibnamefont {Koss}}, \bibinfo {author} {\bibfnamefont {S.}~\bibnamefont
  {Komposch}}, \bibinfo {author} {\bibfnamefont {A.}~\bibnamefont {Kozela}},
  \bibinfo {author} {\bibfnamefont {A.}~\bibnamefont {Kraft}}, \bibinfo
  {author} {\bibfnamefont {J.}~\bibnamefont {Krempel}}, \bibinfo {author}
  {\bibfnamefont {M.}~\bibnamefont {Ku\ifmmode~\acute{z}\else \'{z}\fi{}niak}},
  \bibinfo {author} {\bibfnamefont {B.}~\bibnamefont {Lauss}}, \bibinfo
  {author} {\bibfnamefont {T.}~\bibnamefont {Lefort}}, \bibinfo {author}
  {\bibfnamefont {Y.}~\bibnamefont {Lemi\`ere}}, \bibinfo {author}
  {\bibfnamefont {A.}~\bibnamefont {Leredde}}, \bibinfo {author} {\bibfnamefont
  {P.}~\bibnamefont {Mohanmurthy}}, \bibinfo {author} {\bibfnamefont
  {A.}~\bibnamefont {Mtchedlishvili}}, \bibinfo {author} {\bibfnamefont
  {M.}~\bibnamefont {Musgrave}}, \bibinfo {author} {\bibfnamefont
  {O.}~\bibnamefont {Naviliat-Cuncic}}, \bibinfo {author} {\bibfnamefont
  {D.}~\bibnamefont {Pais}}, \bibinfo {author} {\bibfnamefont {F.~M.}\
  \bibnamefont {Piegsa}}, \bibinfo {author} {\bibfnamefont {E.}~\bibnamefont
  {Pierre}}, \bibinfo {author} {\bibfnamefont {G.}~\bibnamefont {Pignol}},
  \bibinfo {author} {\bibfnamefont {C.}~\bibnamefont {Plonka-Spehr}}, \bibinfo
  {author} {\bibfnamefont {P.~N.}\ \bibnamefont {Prashanth}}, \bibinfo {author}
  {\bibfnamefont {G.}~\bibnamefont {Qu\'em\'ener}}, \bibinfo {author}
  {\bibfnamefont {M.}~\bibnamefont {Rawlik}}, \bibinfo {author} {\bibfnamefont
  {D.}~\bibnamefont {Rebreyend}}, \bibinfo {author} {\bibfnamefont
  {I.}~\bibnamefont {Rien\"acker}}, \bibinfo {author} {\bibfnamefont
  {D.}~\bibnamefont {Ries}}, \bibinfo {author} {\bibfnamefont {S.}~\bibnamefont
  {Roccia}}, \bibinfo {author} {\bibfnamefont {G.}~\bibnamefont {Rogel}},
  \bibinfo {author} {\bibfnamefont {D.}~\bibnamefont {Rozpedzik}}, \bibinfo
  {author} {\bibfnamefont {A.}~\bibnamefont {Schnabel}}, \bibinfo {author}
  {\bibfnamefont {P.}~\bibnamefont {Schmidt-Wellenburg}}, \bibinfo {author}
  {\bibfnamefont {N.}~\bibnamefont {Severijns}}, \bibinfo {author}
  {\bibfnamefont {D.}~\bibnamefont {Shiers}}, \bibinfo {author} {\bibfnamefont
  {R.}~\bibnamefont {Tavakoli~Dinani}}, \bibinfo {author} {\bibfnamefont
  {J.~A.}\ \bibnamefont {Thorne}}, \bibinfo {author} {\bibfnamefont
  {R.}~\bibnamefont {Virot}}, \bibinfo {author} {\bibfnamefont
  {J.}~\bibnamefont {Voigt}}, \bibinfo {author} {\bibfnamefont
  {A.}~\bibnamefont {Weis}}, \bibinfo {author} {\bibfnamefont {E.}~\bibnamefont
  {Wursten}}, \bibinfo {author} {\bibfnamefont {G.}~\bibnamefont {Wyszynski}},
  \bibinfo {author} {\bibfnamefont {J.}~\bibnamefont {Zejma}}, \bibinfo
  {author} {\bibfnamefont {J.}~\bibnamefont {Zenner}},\ and\ \bibinfo {author}
  {\bibfnamefont {G.}~\bibnamefont {Zsigmond}},\ }\bibfield  {title} {\bibinfo
  {title} {Measurement of the permanent electric dipole moment of the
  neutron},\ }\href {https://doi.org/10.1103/PhysRevLett.124.081803} {\bibfield
   {journal} {\bibinfo  {journal} {Phys. Rev. Lett.}\ }\textbf {\bibinfo
  {volume} {124}},\ \bibinfo {pages} {081803} (\bibinfo {year}
  {2020})}\BibitemShut {NoStop}%
\bibitem [{\citenamefont {Sakharov}(1967)}]{Sakharov1967}%
  \BibitemOpen
  \bibfield  {author} {\bibinfo {author} {\bibfnamefont {A.~D.}\ \bibnamefont
  {Sakharov}},\ }\bibfield  {title} {\bibinfo {title} {{Violation of CP
  Invariance, C Asymmetry, and Baryon Asymmetry of the Universe}},\ }\href@noop
  {} {\bibfield  {journal} {\bibinfo  {journal} {JETP Lett.}\ }\textbf
  {\bibinfo {volume} {5}},\ \bibinfo {pages} {27} (\bibinfo {year}
  {1967})}\BibitemShut {NoStop}%
\bibitem [{\citenamefont {Dine}\ and\ \citenamefont
  {Kusenko}(2003)}]{MatterAntimatter:2003}%
  \BibitemOpen
  \bibfield  {author} {\bibinfo {author} {\bibfnamefont {M.}~\bibnamefont
  {Dine}}\ and\ \bibinfo {author} {\bibfnamefont {A.}~\bibnamefont {Kusenko}},\
  }\bibfield  {title} {\bibinfo {title} {Origin of the matter-antimatter
  asymmetry},\ }\href {https://doi.org/10.1103/RevModPhys.76.1} {\bibfield
  {journal} {\bibinfo  {journal} {Rev. Mod. Phys.}\ }\textbf {\bibinfo {volume}
  {76}},\ \bibinfo {pages} {1} (\bibinfo {year} {2003})}\BibitemShut {NoStop}%
\bibitem [{\citenamefont {Drukier}\ \emph {et~al.}(1986)\citenamefont
  {Drukier}, \citenamefont {Freese},\ and\ \citenamefont
  {Spergel}}]{Drukier1986}%
  \BibitemOpen
  \bibfield  {author} {\bibinfo {author} {\bibfnamefont {A.~K.}\ \bibnamefont
  {Drukier}}, \bibinfo {author} {\bibfnamefont {K.}~\bibnamefont {Freese}},\
  and\ \bibinfo {author} {\bibfnamefont {D.~N.}\ \bibnamefont {Spergel}},\
  }\bibfield  {title} {\bibinfo {title} {Detecting cold dark-matter
  candidates},\ }\href {https://doi.org/10.1103/PhysRevD.33.3495} {\bibfield
  {journal} {\bibinfo  {journal} {Phys. Rev. D}\ }\textbf {\bibinfo {volume}
  {33}},\ \bibinfo {pages} {3495} (\bibinfo {year} {1986})}\BibitemShut
  {NoStop}%
\bibitem [{\citenamefont {Ahmed}\ \emph {et~al.}(2011)\citenamefont {Ahmed},
  \citenamefont {Akerib}, \citenamefont {Arrenberg}, \citenamefont {Bailey},
  \citenamefont {Balakishiyeva}, \citenamefont {Baudis}, \citenamefont {Bauer},
  \citenamefont {Brink}, \citenamefont {Bruch}, \citenamefont {Bunker},
  \citenamefont {Cabrera}, \citenamefont {Caldwell}, \citenamefont {Cooley},
  \citenamefont {do~Couto~e Silva}, \citenamefont {Cushman}, \citenamefont
  {Daal}, \citenamefont {DeJongh}, \citenamefont {Di~Stefano}, \citenamefont
  {Dragowsky}, \citenamefont {Duong}, \citenamefont {Fallows}, \citenamefont
  {Figueroa-Feliciano}, \citenamefont {Filippini}, \citenamefont {Fox},
  \citenamefont {Fritts}, \citenamefont {Golwala}, \citenamefont {Hall},
  \citenamefont {Hennings-Yeomans}, \citenamefont {Hertel}, \citenamefont
  {Holmgren}, \citenamefont {Hsu}, \citenamefont {Huber}, \citenamefont
  {Kamaev}, \citenamefont {Kiveni}, \citenamefont {Kos}, \citenamefont {Leman},
  \citenamefont {Liu}, \citenamefont {Mahapatra}, \citenamefont {Mandic},
  \citenamefont {McCarthy}, \citenamefont {Mirabolfathi}, \citenamefont
  {Moore}, \citenamefont {Nelson}, \citenamefont {Ogburn}, \citenamefont
  {Phipps}, \citenamefont {Pyle}, \citenamefont {Qiu}, \citenamefont {Ramberg},
  \citenamefont {Rau}, \citenamefont {Reisetter}, \citenamefont {Resch},
  \citenamefont {Saab}, \citenamefont {Sadoulet}, \citenamefont {Sander},
  \citenamefont {Schnee}, \citenamefont {Seitz}, \citenamefont {Serfass},
  \citenamefont {Sundqvist}, \citenamefont {Tarka}, \citenamefont {Wikus},
  \citenamefont {Yellin}, \citenamefont {Yoo}, \citenamefont {Young},\ and\
  \citenamefont {Zhang}}]{Ahmed2011}%
  \BibitemOpen
  \bibfield  {author} {\bibinfo {author} {\bibfnamefont {Z.}~\bibnamefont
  {Ahmed}}, \bibinfo {author} {\bibfnamefont {D.~S.}\ \bibnamefont {Akerib}},
  \bibinfo {author} {\bibfnamefont {S.}~\bibnamefont {Arrenberg}}, \bibinfo
  {author} {\bibfnamefont {C.~N.}\ \bibnamefont {Bailey}}, \bibinfo {author}
  {\bibfnamefont {D.}~\bibnamefont {Balakishiyeva}}, \bibinfo {author}
  {\bibfnamefont {L.}~\bibnamefont {Baudis}}, \bibinfo {author} {\bibfnamefont
  {D.~A.}\ \bibnamefont {Bauer}}, \bibinfo {author} {\bibfnamefont {P.~L.}\
  \bibnamefont {Brink}}, \bibinfo {author} {\bibfnamefont {T.}~\bibnamefont
  {Bruch}}, \bibinfo {author} {\bibfnamefont {R.}~\bibnamefont {Bunker}},
  \bibinfo {author} {\bibfnamefont {B.}~\bibnamefont {Cabrera}}, \bibinfo
  {author} {\bibfnamefont {D.~O.}\ \bibnamefont {Caldwell}}, \bibinfo {author}
  {\bibfnamefont {J.}~\bibnamefont {Cooley}}, \bibinfo {author} {\bibfnamefont
  {E.}~\bibnamefont {do~Couto~e Silva}}, \bibinfo {author} {\bibfnamefont
  {P.}~\bibnamefont {Cushman}}, \bibinfo {author} {\bibfnamefont
  {M.}~\bibnamefont {Daal}}, \bibinfo {author} {\bibfnamefont {F.}~\bibnamefont
  {DeJongh}}, \bibinfo {author} {\bibfnamefont {P.}~\bibnamefont {Di~Stefano}},
  \bibinfo {author} {\bibfnamefont {M.~R.}\ \bibnamefont {Dragowsky}}, \bibinfo
  {author} {\bibfnamefont {L.}~\bibnamefont {Duong}}, \bibinfo {author}
  {\bibfnamefont {S.}~\bibnamefont {Fallows}}, \bibinfo {author} {\bibfnamefont
  {E.}~\bibnamefont {Figueroa-Feliciano}}, \bibinfo {author} {\bibfnamefont
  {J.}~\bibnamefont {Filippini}}, \bibinfo {author} {\bibfnamefont
  {J.}~\bibnamefont {Fox}}, \bibinfo {author} {\bibfnamefont {M.}~\bibnamefont
  {Fritts}}, \bibinfo {author} {\bibfnamefont {S.~R.}\ \bibnamefont {Golwala}},
  \bibinfo {author} {\bibfnamefont {J.}~\bibnamefont {Hall}}, \bibinfo {author}
  {\bibfnamefont {R.}~\bibnamefont {Hennings-Yeomans}}, \bibinfo {author}
  {\bibfnamefont {S.~A.}\ \bibnamefont {Hertel}}, \bibinfo {author}
  {\bibfnamefont {D.}~\bibnamefont {Holmgren}}, \bibinfo {author}
  {\bibfnamefont {L.}~\bibnamefont {Hsu}}, \bibinfo {author} {\bibfnamefont
  {M.~E.}\ \bibnamefont {Huber}}, \bibinfo {author} {\bibfnamefont
  {O.}~\bibnamefont {Kamaev}}, \bibinfo {author} {\bibfnamefont
  {M.}~\bibnamefont {Kiveni}}, \bibinfo {author} {\bibfnamefont
  {M.}~\bibnamefont {Kos}}, \bibinfo {author} {\bibfnamefont {S.~W.}\
  \bibnamefont {Leman}}, \bibinfo {author} {\bibfnamefont {S.}~\bibnamefont
  {Liu}}, \bibinfo {author} {\bibfnamefont {R.}~\bibnamefont {Mahapatra}},
  \bibinfo {author} {\bibfnamefont {V.}~\bibnamefont {Mandic}}, \bibinfo
  {author} {\bibfnamefont {K.~A.}\ \bibnamefont {McCarthy}}, \bibinfo {author}
  {\bibfnamefont {N.}~\bibnamefont {Mirabolfathi}}, \bibinfo {author}
  {\bibfnamefont {D.}~\bibnamefont {Moore}}, \bibinfo {author} {\bibfnamefont
  {H.}~\bibnamefont {Nelson}}, \bibinfo {author} {\bibfnamefont {R.~W.}\
  \bibnamefont {Ogburn}}, \bibinfo {author} {\bibfnamefont {A.}~\bibnamefont
  {Phipps}}, \bibinfo {author} {\bibfnamefont {M.}~\bibnamefont {Pyle}},
  \bibinfo {author} {\bibfnamefont {X.}~\bibnamefont {Qiu}}, \bibinfo {author}
  {\bibfnamefont {E.}~\bibnamefont {Ramberg}}, \bibinfo {author} {\bibfnamefont
  {W.}~\bibnamefont {Rau}}, \bibinfo {author} {\bibfnamefont {A.}~\bibnamefont
  {Reisetter}}, \bibinfo {author} {\bibfnamefont {R.}~\bibnamefont {Resch}},
  \bibinfo {author} {\bibfnamefont {T.}~\bibnamefont {Saab}}, \bibinfo {author}
  {\bibfnamefont {B.}~\bibnamefont {Sadoulet}}, \bibinfo {author}
  {\bibfnamefont {J.}~\bibnamefont {Sander}}, \bibinfo {author} {\bibfnamefont
  {R.~W.}\ \bibnamefont {Schnee}}, \bibinfo {author} {\bibfnamefont {D.~N.}\
  \bibnamefont {Seitz}}, \bibinfo {author} {\bibfnamefont {B.}~\bibnamefont
  {Serfass}}, \bibinfo {author} {\bibfnamefont {K.~M.}\ \bibnamefont
  {Sundqvist}}, \bibinfo {author} {\bibfnamefont {M.}~\bibnamefont {Tarka}},
  \bibinfo {author} {\bibfnamefont {P.}~\bibnamefont {Wikus}}, \bibinfo
  {author} {\bibfnamefont {S.}~\bibnamefont {Yellin}}, \bibinfo {author}
  {\bibfnamefont {J.}~\bibnamefont {Yoo}}, \bibinfo {author} {\bibfnamefont
  {B.~A.}\ \bibnamefont {Young}},\ and\ \bibinfo {author} {\bibfnamefont
  {J.}~\bibnamefont {Zhang}} (\bibinfo {collaboration} {CDMS Collaboration}),\
  }\bibfield  {title} {\bibinfo {title} {Results from a low-energy analysis of
  the cdms ii germanium data},\ }\href
  {https://doi.org/10.1103/PhysRevLett.106.131302} {\bibfield  {journal}
  {\bibinfo  {journal} {Phys. Rev. Lett.}\ }\textbf {\bibinfo {volume} {106}},\
  \bibinfo {pages} {131302} (\bibinfo {year} {2011})}\BibitemShut {NoStop}%
\bibitem [{\citenamefont {Aprile}\ \emph {et~al.}(2017)\citenamefont {Aprile},
  \citenamefont {Aalbers}, \citenamefont {Agostini}, \citenamefont {Alfonsi},
  \citenamefont {Amaro}, \citenamefont {Anthony}, \citenamefont {Arneodo},
  \citenamefont {Barrow}, \citenamefont {Baudis}, \citenamefont {Bauermeister},
  \citenamefont {Benabderrahmane}, \citenamefont {Berger}, \citenamefont
  {Breur}, \citenamefont {Brown}, \citenamefont {Brown}, \citenamefont {Brown},
  \citenamefont {Bruenner}, \citenamefont {Bruno}, \citenamefont {Budnik},
  \citenamefont {B\"utikofer}, \citenamefont {Calv\'en}, \citenamefont
  {Cardoso}, \citenamefont {Cervantes}, \citenamefont {Cichon}, \citenamefont
  {Coderre}, \citenamefont {Colijn}, \citenamefont {Conrad}, \citenamefont
  {Cussonneau}, \citenamefont {Decowski}, \citenamefont {de~Perio},
  \citenamefont {Di~Gangi}, \citenamefont {Di~Giovanni}, \citenamefont
  {Diglio}, \citenamefont {Eurin}, \citenamefont {Fei}, \citenamefont
  {Ferella}, \citenamefont {Fieguth}, \citenamefont {Fulgione}, \citenamefont
  {Gallo~Rosso}, \citenamefont {Galloway}, \citenamefont {Gao}, \citenamefont
  {Garbini}, \citenamefont {Gardner}, \citenamefont {Geis}, \citenamefont
  {Goetzke}, \citenamefont {Grandi}, \citenamefont {Greene}, \citenamefont
  {Grignon}, \citenamefont {Hasterok}, \citenamefont {Hogenbirk}, \citenamefont
  {Howlett}, \citenamefont {Itay}, \citenamefont {Kaminsky}, \citenamefont
  {Kazama}, \citenamefont {Kessler}, \citenamefont {Kish}, \citenamefont
  {Landsman}, \citenamefont {Lang}, \citenamefont {Lellouch}, \citenamefont
  {Levinson}, \citenamefont {Lin}, \citenamefont {Lindemann}, \citenamefont
  {Lindner}, \citenamefont {Lombardi}, \citenamefont {Lopes}, \citenamefont
  {Manfredini}, \citenamefont {Mari\ifmmode~\mbox{\c{s}}\else \c{s}\fi{}},
  \citenamefont {Marrod\'an~Undagoitia}, \citenamefont {Masbou}, \citenamefont
  {Massoli}, \citenamefont {Masson}, \citenamefont {Mayani}, \citenamefont
  {Messina}, \citenamefont {Micheneau}, \citenamefont {Molinario},
  \citenamefont {Mor\aa{}}, \citenamefont {Murra}, \citenamefont {Naganoma},
  \citenamefont {Ni}, \citenamefont {Oberlack}, \citenamefont {Pakarha},
  \citenamefont {Pelssers}, \citenamefont {Persiani}, \citenamefont {Piastra},
  \citenamefont {Pienaar}, \citenamefont {Pizzella}, \citenamefont {Piro},
  \citenamefont {Plante}, \citenamefont {Priel}, \citenamefont {Rauch},
  \citenamefont {Reichard}, \citenamefont {Reuter}, \citenamefont {Riedel},
  \citenamefont {Rizzo}, \citenamefont {Rosendahl}, \citenamefont {Rupp},
  \citenamefont {Saldanha}, \citenamefont {dos Santos}, \citenamefont
  {Sartorelli}, \citenamefont {Scheibelhut}, \citenamefont {Schindler},
  \citenamefont {Schreiner}, \citenamefont {Schumann}, \citenamefont
  {Scotto~Lavina}, \citenamefont {Selvi}, \citenamefont {Shagin}, \citenamefont
  {Shockley}, \citenamefont {Silva}, \citenamefont {Simgen}, \citenamefont
  {Sivers}, \citenamefont {Stein}, \citenamefont {Thapa}, \citenamefont
  {Thers}, \citenamefont {Tiseni}, \citenamefont {Trinchero}, \citenamefont
  {Tunnell}, \citenamefont {Vargas}, \citenamefont {Upole}, \citenamefont
  {Wang}, \citenamefont {Wang}, \citenamefont {Wei}, \citenamefont
  {Weinheimer}, \citenamefont {Wulf}, \citenamefont {Ye}, \citenamefont
  {Zhang},\ and\ \citenamefont {Zhu}}]{Aprile2017}%
  \BibitemOpen
  \bibfield  {author} {\bibinfo {author} {\bibfnamefont {E.}~\bibnamefont
  {Aprile}}, \bibinfo {author} {\bibfnamefont {J.}~\bibnamefont {Aalbers}},
  \bibinfo {author} {\bibfnamefont {F.}~\bibnamefont {Agostini}}, \bibinfo
  {author} {\bibfnamefont {M.}~\bibnamefont {Alfonsi}}, \bibinfo {author}
  {\bibfnamefont {F.~D.}\ \bibnamefont {Amaro}}, \bibinfo {author}
  {\bibfnamefont {M.}~\bibnamefont {Anthony}}, \bibinfo {author} {\bibfnamefont
  {F.}~\bibnamefont {Arneodo}}, \bibinfo {author} {\bibfnamefont
  {P.}~\bibnamefont {Barrow}}, \bibinfo {author} {\bibfnamefont
  {L.}~\bibnamefont {Baudis}}, \bibinfo {author} {\bibfnamefont
  {B.}~\bibnamefont {Bauermeister}}, \bibinfo {author} {\bibfnamefont {M.~L.}\
  \bibnamefont {Benabderrahmane}}, \bibinfo {author} {\bibfnamefont
  {T.}~\bibnamefont {Berger}}, \bibinfo {author} {\bibfnamefont {P.~A.}\
  \bibnamefont {Breur}}, \bibinfo {author} {\bibfnamefont {A.}~\bibnamefont
  {Brown}}, \bibinfo {author} {\bibfnamefont {A.}~\bibnamefont {Brown}},
  \bibinfo {author} {\bibfnamefont {E.}~\bibnamefont {Brown}}, \bibinfo
  {author} {\bibfnamefont {S.}~\bibnamefont {Bruenner}}, \bibinfo {author}
  {\bibfnamefont {G.}~\bibnamefont {Bruno}}, \bibinfo {author} {\bibfnamefont
  {R.}~\bibnamefont {Budnik}}, \bibinfo {author} {\bibfnamefont
  {L.}~\bibnamefont {B\"utikofer}}, \bibinfo {author} {\bibfnamefont
  {J.}~\bibnamefont {Calv\'en}}, \bibinfo {author} {\bibfnamefont {J.~M.~R.}\
  \bibnamefont {Cardoso}}, \bibinfo {author} {\bibfnamefont {M.}~\bibnamefont
  {Cervantes}}, \bibinfo {author} {\bibfnamefont {D.}~\bibnamefont {Cichon}},
  \bibinfo {author} {\bibfnamefont {D.}~\bibnamefont {Coderre}}, \bibinfo
  {author} {\bibfnamefont {A.~P.}\ \bibnamefont {Colijn}}, \bibinfo {author}
  {\bibfnamefont {J.}~\bibnamefont {Conrad}}, \bibinfo {author} {\bibfnamefont
  {J.~P.}\ \bibnamefont {Cussonneau}}, \bibinfo {author} {\bibfnamefont
  {M.~P.}\ \bibnamefont {Decowski}}, \bibinfo {author} {\bibfnamefont
  {P.}~\bibnamefont {de~Perio}}, \bibinfo {author} {\bibfnamefont
  {P.}~\bibnamefont {Di~Gangi}}, \bibinfo {author} {\bibfnamefont
  {A.}~\bibnamefont {Di~Giovanni}}, \bibinfo {author} {\bibfnamefont
  {S.}~\bibnamefont {Diglio}}, \bibinfo {author} {\bibfnamefont
  {G.}~\bibnamefont {Eurin}}, \bibinfo {author} {\bibfnamefont
  {J.}~\bibnamefont {Fei}}, \bibinfo {author} {\bibfnamefont {A.~D.}\
  \bibnamefont {Ferella}}, \bibinfo {author} {\bibfnamefont {A.}~\bibnamefont
  {Fieguth}}, \bibinfo {author} {\bibfnamefont {W.}~\bibnamefont {Fulgione}},
  \bibinfo {author} {\bibfnamefont {A.}~\bibnamefont {Gallo~Rosso}}, \bibinfo
  {author} {\bibfnamefont {M.}~\bibnamefont {Galloway}}, \bibinfo {author}
  {\bibfnamefont {F.}~\bibnamefont {Gao}}, \bibinfo {author} {\bibfnamefont
  {M.}~\bibnamefont {Garbini}}, \bibinfo {author} {\bibfnamefont
  {R.}~\bibnamefont {Gardner}}, \bibinfo {author} {\bibfnamefont
  {C.}~\bibnamefont {Geis}}, \bibinfo {author} {\bibfnamefont {L.~W.}\
  \bibnamefont {Goetzke}}, \bibinfo {author} {\bibfnamefont {L.}~\bibnamefont
  {Grandi}}, \bibinfo {author} {\bibfnamefont {Z.}~\bibnamefont {Greene}},
  \bibinfo {author} {\bibfnamefont {C.}~\bibnamefont {Grignon}}, \bibinfo
  {author} {\bibfnamefont {C.}~\bibnamefont {Hasterok}}, \bibinfo {author}
  {\bibfnamefont {E.}~\bibnamefont {Hogenbirk}}, \bibinfo {author}
  {\bibfnamefont {J.}~\bibnamefont {Howlett}}, \bibinfo {author} {\bibfnamefont
  {R.}~\bibnamefont {Itay}}, \bibinfo {author} {\bibfnamefont {B.}~\bibnamefont
  {Kaminsky}}, \bibinfo {author} {\bibfnamefont {S.}~\bibnamefont {Kazama}},
  \bibinfo {author} {\bibfnamefont {G.}~\bibnamefont {Kessler}}, \bibinfo
  {author} {\bibfnamefont {A.}~\bibnamefont {Kish}}, \bibinfo {author}
  {\bibfnamefont {H.}~\bibnamefont {Landsman}}, \bibinfo {author}
  {\bibfnamefont {R.~F.}\ \bibnamefont {Lang}}, \bibinfo {author}
  {\bibfnamefont {D.}~\bibnamefont {Lellouch}}, \bibinfo {author}
  {\bibfnamefont {L.}~\bibnamefont {Levinson}}, \bibinfo {author}
  {\bibfnamefont {Q.}~\bibnamefont {Lin}}, \bibinfo {author} {\bibfnamefont
  {S.}~\bibnamefont {Lindemann}}, \bibinfo {author} {\bibfnamefont
  {M.}~\bibnamefont {Lindner}}, \bibinfo {author} {\bibfnamefont
  {F.}~\bibnamefont {Lombardi}}, \bibinfo {author} {\bibfnamefont {J.~A.~M.}\
  \bibnamefont {Lopes}}, \bibinfo {author} {\bibfnamefont {A.}~\bibnamefont
  {Manfredini}}, \bibinfo {author} {\bibfnamefont {I.}~\bibnamefont
  {Mari\ifmmode~\mbox{\c{s}}\else \c{s}\fi{}}}, \bibinfo {author}
  {\bibfnamefont {T.}~\bibnamefont {Marrod\'an~Undagoitia}}, \bibinfo {author}
  {\bibfnamefont {J.}~\bibnamefont {Masbou}}, \bibinfo {author} {\bibfnamefont
  {F.~V.}\ \bibnamefont {Massoli}}, \bibinfo {author} {\bibfnamefont
  {D.}~\bibnamefont {Masson}}, \bibinfo {author} {\bibfnamefont
  {D.}~\bibnamefont {Mayani}}, \bibinfo {author} {\bibfnamefont
  {M.}~\bibnamefont {Messina}}, \bibinfo {author} {\bibfnamefont
  {K.}~\bibnamefont {Micheneau}}, \bibinfo {author} {\bibfnamefont
  {A.}~\bibnamefont {Molinario}}, \bibinfo {author} {\bibfnamefont
  {K.}~\bibnamefont {Mor\aa{}}}, \bibinfo {author} {\bibfnamefont
  {M.}~\bibnamefont {Murra}}, \bibinfo {author} {\bibfnamefont
  {J.}~\bibnamefont {Naganoma}}, \bibinfo {author} {\bibfnamefont
  {K.}~\bibnamefont {Ni}}, \bibinfo {author} {\bibfnamefont {U.}~\bibnamefont
  {Oberlack}}, \bibinfo {author} {\bibfnamefont {P.}~\bibnamefont {Pakarha}},
  \bibinfo {author} {\bibfnamefont {B.}~\bibnamefont {Pelssers}}, \bibinfo
  {author} {\bibfnamefont {R.}~\bibnamefont {Persiani}}, \bibinfo {author}
  {\bibfnamefont {F.}~\bibnamefont {Piastra}}, \bibinfo {author} {\bibfnamefont
  {J.}~\bibnamefont {Pienaar}}, \bibinfo {author} {\bibfnamefont
  {V.}~\bibnamefont {Pizzella}}, \bibinfo {author} {\bibfnamefont {M.-C.}\
  \bibnamefont {Piro}}, \bibinfo {author} {\bibfnamefont {G.}~\bibnamefont
  {Plante}}, \bibinfo {author} {\bibfnamefont {N.}~\bibnamefont {Priel}},
  \bibinfo {author} {\bibfnamefont {L.}~\bibnamefont {Rauch}}, \bibinfo
  {author} {\bibfnamefont {S.}~\bibnamefont {Reichard}}, \bibinfo {author}
  {\bibfnamefont {C.}~\bibnamefont {Reuter}}, \bibinfo {author} {\bibfnamefont
  {B.}~\bibnamefont {Riedel}}, \bibinfo {author} {\bibfnamefont
  {A.}~\bibnamefont {Rizzo}}, \bibinfo {author} {\bibfnamefont
  {S.}~\bibnamefont {Rosendahl}}, \bibinfo {author} {\bibfnamefont
  {N.}~\bibnamefont {Rupp}}, \bibinfo {author} {\bibfnamefont {R.}~\bibnamefont
  {Saldanha}}, \bibinfo {author} {\bibfnamefont {J.~M.~F.}\ \bibnamefont {dos
  Santos}}, \bibinfo {author} {\bibfnamefont {G.}~\bibnamefont {Sartorelli}},
  \bibinfo {author} {\bibfnamefont {M.}~\bibnamefont {Scheibelhut}}, \bibinfo
  {author} {\bibfnamefont {S.}~\bibnamefont {Schindler}}, \bibinfo {author}
  {\bibfnamefont {J.}~\bibnamefont {Schreiner}}, \bibinfo {author}
  {\bibfnamefont {M.}~\bibnamefont {Schumann}}, \bibinfo {author}
  {\bibfnamefont {L.}~\bibnamefont {Scotto~Lavina}}, \bibinfo {author}
  {\bibfnamefont {M.}~\bibnamefont {Selvi}}, \bibinfo {author} {\bibfnamefont
  {P.}~\bibnamefont {Shagin}}, \bibinfo {author} {\bibfnamefont
  {E.}~\bibnamefont {Shockley}}, \bibinfo {author} {\bibfnamefont
  {M.}~\bibnamefont {Silva}}, \bibinfo {author} {\bibfnamefont
  {H.}~\bibnamefont {Simgen}}, \bibinfo {author} {\bibfnamefont {M.~v.}\
  \bibnamefont {Sivers}}, \bibinfo {author} {\bibfnamefont {A.}~\bibnamefont
  {Stein}}, \bibinfo {author} {\bibfnamefont {S.}~\bibnamefont {Thapa}},
  \bibinfo {author} {\bibfnamefont {D.}~\bibnamefont {Thers}}, \bibinfo
  {author} {\bibfnamefont {A.}~\bibnamefont {Tiseni}}, \bibinfo {author}
  {\bibfnamefont {G.}~\bibnamefont {Trinchero}}, \bibinfo {author}
  {\bibfnamefont {C.}~\bibnamefont {Tunnell}}, \bibinfo {author} {\bibfnamefont
  {M.}~\bibnamefont {Vargas}}, \bibinfo {author} {\bibfnamefont
  {N.}~\bibnamefont {Upole}}, \bibinfo {author} {\bibfnamefont
  {H.}~\bibnamefont {Wang}}, \bibinfo {author} {\bibfnamefont {Z.}~\bibnamefont
  {Wang}}, \bibinfo {author} {\bibfnamefont {Y.}~\bibnamefont {Wei}}, \bibinfo
  {author} {\bibfnamefont {C.}~\bibnamefont {Weinheimer}}, \bibinfo {author}
  {\bibfnamefont {J.}~\bibnamefont {Wulf}}, \bibinfo {author} {\bibfnamefont
  {J.}~\bibnamefont {Ye}}, \bibinfo {author} {\bibfnamefont {Y.}~\bibnamefont
  {Zhang}},\ and\ \bibinfo {author} {\bibfnamefont {T.}~\bibnamefont {Zhu}}
  (\bibinfo {collaboration} {XENON Collaboration}),\ }\bibfield  {title}
  {\bibinfo {title} {First dark matter search results from the xenon1t
  experiment},\ }\href {https://doi.org/10.1103/PhysRevLett.119.181301}
  {\bibfield  {journal} {\bibinfo  {journal} {Phys. Rev. Lett.}\ }\textbf
  {\bibinfo {volume} {119}},\ \bibinfo {pages} {181301} (\bibinfo {year}
  {2017})}\BibitemShut {NoStop}%
\bibitem [{\citenamefont {Aalseth}\ \emph {et~al.}(2011)\citenamefont
  {Aalseth}, \citenamefont {Barbeau}, \citenamefont {Bowden}, \citenamefont
  {Cabrera-Palmer}, \citenamefont {Colaresi}, \citenamefont {Collar},
  \citenamefont {Dazeley}, \citenamefont {de~Lurgio}, \citenamefont {Fast},
  \citenamefont {Fields}, \citenamefont {Greenberg}, \citenamefont {Hossbach},
  \citenamefont {Keillor}, \citenamefont {Kephart}, \citenamefont {Marino},
  \citenamefont {Miley}, \citenamefont {Miller}, \citenamefont {Orrell},
  \citenamefont {Radford}, \citenamefont {Reyna}, \citenamefont {Tench},
  \citenamefont {Van~Wechel}, \citenamefont {Wilkerson},\ and\ \citenamefont
  {Yocum}}]{Aalseth2011}%
  \BibitemOpen
  \bibfield  {author} {\bibinfo {author} {\bibfnamefont {C.~E.}\ \bibnamefont
  {Aalseth}}, \bibinfo {author} {\bibfnamefont {P.~S.}\ \bibnamefont
  {Barbeau}}, \bibinfo {author} {\bibfnamefont {N.~S.}\ \bibnamefont {Bowden}},
  \bibinfo {author} {\bibfnamefont {B.}~\bibnamefont {Cabrera-Palmer}},
  \bibinfo {author} {\bibfnamefont {J.}~\bibnamefont {Colaresi}}, \bibinfo
  {author} {\bibfnamefont {J.~I.}\ \bibnamefont {Collar}}, \bibinfo {author}
  {\bibfnamefont {S.}~\bibnamefont {Dazeley}}, \bibinfo {author} {\bibfnamefont
  {P.}~\bibnamefont {de~Lurgio}}, \bibinfo {author} {\bibfnamefont {J.~E.}\
  \bibnamefont {Fast}}, \bibinfo {author} {\bibfnamefont {N.}~\bibnamefont
  {Fields}}, \bibinfo {author} {\bibfnamefont {C.~H.}\ \bibnamefont
  {Greenberg}}, \bibinfo {author} {\bibfnamefont {T.~W.}\ \bibnamefont
  {Hossbach}}, \bibinfo {author} {\bibfnamefont {M.~E.}\ \bibnamefont
  {Keillor}}, \bibinfo {author} {\bibfnamefont {J.~D.}\ \bibnamefont
  {Kephart}}, \bibinfo {author} {\bibfnamefont {M.~G.}\ \bibnamefont {Marino}},
  \bibinfo {author} {\bibfnamefont {H.~S.}\ \bibnamefont {Miley}}, \bibinfo
  {author} {\bibfnamefont {M.~L.}\ \bibnamefont {Miller}}, \bibinfo {author}
  {\bibfnamefont {J.~L.}\ \bibnamefont {Orrell}}, \bibinfo {author}
  {\bibfnamefont {D.~C.}\ \bibnamefont {Radford}}, \bibinfo {author}
  {\bibfnamefont {D.}~\bibnamefont {Reyna}}, \bibinfo {author} {\bibfnamefont
  {O.}~\bibnamefont {Tench}}, \bibinfo {author} {\bibfnamefont {T.~D.}\
  \bibnamefont {Van~Wechel}}, \bibinfo {author} {\bibfnamefont {J.~F.}\
  \bibnamefont {Wilkerson}},\ and\ \bibinfo {author} {\bibfnamefont {K.~M.}\
  \bibnamefont {Yocum}} (\bibinfo {collaboration} {CoGeNT Collaboration}),\
  }\bibfield  {title} {\bibinfo {title} {Results from a search for light-mass
  dark matter with a $p$-type point contact germanium detector},\ }\href
  {https://doi.org/10.1103/PhysRevLett.106.131301} {\bibfield  {journal}
  {\bibinfo  {journal} {Phys. Rev. Lett.}\ }\textbf {\bibinfo {volume} {106}},\
  \bibinfo {pages} {131301} (\bibinfo {year} {2011})}\BibitemShut {NoStop}%
\bibitem [{\citenamefont {Adhikari}\ \emph {et~al.}(2018)\citenamefont
  {Adhikari}, \citenamefont {Adhikari}, \citenamefont {de~Souza}, \citenamefont
  {Carlin}, \citenamefont {Choi}, \citenamefont {Djamal}, \citenamefont
  {Ezeribe}, \citenamefont {Ha}, \citenamefont {Hahn}, \citenamefont {Hubbard}
  \emph {et~al.}}]{adhikari2019experiment}%
  \BibitemOpen
  \bibfield  {author} {\bibinfo {author} {\bibfnamefont {G.}~\bibnamefont
  {Adhikari}}, \bibinfo {author} {\bibfnamefont {P.}~\bibnamefont {Adhikari}},
  \bibinfo {author} {\bibfnamefont {E.~B.}\ \bibnamefont {de~Souza}}, \bibinfo
  {author} {\bibfnamefont {N.}~\bibnamefont {Carlin}}, \bibinfo {author}
  {\bibfnamefont {S.}~\bibnamefont {Choi}}, \bibinfo {author} {\bibfnamefont
  {M.}~\bibnamefont {Djamal}}, \bibinfo {author} {\bibfnamefont {A.~C.}\
  \bibnamefont {Ezeribe}}, \bibinfo {author} {\bibfnamefont {C.~H.}\
  \bibnamefont {Ha}}, \bibinfo {author} {\bibfnamefont {I.}~\bibnamefont
  {Hahn}}, \bibinfo {author} {\bibfnamefont {A.~J.}\ \bibnamefont {Hubbard}},
  \emph {et~al.} (\bibinfo {collaboration} {COSINE-100 Collaboration}),\
  }\bibfield  {title} {\bibinfo {title} {An experiment to search for dark
  matter interactions using sodium iodide detectors},\ }\href
  {https://doi.org/10.1038/s41586-018-0739-1} {\bibfield  {journal} {\bibinfo
  {journal} {Nature}\ }\textbf {\bibinfo {volume} {564}},\ \bibinfo {pages}
  {83} (\bibinfo {year} {2018})}\BibitemShut {NoStop}%
\bibitem [{\citenamefont {Kane}\ and\ \citenamefont
  {Watson}(2008)}]{kane2008dark}%
  \BibitemOpen
  \bibfield  {author} {\bibinfo {author} {\bibfnamefont {G.}~\bibnamefont
  {Kane}}\ and\ \bibinfo {author} {\bibfnamefont {S.}~\bibnamefont {Watson}},\
  }\bibfield  {title} {\bibinfo {title} {Dark matter and {LHC}: What is the
  connection?},\ }\href {https://doi.org/10.1142/S0217732308028314} {\bibfield
  {journal} {\bibinfo  {journal} {Mod. Phys. Lett. A}\ }\textbf {\bibinfo
  {volume} {23}},\ \bibinfo {pages} {2103} (\bibinfo {year}
  {2008})}\BibitemShut {NoStop}%
\bibitem [{\citenamefont {Preskill}\ \emph {et~al.}(1983)\citenamefont
  {Preskill}, \citenamefont {Wise},\ and\ \citenamefont
  {Wilczek}}]{preskill1983cosmology}%
  \BibitemOpen
  \bibfield  {author} {\bibinfo {author} {\bibfnamefont {J.}~\bibnamefont
  {Preskill}}, \bibinfo {author} {\bibfnamefont {M.~B.}\ \bibnamefont {Wise}},\
  and\ \bibinfo {author} {\bibfnamefont {F.}~\bibnamefont {Wilczek}},\
  }\bibfield  {title} {\bibinfo {title} {Cosmology of the invisible axion},\
  }\href@noop {} {\bibfield  {journal} {\bibinfo  {journal} {Phys.\ Lett.\ B}\
  }\textbf {\bibinfo {volume} {120}},\ \bibinfo {pages} {127} (\bibinfo {year}
  {1983})}\BibitemShut {NoStop}%
\bibitem [{\citenamefont {Abbott}\ and\ \citenamefont
  {Sikivie}(1983)}]{abbott1983cosmological}%
  \BibitemOpen
  \bibfield  {author} {\bibinfo {author} {\bibfnamefont {L.~F.}\ \bibnamefont
  {Abbott}}\ and\ \bibinfo {author} {\bibfnamefont {P.}~\bibnamefont
  {Sikivie}},\ }\bibfield  {title} {\bibinfo {title} {A cosmological bound on
  the invisible axion},\ }\href@noop {} {\bibfield  {journal} {\bibinfo
  {journal} {Phys.\ Lett.\ B}\ }\textbf {\bibinfo {volume} {120}},\ \bibinfo
  {pages} {133} (\bibinfo {year} {1983})}\BibitemShut {NoStop}%
\bibitem [{\citenamefont {Dine}\ and\ \citenamefont
  {Fischler}(1983)}]{dine1983not}%
  \BibitemOpen
  \bibfield  {author} {\bibinfo {author} {\bibfnamefont {M.}~\bibnamefont
  {Dine}}\ and\ \bibinfo {author} {\bibfnamefont {W.}~\bibnamefont
  {Fischler}},\ }\bibfield  {title} {\bibinfo {title} {The not-so-harmless
  axion},\ }\href@noop {} {\bibfield  {journal} {\bibinfo  {journal} {Phys.\
  Lett.\ B}\ }\textbf {\bibinfo {volume} {120}},\ \bibinfo {pages} {137}
  (\bibinfo {year} {1983})}\BibitemShut {NoStop}%
\bibitem [{\citenamefont {Weinberg}(1978)}]{weinberg1978new}%
  \BibitemOpen
  \bibfield  {author} {\bibinfo {author} {\bibfnamefont {S.}~\bibnamefont
  {Weinberg}},\ }\bibfield  {title} {\bibinfo {title} {A new light boson?},\
  }\href {https://doi.org/10.1103/PhysRevLett.40.223} {\bibfield  {journal}
  {\bibinfo  {journal} {Phys.\ Rev.\ Lett.}\ }\textbf {\bibinfo {volume}
  {40}},\ \bibinfo {pages} {223} (\bibinfo {year} {1978})}\BibitemShut
  {NoStop}%
\bibitem [{\citenamefont {Wilczek}(1978)}]{wilczek1978problem}%
  \BibitemOpen
  \bibfield  {author} {\bibinfo {author} {\bibfnamefont {F.}~\bibnamefont
  {Wilczek}},\ }\bibfield  {title} {\bibinfo {title} {Problem of strong
  $\mathcal{P}$ and $\mathcal{T}$ invariance in the presence of instantons},\
  }\href {https://doi.org/10.1103/PhysRevLett.40.279} {\bibfield  {journal}
  {\bibinfo  {journal} {Phys.\ Rev.\ Lett.}\ }\textbf {\bibinfo {volume}
  {40}},\ \bibinfo {pages} {279} (\bibinfo {year} {1978})}\BibitemShut
  {NoStop}%
\bibitem [{\citenamefont {Peccei}\ and\ \citenamefont
  {Quinn}(1977)}]{peccei1977cp}%
  \BibitemOpen
  \bibfield  {author} {\bibinfo {author} {\bibfnamefont {R.~D.}\ \bibnamefont
  {Peccei}}\ and\ \bibinfo {author} {\bibfnamefont {H.~R.}\ \bibnamefont
  {Quinn}},\ }\bibfield  {title} {\bibinfo {title} {$\mathrm{CP}$ conservation
  in the presence of pseudoparticles},\ }\href
  {https://doi.org/10.1103/PhysRevLett.38.1440} {\bibfield  {journal} {\bibinfo
   {journal} {Phys. Rev. Lett.}\ }\textbf {\bibinfo {volume} {38}},\ \bibinfo
  {pages} {1440} (\bibinfo {year} {1977})}\BibitemShut {NoStop}%
\bibitem [{\citenamefont {Kim}\ and\ \citenamefont {Carosi}(2010)}]{Kim:2010}%
  \BibitemOpen
  \bibfield  {author} {\bibinfo {author} {\bibfnamefont {J.~E.}\ \bibnamefont
  {Kim}}\ and\ \bibinfo {author} {\bibfnamefont {G.}~\bibnamefont {Carosi}},\
  }\bibfield  {title} {\bibinfo {title} {Axions and the strong $cp$ problem},\
  }\href {https://doi.org/10.1103/RevModPhys.82.557} {\bibfield  {journal}
  {\bibinfo  {journal} {Rev. Mod. Phys.}\ }\textbf {\bibinfo {volume} {82}},\
  \bibinfo {pages} {557} (\bibinfo {year} {2010})}\BibitemShut {NoStop}%
\bibitem [{\citenamefont {O'Hare}\ and\ \citenamefont
  {Vitagliano}(2020)}]{Hare:2020}%
  \BibitemOpen
  \bibfield  {author} {\bibinfo {author} {\bibfnamefont {C.~A.~J.}\
  \bibnamefont {O'Hare}}\ and\ \bibinfo {author} {\bibfnamefont
  {E.}~\bibnamefont {Vitagliano}},\ }\bibfield  {title} {\bibinfo {title}
  {Cornering the axion with $cp$-violating interactions},\ }\href@noop {}
  {\bibfield  {journal} {\bibinfo  {journal} {Phys. Rev. D}\ }\textbf {\bibinfo
  {volume} {102}},\ \bibinfo {pages} {115026} (\bibinfo {year}
  {2020})}\BibitemShut {NoStop}%
\bibitem [{\citenamefont {{Di Luzio}}\ \emph {et~al.}(2020)\citenamefont {{Di
  Luzio}}, \citenamefont {Giannotti}, \citenamefont {Nardi},\ and\
  \citenamefont {Visinelli}}]{QCDaxion:2020}%
  \BibitemOpen
  \bibfield  {author} {\bibinfo {author} {\bibfnamefont {L.}~\bibnamefont {{Di
  Luzio}}}, \bibinfo {author} {\bibfnamefont {M.}~\bibnamefont {Giannotti}},
  \bibinfo {author} {\bibfnamefont {E.}~\bibnamefont {Nardi}},\ and\ \bibinfo
  {author} {\bibfnamefont {L.}~\bibnamefont {Visinelli}},\ }\bibfield  {title}
  {\bibinfo {title} {The landscape of qcd axion models},\ }\href
  {https://doi.org/https://doi.org/10.1016/j.physrep.2020.06.002} {\bibfield
  {journal} {\bibinfo  {journal} {Phys. Rep.}\ }\textbf {\bibinfo {volume}
  {870}},\ \bibinfo {pages} {1} (\bibinfo {year} {2020})},\ \bibinfo {note}
  {the landscape of QCD axion models}\BibitemShut {NoStop}%
\bibitem [{\citenamefont {Youdin}\ \emph {et~al.}(1996)\citenamefont {Youdin},
  \citenamefont {Krause~Jr.}, \citenamefont {Jagannathan}, \citenamefont
  {Hunter},\ and\ \citenamefont {Lamoreaux}}]{youdin1996limits}%
  \BibitemOpen
  \bibfield  {author} {\bibinfo {author} {\bibfnamefont {A.~N.}\ \bibnamefont
  {Youdin}}, \bibinfo {author} {\bibfnamefont {D.}~\bibnamefont {Krause~Jr.}},
  \bibinfo {author} {\bibfnamefont {K.}~\bibnamefont {Jagannathan}}, \bibinfo
  {author} {\bibfnamefont {L.~R.}\ \bibnamefont {Hunter}},\ and\ \bibinfo
  {author} {\bibfnamefont {S.~K.}\ \bibnamefont {Lamoreaux}},\ }\bibfield
  {title} {\bibinfo {title} {Limits on spin-mass couplings within the axion
  window},\ }\href {https://doi.org/10.1103/PhysRevLett.77.2170} {\bibfield
  {journal} {\bibinfo  {journal} {Phys.\ Rev.\ Lett.}\ }\textbf {\bibinfo
  {volume} {77}},\ \bibinfo {pages} {2170} (\bibinfo {year}
  {1996})}\BibitemShut {NoStop}%
\bibitem [{\citenamefont {Ni}\ \emph {et~al.}(1999)\citenamefont {Ni},
  \citenamefont {Pan}, \citenamefont {Yeh}, \citenamefont {Hou},\ and\
  \citenamefont {Wan}}]{ni1999search}%
  \BibitemOpen
  \bibfield  {author} {\bibinfo {author} {\bibfnamefont {W.-T.}\ \bibnamefont
  {Ni}}, \bibinfo {author} {\bibfnamefont {S.-s.}\ \bibnamefont {Pan}},
  \bibinfo {author} {\bibfnamefont {H.-C.}\ \bibnamefont {Yeh}}, \bibinfo
  {author} {\bibfnamefont {L.-S.}\ \bibnamefont {Hou}},\ and\ \bibinfo {author}
  {\bibfnamefont {J.}~\bibnamefont {Wan}},\ }\bibfield  {title} {\bibinfo
  {title} {Search for an axionlike spin coupling using a paramagnetic salt with
  a dc squid},\ }\href {https://doi.org/10.1103/PhysRevLett.82.2439} {\bibfield
   {journal} {\bibinfo  {journal} {Phys.\ Rev.\ Lett.}\ }\textbf {\bibinfo
  {volume} {82}},\ \bibinfo {pages} {2439} (\bibinfo {year}
  {1999})}\BibitemShut {NoStop}%
\bibitem [{\citenamefont {Duffy}\ \emph {et~al.}(2006)\citenamefont {Duffy},
  \citenamefont {Sikivie}, \citenamefont {Tanner}, \citenamefont {Asztalos},
  \citenamefont {Hagmann}, \citenamefont {Kinion}, \citenamefont {Rosenberg},
  \citenamefont {van Bibber}, \citenamefont {Yu},\ and\ \citenamefont
  {Bradley}}]{Duffy:2006}%
  \BibitemOpen
  \bibfield  {author} {\bibinfo {author} {\bibfnamefont {L.~D.}\ \bibnamefont
  {Duffy}}, \bibinfo {author} {\bibfnamefont {P.}~\bibnamefont {Sikivie}},
  \bibinfo {author} {\bibfnamefont {D.~B.}\ \bibnamefont {Tanner}}, \bibinfo
  {author} {\bibfnamefont {S.~J.}\ \bibnamefont {Asztalos}}, \bibinfo {author}
  {\bibfnamefont {C.}~\bibnamefont {Hagmann}}, \bibinfo {author} {\bibfnamefont
  {D.}~\bibnamefont {Kinion}}, \bibinfo {author} {\bibfnamefont {L.~J.}\
  \bibnamefont {Rosenberg}}, \bibinfo {author} {\bibfnamefont {K.}~\bibnamefont
  {van Bibber}}, \bibinfo {author} {\bibfnamefont {D.~B.}\ \bibnamefont {Yu}},\
  and\ \bibinfo {author} {\bibfnamefont {R.~F.}\ \bibnamefont {Bradley}},\
  }\bibfield  {title} {\bibinfo {title} {High resolution search for dark-matter
  axions},\ }\href {https://doi.org/10.1103/PhysRevD.74.012006} {\bibfield
  {journal} {\bibinfo  {journal} {Phys. Rev. D}\ }\textbf {\bibinfo {volume}
  {74}},\ \bibinfo {pages} {012006} (\bibinfo {year} {2006})}\BibitemShut
  {NoStop}%
\bibitem [{\citenamefont {Zavattini}\ \emph {et~al.}(2006)\citenamefont
  {Zavattini}, \citenamefont {Zavattini}, \citenamefont {Ruoso}, \citenamefont
  {Polacco}, \citenamefont {Milotti}, \citenamefont {Karuza}, \citenamefont
  {Gastaldi}, \citenamefont {Di~Domenico}, \citenamefont {Della~Valle},
  \citenamefont {Cimino}, \citenamefont {Carusotto}, \citenamefont
  {Cantatore},\ and\ \citenamefont {Bregant}}]{Zavattini:2006}%
  \BibitemOpen
  \bibfield  {author} {\bibinfo {author} {\bibfnamefont {E.}~\bibnamefont
  {Zavattini}}, \bibinfo {author} {\bibfnamefont {G.}~\bibnamefont
  {Zavattini}}, \bibinfo {author} {\bibfnamefont {G.}~\bibnamefont {Ruoso}},
  \bibinfo {author} {\bibfnamefont {E.}~\bibnamefont {Polacco}}, \bibinfo
  {author} {\bibfnamefont {E.}~\bibnamefont {Milotti}}, \bibinfo {author}
  {\bibfnamefont {M.}~\bibnamefont {Karuza}}, \bibinfo {author} {\bibfnamefont
  {U.}~\bibnamefont {Gastaldi}}, \bibinfo {author} {\bibfnamefont
  {G.}~\bibnamefont {Di~Domenico}}, \bibinfo {author} {\bibfnamefont
  {F.}~\bibnamefont {Della~Valle}}, \bibinfo {author} {\bibfnamefont
  {R.}~\bibnamefont {Cimino}}, \bibinfo {author} {\bibfnamefont
  {S.}~\bibnamefont {Carusotto}}, \bibinfo {author} {\bibfnamefont
  {G.}~\bibnamefont {Cantatore}},\ and\ \bibinfo {author} {\bibfnamefont
  {M.}~\bibnamefont {Bregant}} (\bibinfo {collaboration} {PVLAS
  Collaboration}),\ }\bibfield  {title} {\bibinfo {title} {Experimental
  observation of optical rotation generated in vacuum by a magnetic field},\
  }\href {https://doi.org/10.1103/PhysRevLett.96.110406} {\bibfield  {journal}
  {\bibinfo  {journal} {Phys. Rev. Lett.}\ }\textbf {\bibinfo {volume} {96}},\
  \bibinfo {pages} {110406} (\bibinfo {year} {2006})}\BibitemShut {NoStop}%
\bibitem [{\citenamefont {Hammond}\ \emph {et~al.}(2007)\citenamefont
  {Hammond}, \citenamefont {Speake}, \citenamefont {Trenkel},\ and\
  \citenamefont {Pat{\'o}n}}]{hammond2007new}%
  \BibitemOpen
  \bibfield  {author} {\bibinfo {author} {\bibfnamefont {G.~D.}\ \bibnamefont
  {Hammond}}, \bibinfo {author} {\bibfnamefont {C.~C.}\ \bibnamefont {Speake}},
  \bibinfo {author} {\bibfnamefont {C.}~\bibnamefont {Trenkel}},\ and\ \bibinfo
  {author} {\bibfnamefont {A.~P.}\ \bibnamefont {Pat{\'o}n}},\ }\bibfield
  {title} {\bibinfo {title} {New constraints on short-range forces coupling
  mass to intrinsic spin},\ }\href
  {https://doi.org/10.1103/PhysRevLett.98.081101} {\bibfield  {journal}
  {\bibinfo  {journal} {Phys.\ Rev.\ Lett.}\ }\textbf {\bibinfo {volume}
  {98}},\ \bibinfo {pages} {081101} (\bibinfo {year} {2007})}\BibitemShut
  {NoStop}%
\bibitem [{\citenamefont {Hoedl}\ \emph {et~al.}(2011)\citenamefont {Hoedl},
  \citenamefont {Fleischer}, \citenamefont {Adelberger},\ and\ \citenamefont
  {Heckel}}]{hoedl2011improved}%
  \BibitemOpen
  \bibfield  {author} {\bibinfo {author} {\bibfnamefont {S.}~\bibnamefont
  {Hoedl}}, \bibinfo {author} {\bibfnamefont {F.}~\bibnamefont {Fleischer}},
  \bibinfo {author} {\bibfnamefont {E.}~\bibnamefont {Adelberger}},\ and\
  \bibinfo {author} {\bibfnamefont {B.}~\bibnamefont {Heckel}},\ }\bibfield
  {title} {\bibinfo {title} {Improved constraints on an axion-mediated force},\
  }\href {https://doi.org/10.1103/PhysRevLett.106.041801} {\bibfield  {journal}
  {\bibinfo  {journal} {Phys.\ Rev.\ Lett.}\ }\textbf {\bibinfo {volume}
  {106}},\ \bibinfo {pages} {041801} (\bibinfo {year} {2011})}\BibitemShut
  {NoStop}%
\bibitem [{\citenamefont {Barth}\ \emph {et~al.}(2013)\citenamefont {Barth},
  \citenamefont {Belov}, \citenamefont {Beltran}, \citenamefont {Bräuninger},
  \citenamefont {Carmona}, \citenamefont {Collar}, \citenamefont {Dafni},
  \citenamefont {Davenport}, \citenamefont {Lella}, \citenamefont
  {Eleftheriadis}, \citenamefont {Englhauser}, \citenamefont {Fanourakis},
  \citenamefont {Ferrer-Ribas}, \citenamefont {Fischer}, \citenamefont {Franz},
  \citenamefont {Friedrich}, \citenamefont {Galán}, \citenamefont {García},
  \citenamefont {Geralis}, \citenamefont {Giomataris}, \citenamefont
  {Gninenko}, \citenamefont {Gómez}, \citenamefont {Hasinoff}, \citenamefont
  {Heinsius}, \citenamefont {Hoffmann}, \citenamefont {Irastorza},
  \citenamefont {Jacoby}, \citenamefont {Jakovčić}, \citenamefont {Kang},
  \citenamefont {Königsmann}, \citenamefont {Kotthaus}, \citenamefont
  {Kousouris}, \citenamefont {Krčmar}, \citenamefont {Kuster}, \citenamefont
  {Lakić}, \citenamefont {Liolios}, \citenamefont {Ljubičić}, \citenamefont
  {Lutz}, \citenamefont {Luzón}, \citenamefont {Miller}, \citenamefont
  {Papaevangelou}, \citenamefont {Pivovaroff}, \citenamefont {Raffelt},
  \citenamefont {Redondo}, \citenamefont {Riege}, \citenamefont {Rodríguez},
  \citenamefont {Ruz}, \citenamefont {Savvidis}, \citenamefont {Semertzidis},
  \citenamefont {Stewart}, \citenamefont {Bibber}, \citenamefont {Vieira},
  \citenamefont {Villar}, \citenamefont {Vogel}, \citenamefont {Walckiers},\
  and\ \citenamefont {Zioutas}}]{barth2013cast}%
  \BibitemOpen
  \bibfield  {author} {\bibinfo {author} {\bibfnamefont {K.}~\bibnamefont
  {Barth}}, \bibinfo {author} {\bibfnamefont {A.}~\bibnamefont {Belov}},
  \bibinfo {author} {\bibfnamefont {B.}~\bibnamefont {Beltran}}, \bibinfo
  {author} {\bibfnamefont {H.}~\bibnamefont {Bräuninger}}, \bibinfo {author}
  {\bibfnamefont {J.}~\bibnamefont {Carmona}}, \bibinfo {author} {\bibfnamefont
  {J.}~\bibnamefont {Collar}}, \bibinfo {author} {\bibfnamefont
  {T.}~\bibnamefont {Dafni}}, \bibinfo {author} {\bibfnamefont
  {M.}~\bibnamefont {Davenport}}, \bibinfo {author} {\bibfnamefont {L.~D.}\
  \bibnamefont {Lella}}, \bibinfo {author} {\bibfnamefont {C.}~\bibnamefont
  {Eleftheriadis}}, \bibinfo {author} {\bibfnamefont {J.}~\bibnamefont
  {Englhauser}}, \bibinfo {author} {\bibfnamefont {G.}~\bibnamefont
  {Fanourakis}}, \bibinfo {author} {\bibfnamefont {E.}~\bibnamefont
  {Ferrer-Ribas}}, \bibinfo {author} {\bibfnamefont {H.}~\bibnamefont
  {Fischer}}, \bibinfo {author} {\bibfnamefont {J.}~\bibnamefont {Franz}},
  \bibinfo {author} {\bibfnamefont {P.}~\bibnamefont {Friedrich}}, \bibinfo
  {author} {\bibfnamefont {J.}~\bibnamefont {Galán}}, \bibinfo {author}
  {\bibfnamefont {J.}~\bibnamefont {García}}, \bibinfo {author} {\bibfnamefont
  {T.}~\bibnamefont {Geralis}}, \bibinfo {author} {\bibfnamefont
  {I.}~\bibnamefont {Giomataris}}, \bibinfo {author} {\bibfnamefont
  {S.}~\bibnamefont {Gninenko}}, \bibinfo {author} {\bibfnamefont
  {H.}~\bibnamefont {Gómez}}, \bibinfo {author} {\bibfnamefont
  {M.}~\bibnamefont {Hasinoff}}, \bibinfo {author} {\bibfnamefont
  {F.}~\bibnamefont {Heinsius}}, \bibinfo {author} {\bibfnamefont
  {D.}~\bibnamefont {Hoffmann}}, \bibinfo {author} {\bibfnamefont
  {I.}~\bibnamefont {Irastorza}}, \bibinfo {author} {\bibfnamefont
  {J.}~\bibnamefont {Jacoby}}, \bibinfo {author} {\bibfnamefont
  {K.}~\bibnamefont {Jakovčić}}, \bibinfo {author} {\bibfnamefont
  {D.}~\bibnamefont {Kang}}, \bibinfo {author} {\bibfnamefont {K.}~\bibnamefont
  {Königsmann}}, \bibinfo {author} {\bibfnamefont {R.}~\bibnamefont
  {Kotthaus}}, \bibinfo {author} {\bibfnamefont {K.}~\bibnamefont {Kousouris}},
  \bibinfo {author} {\bibfnamefont {M.}~\bibnamefont {Krčmar}}, \bibinfo
  {author} {\bibfnamefont {M.}~\bibnamefont {Kuster}}, \bibinfo {author}
  {\bibfnamefont {B.}~\bibnamefont {Lakić}}, \bibinfo {author} {\bibfnamefont
  {A.}~\bibnamefont {Liolios}}, \bibinfo {author} {\bibfnamefont
  {A.}~\bibnamefont {Ljubičić}}, \bibinfo {author} {\bibfnamefont
  {G.}~\bibnamefont {Lutz}}, \bibinfo {author} {\bibfnamefont {G.}~\bibnamefont
  {Luzón}}, \bibinfo {author} {\bibfnamefont {D.}~\bibnamefont {Miller}},
  \bibinfo {author} {\bibfnamefont {T.}~\bibnamefont {Papaevangelou}}, \bibinfo
  {author} {\bibfnamefont {M.}~\bibnamefont {Pivovaroff}}, \bibinfo {author}
  {\bibfnamefont {G.}~\bibnamefont {Raffelt}}, \bibinfo {author} {\bibfnamefont
  {J.}~\bibnamefont {Redondo}}, \bibinfo {author} {\bibfnamefont
  {H.}~\bibnamefont {Riege}}, \bibinfo {author} {\bibfnamefont
  {A.}~\bibnamefont {Rodríguez}}, \bibinfo {author} {\bibfnamefont
  {J.}~\bibnamefont {Ruz}}, \bibinfo {author} {\bibfnamefont {I.}~\bibnamefont
  {Savvidis}}, \bibinfo {author} {\bibfnamefont {Y.}~\bibnamefont
  {Semertzidis}}, \bibinfo {author} {\bibfnamefont {L.}~\bibnamefont
  {Stewart}}, \bibinfo {author} {\bibfnamefont {K.~V.}\ \bibnamefont {Bibber}},
  \bibinfo {author} {\bibfnamefont {J.}~\bibnamefont {Vieira}}, \bibinfo
  {author} {\bibfnamefont {J.}~\bibnamefont {Villar}}, \bibinfo {author}
  {\bibfnamefont {J.}~\bibnamefont {Vogel}}, \bibinfo {author} {\bibfnamefont
  {L.}~\bibnamefont {Walckiers}},\ and\ \bibinfo {author} {\bibfnamefont
  {K.}~\bibnamefont {Zioutas}},\ }\bibfield  {title} {\bibinfo {title} {{CAST}
  constraints on the axion-electron coupling},\ }\href
  {https://doi.org/10.1088/1475-7516/2013/05/010} {\bibfield  {journal}
  {\bibinfo  {journal} {J. Cosmol. Astropart. Phys.}\ }\textbf {\bibinfo
  {volume} {2013}}\bibinfo  {number} { (05)},\ \bibinfo {pages}
  {010}}\BibitemShut {NoStop}%
\bibitem [{\citenamefont {Pugnat}\ \emph {et~al.}(2014)\citenamefont {Pugnat},
  \citenamefont {Ballou}, \citenamefont {Schott}, \citenamefont {Husek},
  \citenamefont {Sulc}, \citenamefont {Deferne}, \citenamefont {Duvillaret},
  \citenamefont {Finger}, \citenamefont {Flekova}, \citenamefont {Hosek} \emph
  {et~al.}}]{pugnat2014search}%
  \BibitemOpen
\bibfield  {number} {  }\bibfield  {author} {\bibinfo {author} {\bibfnamefont
  {P.}~\bibnamefont {Pugnat}}, \bibinfo {author} {\bibfnamefont
  {R.}~\bibnamefont {Ballou}}, \bibinfo {author} {\bibfnamefont
  {M.}~\bibnamefont {Schott}}, \bibinfo {author} {\bibfnamefont
  {T.}~\bibnamefont {Husek}}, \bibinfo {author} {\bibfnamefont
  {M.}~\bibnamefont {Sulc}}, \bibinfo {author} {\bibfnamefont {G.}~\bibnamefont
  {Deferne}}, \bibinfo {author} {\bibfnamefont {L.}~\bibnamefont {Duvillaret}},
  \bibinfo {author} {\bibfnamefont {M.}~\bibnamefont {Finger}}, \bibinfo
  {author} {\bibfnamefont {L.}~\bibnamefont {Flekova}}, \bibinfo {author}
  {\bibfnamefont {J.}~\bibnamefont {Hosek}}, \emph {et~al.},\ }\bibfield
  {title} {\bibinfo {title} {Search for weakly interacting sub-e{V} particles
  with the {OSQAR} laser-based experiment: results and perspectives},\ }\href
  {https://doi.org/10.1140/epjc/s10052-014-3027-8} {\bibfield  {journal}
  {\bibinfo  {journal} {Eur. Phys. J. C}\ }\textbf {\bibinfo {volume} {74}},\
  \bibinfo {pages} {3027} (\bibinfo {year} {2014})}\BibitemShut {NoStop}%
\bibitem [{\citenamefont {Flambaum}\ \emph {et~al.}(2018)\citenamefont
  {Flambaum}, \citenamefont {Tran~Tan}, \citenamefont {Samsonov}, \citenamefont
  {Stadnik},\ and\ \citenamefont {Budker}}]{flambaum2018resonant}%
  \BibitemOpen
  \bibfield  {author} {\bibinfo {author} {\bibfnamefont {V.}~\bibnamefont
  {Flambaum}}, \bibinfo {author} {\bibfnamefont {H.}~\bibnamefont {Tran~Tan}},
  \bibinfo {author} {\bibfnamefont {I.}~\bibnamefont {Samsonov}}, \bibinfo
  {author} {\bibfnamefont {Y.}~\bibnamefont {Stadnik}},\ and\ \bibinfo {author}
  {\bibfnamefont {D.}~\bibnamefont {Budker}},\ }\bibfield  {title} {\bibinfo
  {title} {Resonant detection and production of axions with atoms},\ }\href
  {https://doi.org/10.1142/S0217751X1844030X} {\bibfield  {journal} {\bibinfo
  {journal} {Int. J. Mod. Phys. A}\ }\textbf {\bibinfo {volume} {33}},\
  \bibinfo {pages} {1844030} (\bibinfo {year} {2018})}\BibitemShut {NoStop}%
\bibitem [{\citenamefont {Roussy}\ \emph {et~al.}(2021)\citenamefont {Roussy},
  \citenamefont {Palken}, \citenamefont {Cairncross}, \citenamefont {Brubaker},
  \citenamefont {Gresh}, \citenamefont {Grau}, \citenamefont {Cossel},
  \citenamefont {Ng}, \citenamefont {Shagam}, \citenamefont {Zhou},
  \citenamefont {Flambaum}, \citenamefont {Lehnert}, \citenamefont {Ye},\ and\
  \citenamefont {Cornell}}]{Roussy:2021}%
  \BibitemOpen
  \bibfield  {author} {\bibinfo {author} {\bibfnamefont {T.~S.}\ \bibnamefont
  {Roussy}}, \bibinfo {author} {\bibfnamefont {D.~A.}\ \bibnamefont {Palken}},
  \bibinfo {author} {\bibfnamefont {W.~B.}\ \bibnamefont {Cairncross}},
  \bibinfo {author} {\bibfnamefont {B.~M.}\ \bibnamefont {Brubaker}}, \bibinfo
  {author} {\bibfnamefont {D.~N.}\ \bibnamefont {Gresh}}, \bibinfo {author}
  {\bibfnamefont {M.}~\bibnamefont {Grau}}, \bibinfo {author} {\bibfnamefont
  {K.~C.}\ \bibnamefont {Cossel}}, \bibinfo {author} {\bibfnamefont {K.~B.}\
  \bibnamefont {Ng}}, \bibinfo {author} {\bibfnamefont {Y.}~\bibnamefont
  {Shagam}}, \bibinfo {author} {\bibfnamefont {Y.}~\bibnamefont {Zhou}},
  \bibinfo {author} {\bibfnamefont {V.~V.}\ \bibnamefont {Flambaum}}, \bibinfo
  {author} {\bibfnamefont {K.~W.}\ \bibnamefont {Lehnert}}, \bibinfo {author}
  {\bibfnamefont {J.}~\bibnamefont {Ye}},\ and\ \bibinfo {author}
  {\bibfnamefont {E.~A.}\ \bibnamefont {Cornell}},\ }\bibfield  {title}
  {\bibinfo {title} {Experimental constraint on axionlike particles over seven
  orders of magnitude in mass},\ }\href
  {https://doi.org/10.1103/PhysRevLett.126.171301} {\bibfield  {journal}
  {\bibinfo  {journal} {Phys.\ Rev.\ Lett.}\ }\textbf {\bibinfo {volume}
  {126}},\ \bibinfo {pages} {171301} (\bibinfo {year} {2021})}\BibitemShut
  {NoStop}%
\bibitem [{\citenamefont {Aybas}\ \emph {et~al.}(2021)\citenamefont {Aybas},
  \citenamefont {Adam}, \citenamefont {Blumenthal}, \citenamefont {Gramolin},
  \citenamefont {Johnson}, \citenamefont {Kleyheeg}, \citenamefont {Afach},
  \citenamefont {Blanchard}, \citenamefont {Centers}, \citenamefont {Garcon},
  \citenamefont {Engler}, \citenamefont {Figueroa}, \citenamefont {Sendra},
  \citenamefont {Wickenbrock}, \citenamefont {Lawson}, \citenamefont {Wang},
  \citenamefont {Wu}, \citenamefont {Luo}, \citenamefont {Mani}, \citenamefont
  {Mauskopf}, \citenamefont {Graham}, \citenamefont {Rajendran}, \citenamefont
  {Kimball}, \citenamefont {Budker},\ and\ \citenamefont
  {Sushkov}}]{aybas2021}%
  \BibitemOpen
  \bibfield  {author} {\bibinfo {author} {\bibfnamefont {D.}~\bibnamefont
  {Aybas}}, \bibinfo {author} {\bibfnamefont {J.}~\bibnamefont {Adam}},
  \bibinfo {author} {\bibfnamefont {E.}~\bibnamefont {Blumenthal}}, \bibinfo
  {author} {\bibfnamefont {A.~V.}\ \bibnamefont {Gramolin}}, \bibinfo {author}
  {\bibfnamefont {D.}~\bibnamefont {Johnson}}, \bibinfo {author} {\bibfnamefont
  {A.}~\bibnamefont {Kleyheeg}}, \bibinfo {author} {\bibfnamefont
  {S.}~\bibnamefont {Afach}}, \bibinfo {author} {\bibfnamefont {J.~W.}\
  \bibnamefont {Blanchard}}, \bibinfo {author} {\bibfnamefont {G.~P.}\
  \bibnamefont {Centers}}, \bibinfo {author} {\bibfnamefont {A.}~\bibnamefont
  {Garcon}}, \bibinfo {author} {\bibfnamefont {M.}~\bibnamefont {Engler}},
  \bibinfo {author} {\bibfnamefont {N.~L.}\ \bibnamefont {Figueroa}}, \bibinfo
  {author} {\bibfnamefont {M.~G.}\ \bibnamefont {Sendra}}, \bibinfo {author}
  {\bibfnamefont {A.}~\bibnamefont {Wickenbrock}}, \bibinfo {author}
  {\bibfnamefont {M.}~\bibnamefont {Lawson}}, \bibinfo {author} {\bibfnamefont
  {T.}~\bibnamefont {Wang}}, \bibinfo {author} {\bibfnamefont {T.}~\bibnamefont
  {Wu}}, \bibinfo {author} {\bibfnamefont {H.}~\bibnamefont {Luo}}, \bibinfo
  {author} {\bibfnamefont {H.}~\bibnamefont {Mani}}, \bibinfo {author}
  {\bibfnamefont {P.}~\bibnamefont {Mauskopf}}, \bibinfo {author}
  {\bibfnamefont {P.~W.}\ \bibnamefont {Graham}}, \bibinfo {author}
  {\bibfnamefont {S.}~\bibnamefont {Rajendran}}, \bibinfo {author}
  {\bibfnamefont {D.~F.~J.}\ \bibnamefont {Kimball}}, \bibinfo {author}
  {\bibfnamefont {D.}~\bibnamefont {Budker}},\ and\ \bibinfo {author}
  {\bibfnamefont {A.~O.}\ \bibnamefont {Sushkov}},\ }\bibfield  {title}
  {\bibinfo {title} {Search for axionlike dark matter using solid-state nuclear
  magnetic resonance},\ }\href {https://doi.org/10.1103/PhysRevLett.126.141802}
  {\bibfield  {journal} {\bibinfo  {journal} {Phys. Rev. Lett.}\ }\textbf
  {\bibinfo {volume} {126}},\ \bibinfo {pages} {141802} (\bibinfo {year}
  {2021})}\BibitemShut {NoStop}%
\bibitem [{\citenamefont {Adams}\ \emph {et~al.}(2023)\citenamefont {Adams},
  \citenamefont {Aggarwal}, \citenamefont {Agrawal}, \citenamefont
  {Balafendiev}, \citenamefont {Bartram}, \citenamefont {Baryakhtar},
  \citenamefont {Bekker}, \citenamefont {Belov}, \citenamefont {Berggren},
  \citenamefont {Berlin}, \citenamefont {Boutan}, \citenamefont {Bowring},
  \citenamefont {Budker}, \citenamefont {Caldwell}, \citenamefont {Carenza},
  \citenamefont {Carosi}, \citenamefont {Cervantes}, \citenamefont
  {Chakrabarty}, \citenamefont {Chaudhuri}, \citenamefont {Chen}, \citenamefont
  {Cheong}, \citenamefont {Chou}, \citenamefont {Co}, \citenamefont {Conrad},
  \citenamefont {Croon}, \citenamefont {D'Agnolo}, \citenamefont {Demarteau},
  \citenamefont {DePorzio}, \citenamefont {Descalle}, \citenamefont {Desch},
  \citenamefont {Luzio}, \citenamefont {Diaz-Morcillo}, \citenamefont {Dona},
  \citenamefont {Drachnev}, \citenamefont {Droster}, \citenamefont {Du},
  \citenamefont {Dunne}, \citenamefont {Döbrich}, \citenamefont {Ellis},
  \citenamefont {Essig}, \citenamefont {Fan}, \citenamefont {Foster},
  \citenamefont {Fry}, \citenamefont {Rosso}, \citenamefont {Barceló},
  \citenamefont {Irastorza}, \citenamefont {Gardner}, \citenamefont {Geraci},
  \citenamefont {Ghosh}, \citenamefont {Giaccone}, \citenamefont {Giannotti},
  \citenamefont {Gimeno}, \citenamefont {Grin}, \citenamefont {Grote},
  \citenamefont {Guzzetti}, \citenamefont {Awida}, \citenamefont {Henning},
  \citenamefont {Hoof}, \citenamefont {Hoshino}, \citenamefont {Irsic},
  \citenamefont {Irwin}, \citenamefont {Jackson}, \citenamefont {Kimball},
  \citenamefont {Jaeckel}, \citenamefont {Jakovcic}, \citenamefont {Jewell},
  \citenamefont {Kagan}, \citenamefont {Kahn}, \citenamefont {Khatiwada},
  \citenamefont {Knirck}, \citenamefont {Kovachy}, \citenamefont {Krueger},
  \citenamefont {Kuenstner}, \citenamefont {Kurinsky}, \citenamefont {Leane},
  \citenamefont {Leder}, \citenamefont {Lee}, \citenamefont {Lehnert},
  \citenamefont {Lentz}, \citenamefont {Lewis}, \citenamefont {Liu},
  \citenamefont {Lynn}, \citenamefont {Majorovits}, \citenamefont {Marsh},
  \citenamefont {Maruyama}, \citenamefont {McAllister}, \citenamefont {Millar},
  \citenamefont {Miller}, \citenamefont {Mitchell}, \citenamefont {Morampudi},
  \citenamefont {Mueller}, \citenamefont {Nagaitsev}, \citenamefont {Nardi},
  \citenamefont {Noroozian}, \citenamefont {O'Hare}, \citenamefont {Oblath},
  \citenamefont {Ouellet}, \citenamefont {Pappas}, \citenamefont {Peiris},
  \citenamefont {Perez}, \citenamefont {Phipps}, \citenamefont {Pivovaroff},
  \citenamefont {Quílez}, \citenamefont {Rapidis}, \citenamefont {Robles},
  \citenamefont {Rogers}, \citenamefont {Rudolph}, \citenamefont {Ruz},
  \citenamefont {Rybka}, \citenamefont {Safdari}, \citenamefont {Safdi},
  \citenamefont {Safronova}, \citenamefont {Salemi}, \citenamefont {Schuster},
  \citenamefont {Schwartzman}, \citenamefont {Shu}, \citenamefont
  {Simanovskaia}, \citenamefont {Singh}, \citenamefont {Singh}, \citenamefont
  {Sinha}, \citenamefont {Sinnis}, \citenamefont {Siodlaczek}, \citenamefont
  {Smith}, \citenamefont {Snow}, \citenamefont {Sokolov}, \citenamefont
  {Sonnenschein}, \citenamefont {Speller}, \citenamefont {Stadnik},
  \citenamefont {Sun}, \citenamefont {Sushkov}, \citenamefont {Tait},
  \citenamefont {Takhistov}, \citenamefont {Tanner}, \citenamefont {Tavecchio},
  \citenamefont {Temples}, \citenamefont {Thomas}, \citenamefont {Tobar},
  \citenamefont {Toro}, \citenamefont {Tsai}, \citenamefont {van Assendelft},
  \citenamefont {van Bibber}, \citenamefont {Vandegar}, \citenamefont
  {Visinelli}, \citenamefont {Vitagliano}, \citenamefont {Vogel}, \citenamefont
  {Wang}, \citenamefont {Wickenbrock}, \citenamefont {Winslow}, \citenamefont
  {Withington}, \citenamefont {Wooten}, \citenamefont {Yang}, \citenamefont
  {Young}, \citenamefont {Yu}, \citenamefont {Zhou},\ and\ \citenamefont
  {Zhou}}]{adams2023axion}%
  \BibitemOpen
  \bibfield  {author} {\bibinfo {author} {\bibfnamefont {C.~B.}\ \bibnamefont
  {Adams}}, \bibinfo {author} {\bibfnamefont {N.}~\bibnamefont {Aggarwal}},
  \bibinfo {author} {\bibfnamefont {A.}~\bibnamefont {Agrawal}}, \bibinfo
  {author} {\bibfnamefont {R.}~\bibnamefont {Balafendiev}}, \bibinfo {author}
  {\bibfnamefont {C.}~\bibnamefont {Bartram}}, \bibinfo {author} {\bibfnamefont
  {M.}~\bibnamefont {Baryakhtar}}, \bibinfo {author} {\bibfnamefont
  {H.}~\bibnamefont {Bekker}}, \bibinfo {author} {\bibfnamefont
  {P.}~\bibnamefont {Belov}}, \bibinfo {author} {\bibfnamefont {K.~K.}\
  \bibnamefont {Berggren}}, \bibinfo {author} {\bibfnamefont {A.}~\bibnamefont
  {Berlin}}, \bibinfo {author} {\bibfnamefont {C.}~\bibnamefont {Boutan}},
  \bibinfo {author} {\bibfnamefont {D.}~\bibnamefont {Bowring}}, \bibinfo
  {author} {\bibfnamefont {D.}~\bibnamefont {Budker}}, \bibinfo {author}
  {\bibfnamefont {A.}~\bibnamefont {Caldwell}}, \bibinfo {author}
  {\bibfnamefont {P.}~\bibnamefont {Carenza}}, \bibinfo {author} {\bibfnamefont
  {G.}~\bibnamefont {Carosi}}, \bibinfo {author} {\bibfnamefont
  {R.}~\bibnamefont {Cervantes}}, \bibinfo {author} {\bibfnamefont {S.~S.}\
  \bibnamefont {Chakrabarty}}, \bibinfo {author} {\bibfnamefont
  {S.}~\bibnamefont {Chaudhuri}}, \bibinfo {author} {\bibfnamefont {T.~Y.}\
  \bibnamefont {Chen}}, \bibinfo {author} {\bibfnamefont {S.}~\bibnamefont
  {Cheong}}, \bibinfo {author} {\bibfnamefont {A.}~\bibnamefont {Chou}},
  \bibinfo {author} {\bibfnamefont {R.~T.}\ \bibnamefont {Co}}, \bibinfo
  {author} {\bibfnamefont {J.}~\bibnamefont {Conrad}}, \bibinfo {author}
  {\bibfnamefont {D.}~\bibnamefont {Croon}}, \bibinfo {author} {\bibfnamefont
  {R.~T.}\ \bibnamefont {D'Agnolo}}, \bibinfo {author} {\bibfnamefont
  {M.}~\bibnamefont {Demarteau}}, \bibinfo {author} {\bibfnamefont
  {N.}~\bibnamefont {DePorzio}}, \bibinfo {author} {\bibfnamefont
  {M.}~\bibnamefont {Descalle}}, \bibinfo {author} {\bibfnamefont
  {K.}~\bibnamefont {Desch}}, \bibinfo {author} {\bibfnamefont {L.~D.}\
  \bibnamefont {Luzio}}, \bibinfo {author} {\bibfnamefont {A.}~\bibnamefont
  {Diaz-Morcillo}}, \bibinfo {author} {\bibfnamefont {K.}~\bibnamefont {Dona}},
  \bibinfo {author} {\bibfnamefont {I.~S.}\ \bibnamefont {Drachnev}}, \bibinfo
  {author} {\bibfnamefont {A.}~\bibnamefont {Droster}}, \bibinfo {author}
  {\bibfnamefont {N.}~\bibnamefont {Du}}, \bibinfo {author} {\bibfnamefont
  {K.}~\bibnamefont {Dunne}}, \bibinfo {author} {\bibfnamefont
  {B.}~\bibnamefont {Döbrich}}, \bibinfo {author} {\bibfnamefont {S.~A.~R.}\
  \bibnamefont {Ellis}}, \bibinfo {author} {\bibfnamefont {R.}~\bibnamefont
  {Essig}}, \bibinfo {author} {\bibfnamefont {J.}~\bibnamefont {Fan}}, \bibinfo
  {author} {\bibfnamefont {J.~W.}\ \bibnamefont {Foster}}, \bibinfo {author}
  {\bibfnamefont {J.~T.}\ \bibnamefont {Fry}}, \bibinfo {author} {\bibfnamefont
  {A.~G.}\ \bibnamefont {Rosso}}, \bibinfo {author} {\bibfnamefont {J.~M.~G.}\
  \bibnamefont {Barceló}}, \bibinfo {author} {\bibfnamefont {I.~G.}\
  \bibnamefont {Irastorza}}, \bibinfo {author} {\bibfnamefont {S.}~\bibnamefont
  {Gardner}}, \bibinfo {author} {\bibfnamefont {A.~A.}\ \bibnamefont {Geraci}},
  \bibinfo {author} {\bibfnamefont {S.}~\bibnamefont {Ghosh}}, \bibinfo
  {author} {\bibfnamefont {B.}~\bibnamefont {Giaccone}}, \bibinfo {author}
  {\bibfnamefont {M.}~\bibnamefont {Giannotti}}, \bibinfo {author}
  {\bibfnamefont {B.}~\bibnamefont {Gimeno}}, \bibinfo {author} {\bibfnamefont
  {D.}~\bibnamefont {Grin}}, \bibinfo {author} {\bibfnamefont {H.}~\bibnamefont
  {Grote}}, \bibinfo {author} {\bibfnamefont {M.}~\bibnamefont {Guzzetti}},
  \bibinfo {author} {\bibfnamefont {M.~H.}\ \bibnamefont {Awida}}, \bibinfo
  {author} {\bibfnamefont {R.}~\bibnamefont {Henning}}, \bibinfo {author}
  {\bibfnamefont {S.}~\bibnamefont {Hoof}}, \bibinfo {author} {\bibfnamefont
  {G.}~\bibnamefont {Hoshino}}, \bibinfo {author} {\bibfnamefont
  {V.}~\bibnamefont {Irsic}}, \bibinfo {author} {\bibfnamefont {K.~D.}\
  \bibnamefont {Irwin}}, \bibinfo {author} {\bibfnamefont {H.}~\bibnamefont
  {Jackson}}, \bibinfo {author} {\bibfnamefont {D.~F.~J.}\ \bibnamefont
  {Kimball}}, \bibinfo {author} {\bibfnamefont {J.}~\bibnamefont {Jaeckel}},
  \bibinfo {author} {\bibfnamefont {K.}~\bibnamefont {Jakovcic}}, \bibinfo
  {author} {\bibfnamefont {M.~J.}\ \bibnamefont {Jewell}}, \bibinfo {author}
  {\bibfnamefont {M.}~\bibnamefont {Kagan}}, \bibinfo {author} {\bibfnamefont
  {Y.}~\bibnamefont {Kahn}}, \bibinfo {author} {\bibfnamefont {R.}~\bibnamefont
  {Khatiwada}}, \bibinfo {author} {\bibfnamefont {S.}~\bibnamefont {Knirck}},
  \bibinfo {author} {\bibfnamefont {T.}~\bibnamefont {Kovachy}}, \bibinfo
  {author} {\bibfnamefont {P.}~\bibnamefont {Krueger}}, \bibinfo {author}
  {\bibfnamefont {S.~E.}\ \bibnamefont {Kuenstner}}, \bibinfo {author}
  {\bibfnamefont {N.~A.}\ \bibnamefont {Kurinsky}}, \bibinfo {author}
  {\bibfnamefont {R.~K.}\ \bibnamefont {Leane}}, \bibinfo {author}
  {\bibfnamefont {A.~F.}\ \bibnamefont {Leder}}, \bibinfo {author}
  {\bibfnamefont {C.}~\bibnamefont {Lee}}, \bibinfo {author} {\bibfnamefont
  {K.~W.}\ \bibnamefont {Lehnert}}, \bibinfo {author} {\bibfnamefont {E.~W.}\
  \bibnamefont {Lentz}}, \bibinfo {author} {\bibfnamefont {S.~M.}\ \bibnamefont
  {Lewis}}, \bibinfo {author} {\bibfnamefont {J.}~\bibnamefont {Liu}}, \bibinfo
  {author} {\bibfnamefont {M.}~\bibnamefont {Lynn}}, \bibinfo {author}
  {\bibfnamefont {B.}~\bibnamefont {Majorovits}}, \bibinfo {author}
  {\bibfnamefont {D.~J.~E.}\ \bibnamefont {Marsh}}, \bibinfo {author}
  {\bibfnamefont {R.~H.}\ \bibnamefont {Maruyama}}, \bibinfo {author}
  {\bibfnamefont {B.~T.}\ \bibnamefont {McAllister}}, \bibinfo {author}
  {\bibfnamefont {A.~J.}\ \bibnamefont {Millar}}, \bibinfo {author}
  {\bibfnamefont {D.~W.}\ \bibnamefont {Miller}}, \bibinfo {author}
  {\bibfnamefont {J.}~\bibnamefont {Mitchell}}, \bibinfo {author}
  {\bibfnamefont {S.}~\bibnamefont {Morampudi}}, \bibinfo {author}
  {\bibfnamefont {G.}~\bibnamefont {Mueller}}, \bibinfo {author} {\bibfnamefont
  {S.}~\bibnamefont {Nagaitsev}}, \bibinfo {author} {\bibfnamefont
  {E.}~\bibnamefont {Nardi}}, \bibinfo {author} {\bibfnamefont
  {O.}~\bibnamefont {Noroozian}}, \bibinfo {author} {\bibfnamefont {C.~A.~J.}\
  \bibnamefont {O'Hare}}, \bibinfo {author} {\bibfnamefont {N.~S.}\
  \bibnamefont {Oblath}}, \bibinfo {author} {\bibfnamefont {J.~L.}\
  \bibnamefont {Ouellet}}, \bibinfo {author} {\bibfnamefont {K.~M.~W.}\
  \bibnamefont {Pappas}}, \bibinfo {author} {\bibfnamefont {H.~V.}\
  \bibnamefont {Peiris}}, \bibinfo {author} {\bibfnamefont {K.}~\bibnamefont
  {Perez}}, \bibinfo {author} {\bibfnamefont {A.}~\bibnamefont {Phipps}},
  \bibinfo {author} {\bibfnamefont {M.~J.}\ \bibnamefont {Pivovaroff}},
  \bibinfo {author} {\bibfnamefont {P.}~\bibnamefont {Quílez}}, \bibinfo
  {author} {\bibfnamefont {N.~M.}\ \bibnamefont {Rapidis}}, \bibinfo {author}
  {\bibfnamefont {V.~H.}\ \bibnamefont {Robles}}, \bibinfo {author}
  {\bibfnamefont {K.~K.}\ \bibnamefont {Rogers}}, \bibinfo {author}
  {\bibfnamefont {J.}~\bibnamefont {Rudolph}}, \bibinfo {author} {\bibfnamefont
  {J.}~\bibnamefont {Ruz}}, \bibinfo {author} {\bibfnamefont {G.}~\bibnamefont
  {Rybka}}, \bibinfo {author} {\bibfnamefont {M.}~\bibnamefont {Safdari}},
  \bibinfo {author} {\bibfnamefont {B.~R.}\ \bibnamefont {Safdi}}, \bibinfo
  {author} {\bibfnamefont {M.~S.}\ \bibnamefont {Safronova}}, \bibinfo {author}
  {\bibfnamefont {C.~P.}\ \bibnamefont {Salemi}}, \bibinfo {author}
  {\bibfnamefont {P.}~\bibnamefont {Schuster}}, \bibinfo {author}
  {\bibfnamefont {A.}~\bibnamefont {Schwartzman}}, \bibinfo {author}
  {\bibfnamefont {J.}~\bibnamefont {Shu}}, \bibinfo {author} {\bibfnamefont
  {M.}~\bibnamefont {Simanovskaia}}, \bibinfo {author} {\bibfnamefont
  {J.}~\bibnamefont {Singh}}, \bibinfo {author} {\bibfnamefont
  {S.}~\bibnamefont {Singh}}, \bibinfo {author} {\bibfnamefont
  {K.}~\bibnamefont {Sinha}}, \bibinfo {author} {\bibfnamefont {J.~T.}\
  \bibnamefont {Sinnis}}, \bibinfo {author} {\bibfnamefont {M.}~\bibnamefont
  {Siodlaczek}}, \bibinfo {author} {\bibfnamefont {M.~S.}\ \bibnamefont
  {Smith}}, \bibinfo {author} {\bibfnamefont {W.~M.}\ \bibnamefont {Snow}},
  \bibinfo {author} {\bibfnamefont {A.~V.}\ \bibnamefont {Sokolov}}, \bibinfo
  {author} {\bibfnamefont {A.}~\bibnamefont {Sonnenschein}}, \bibinfo {author}
  {\bibfnamefont {D.~H.}\ \bibnamefont {Speller}}, \bibinfo {author}
  {\bibfnamefont {Y.~V.}\ \bibnamefont {Stadnik}}, \bibinfo {author}
  {\bibfnamefont {C.}~\bibnamefont {Sun}}, \bibinfo {author} {\bibfnamefont
  {A.~O.}\ \bibnamefont {Sushkov}}, \bibinfo {author} {\bibfnamefont
  {T.~M.~P.}\ \bibnamefont {Tait}}, \bibinfo {author} {\bibfnamefont
  {V.}~\bibnamefont {Takhistov}}, \bibinfo {author} {\bibfnamefont {D.~B.}\
  \bibnamefont {Tanner}}, \bibinfo {author} {\bibfnamefont {F.}~\bibnamefont
  {Tavecchio}}, \bibinfo {author} {\bibfnamefont {D.~J.}\ \bibnamefont
  {Temples}}, \bibinfo {author} {\bibfnamefont {J.~H.}\ \bibnamefont {Thomas}},
  \bibinfo {author} {\bibfnamefont {M.~E.}\ \bibnamefont {Tobar}}, \bibinfo
  {author} {\bibfnamefont {N.}~\bibnamefont {Toro}}, \bibinfo {author}
  {\bibfnamefont {Y.~D.}\ \bibnamefont {Tsai}}, \bibinfo {author}
  {\bibfnamefont {E.~C.}\ \bibnamefont {van Assendelft}}, \bibinfo {author}
  {\bibfnamefont {K.}~\bibnamefont {van Bibber}}, \bibinfo {author}
  {\bibfnamefont {M.}~\bibnamefont {Vandegar}}, \bibinfo {author}
  {\bibfnamefont {L.}~\bibnamefont {Visinelli}}, \bibinfo {author}
  {\bibfnamefont {E.}~\bibnamefont {Vitagliano}}, \bibinfo {author}
  {\bibfnamefont {J.~K.}\ \bibnamefont {Vogel}}, \bibinfo {author}
  {\bibfnamefont {Z.}~\bibnamefont {Wang}}, \bibinfo {author} {\bibfnamefont
  {A.}~\bibnamefont {Wickenbrock}}, \bibinfo {author} {\bibfnamefont
  {L.}~\bibnamefont {Winslow}}, \bibinfo {author} {\bibfnamefont
  {S.}~\bibnamefont {Withington}}, \bibinfo {author} {\bibfnamefont
  {M.}~\bibnamefont {Wooten}}, \bibinfo {author} {\bibfnamefont
  {J.}~\bibnamefont {Yang}}, \bibinfo {author} {\bibfnamefont {B.~A.}\
  \bibnamefont {Young}}, \bibinfo {author} {\bibfnamefont {F.}~\bibnamefont
  {Yu}}, \bibinfo {author} {\bibfnamefont {K.}~\bibnamefont {Zhou}},\ and\
  \bibinfo {author} {\bibfnamefont {T.}~\bibnamefont {Zhou}},\ }\href@noop {}
  {\bibinfo {title} {Axion dark matter}} (\bibinfo {year} {2023}),\ \Eprint
  {https://arxiv.org/abs/2203.14923} {arXiv:2203.14923 [hep-ex]} \BibitemShut
  {NoStop}%
\bibitem [{\citenamefont {Arza}\ and\ \citenamefont
  {Todarello}(2021)}]{sym13112150}%
  \BibitemOpen
  \bibfield  {author} {\bibinfo {author} {\bibfnamefont {A.}~\bibnamefont
  {Arza}}\ and\ \bibinfo {author} {\bibfnamefont {E.}~\bibnamefont
  {Todarello}},\ }\bibfield  {title} {\bibinfo {title} {The echo method for
  axion dark matter detection},\ }\bibfield  {journal} {\bibinfo  {journal}
  {Symmetry}\ }\textbf {\bibinfo {volume} {13}},\ \href
  {https://doi.org/10.3390/sym13112150} {10.3390/sym13112150} (\bibinfo {year}
  {2021})\BibitemShut {NoStop}%
\bibitem [{\citenamefont {Wang}\ \emph {et~al.}(2022)\citenamefont {Wang},
  \citenamefont {Su}, \citenamefont {Jiang}, \citenamefont {Huang},
  \citenamefont {Qin}, \citenamefont {Guo}, \citenamefont {Wang}, \citenamefont
  {Hu}, \citenamefont {Ji}, \citenamefont {Fadeev}, \citenamefont {Peng},\ and\
  \citenamefont {Budker}}]{Budker:2023a}%
  \BibitemOpen
  \bibfield  {author} {\bibinfo {author} {\bibfnamefont {Y.}~\bibnamefont
  {Wang}}, \bibinfo {author} {\bibfnamefont {H.}~\bibnamefont {Su}}, \bibinfo
  {author} {\bibfnamefont {M.}~\bibnamefont {Jiang}}, \bibinfo {author}
  {\bibfnamefont {Y.}~\bibnamefont {Huang}}, \bibinfo {author} {\bibfnamefont
  {Y.}~\bibnamefont {Qin}}, \bibinfo {author} {\bibfnamefont {C.}~\bibnamefont
  {Guo}}, \bibinfo {author} {\bibfnamefont {Z.}~\bibnamefont {Wang}}, \bibinfo
  {author} {\bibfnamefont {D.}~\bibnamefont {Hu}}, \bibinfo {author}
  {\bibfnamefont {W.}~\bibnamefont {Ji}}, \bibinfo {author} {\bibfnamefont
  {P.}~\bibnamefont {Fadeev}}, \bibinfo {author} {\bibfnamefont
  {X.}~\bibnamefont {Peng}},\ and\ \bibinfo {author} {\bibfnamefont
  {D.}~\bibnamefont {Budker}},\ }\bibfield  {title} {\bibinfo {title} {Limits
  on axions and axionlike particles within the axion window using a spin-based
  amplifier},\ }\href {https://doi.org/10.1103/PhysRevLett.129.051801}
  {\bibfield  {journal} {\bibinfo  {journal} {Phys. Rev. Lett.}\ }\textbf
  {\bibinfo {volume} {129}},\ \bibinfo {pages} {051801} (\bibinfo {year}
  {2022})}\BibitemShut {NoStop}%
\bibitem [{\citenamefont {Tobar}\ \emph {et~al.}(2022)\citenamefont {Tobar},
  \citenamefont {Thomson}, \citenamefont {Campbell}, \citenamefont {Quiskamp},
  \citenamefont {Bourhill}, \citenamefont {McAllister}, \citenamefont
  {Ivanov},\ and\ \citenamefont {Goryachev}}]{sym14102165}%
  \BibitemOpen
  \bibfield  {author} {\bibinfo {author} {\bibfnamefont {M.~E.}\ \bibnamefont
  {Tobar}}, \bibinfo {author} {\bibfnamefont {C.~A.}\ \bibnamefont {Thomson}},
  \bibinfo {author} {\bibfnamefont {W.~M.}\ \bibnamefont {Campbell}}, \bibinfo
  {author} {\bibfnamefont {A.}~\bibnamefont {Quiskamp}}, \bibinfo {author}
  {\bibfnamefont {J.~F.}\ \bibnamefont {Bourhill}}, \bibinfo {author}
  {\bibfnamefont {B.~T.}\ \bibnamefont {McAllister}}, \bibinfo {author}
  {\bibfnamefont {E.~N.}\ \bibnamefont {Ivanov}},\ and\ \bibinfo {author}
  {\bibfnamefont {M.}~\bibnamefont {Goryachev}},\ }\bibfield  {title} {\bibinfo
  {title} {Comparing instrument spectral sensitivity of dissimilar
  electromagnetic haloscopes to axion dark matter and high frequency
  gravitational waves},\ }\bibfield  {journal} {\bibinfo  {journal} {Symmetry}\
  }\textbf {\bibinfo {volume} {14}},\ \href
  {https://doi.org/10.3390/sym14102165} {10.3390/sym14102165} (\bibinfo {year}
  {2022})\BibitemShut {NoStop}%
\bibitem [{\citenamefont {Zhang}(2020)}]{sym12010025}%
  \BibitemOpen
  \bibfield  {author} {\bibinfo {author} {\bibfnamefont {H.}~\bibnamefont
  {Zhang}},\ }\bibfield  {title} {\bibinfo {title} {Axion stars},\ }\bibfield
  {journal} {\bibinfo  {journal} {Symmetry}\ }\textbf {\bibinfo {volume}
  {12}},\ \href {https://doi.org/10.3390/sym12010025} {10.3390/sym12010025}
  (\bibinfo {year} {2020})\BibitemShut {NoStop}%
\bibitem [{\citenamefont {Graham}\ and\ \citenamefont
  {Rajendran}(2011)}]{Graham:2011}%
  \BibitemOpen
  \bibfield  {author} {\bibinfo {author} {\bibfnamefont {P.~W.}\ \bibnamefont
  {Graham}}\ and\ \bibinfo {author} {\bibfnamefont {S.}~\bibnamefont
  {Rajendran}},\ }\bibfield  {title} {\bibinfo {title} {Axion dark matter
  detection with cold molecules},\ }\href
  {https://doi.org/10.1103/PhysRevD.84.055013} {\bibfield  {journal} {\bibinfo
  {journal} {Phys.\ Rev.\ D}\ }\textbf {\bibinfo {volume} {84}},\ \bibinfo
  {pages} {055013} (\bibinfo {year} {2011})}\BibitemShut {NoStop}%
\bibitem [{\citenamefont {Graham}\ and\ \citenamefont
  {Rajendran}(2013)}]{graham2013new}%
  \BibitemOpen
  \bibfield  {author} {\bibinfo {author} {\bibfnamefont {P.~W.}\ \bibnamefont
  {Graham}}\ and\ \bibinfo {author} {\bibfnamefont {S.}~\bibnamefont
  {Rajendran}},\ }\bibfield  {title} {\bibinfo {title} {New observables for
  direct detection of axion dark matter},\ }\href
  {https://doi.org/10.1103/PhysRevD.88.035023} {\bibfield  {journal} {\bibinfo
  {journal} {Phys. Rev. D}\ }\textbf {\bibinfo {volume} {88}},\ \bibinfo
  {pages} {035023} (\bibinfo {year} {2013})}\BibitemShut {NoStop}%
\bibitem [{\citenamefont {Budker}\ \emph {et~al.}(2014)\citenamefont {Budker},
  \citenamefont {Graham}, \citenamefont {Ledbetter}, \citenamefont
  {Rajendran},\ and\ \citenamefont {Sushkov}}]{budker2014proposal}%
  \BibitemOpen
  \bibfield  {author} {\bibinfo {author} {\bibfnamefont {D.}~\bibnamefont
  {Budker}}, \bibinfo {author} {\bibfnamefont {P.~W.}\ \bibnamefont {Graham}},
  \bibinfo {author} {\bibfnamefont {M.}~\bibnamefont {Ledbetter}}, \bibinfo
  {author} {\bibfnamefont {S.}~\bibnamefont {Rajendran}},\ and\ \bibinfo
  {author} {\bibfnamefont {A.~O.}\ \bibnamefont {Sushkov}},\ }\bibfield
  {title} {\bibinfo {title} {Proposal for a cosmic axion spin precession
  experiment ({CASPEr})},\ }\href {https://doi.org/10.1103/PhysRevX.4.021030}
  {\bibfield  {journal} {\bibinfo  {journal} {Phys. Rev. X}\ }\textbf {\bibinfo
  {volume} {4}},\ \bibinfo {pages} {021030} (\bibinfo {year}
  {2014})}\BibitemShut {NoStop}%
\bibitem [{\citenamefont {Skripnikov}\ and\ \citenamefont
  {Titov}(2016)}]{Skripnikov:16a}%
  \BibitemOpen
  \bibfield  {author} {\bibinfo {author} {\bibfnamefont {L.~V.}\ \bibnamefont
  {Skripnikov}}\ and\ \bibinfo {author} {\bibfnamefont {A.~V.}\ \bibnamefont
  {Titov}},\ }\bibfield  {title} {\bibinfo {title} {Lcao-based theoretical
  study of {PbTiO$_3$} crystal to search for parity and time reversal violating
  interaction in solids},\ }\href
  {https://doi.org/http://dx.doi.org/10.1063/1.4959973} {\bibfield  {journal}
  {\bibinfo  {journal} {J.\ Chem.\ Phys.}\ }\textbf {\bibinfo {volume} {145}},\
  \bibinfo {eid} {054115} (\bibinfo {year} {2016})}\BibitemShut {NoStop}%
\bibitem [{\citenamefont {Stadnik}\ \emph {et~al.}(2018)\citenamefont
  {Stadnik}, \citenamefont {Dzuba},\ and\ \citenamefont
  {Flambaum}}]{Stadnik:2018}%
  \BibitemOpen
  \bibfield  {author} {\bibinfo {author} {\bibfnamefont {Y.~V.}\ \bibnamefont
  {Stadnik}}, \bibinfo {author} {\bibfnamefont {V.~A.}\ \bibnamefont {Dzuba}},\
  and\ \bibinfo {author} {\bibfnamefont {V.~V.}\ \bibnamefont {Flambaum}},\
  }\bibfield  {title} {\bibinfo {title} {Improved limits on
  axionlike-particle-mediated {$P$},{$T$}-violating interactions between
  electrons and nucleons from electric dipole moments of atoms and molecules},\
  }\href {https://doi.org/10.1103/PhysRevLett.120.013202} {\bibfield  {journal}
  {\bibinfo  {journal} {Phys.\ Rev.\ Lett.}\ }\textbf {\bibinfo {volume}
  {120}},\ \bibinfo {pages} {013202} (\bibinfo {year} {2018})},\ \bibinfo
  {note} {see also https://arxiv.org/abs/1708.00486v3 (2020)}\BibitemShut
  {NoStop}%
\bibitem [{\citenamefont {Maison}\ \emph
  {et~al.}(2021{\natexlab{a}})\citenamefont {Maison}, \citenamefont {Flambaum},
  \citenamefont {Hutzler},\ and\ \citenamefont {Skripnikov}}]{Maison:2021}%
  \BibitemOpen
  \bibfield  {author} {\bibinfo {author} {\bibfnamefont {D.~E.}\ \bibnamefont
  {Maison}}, \bibinfo {author} {\bibfnamefont {V.~V.}\ \bibnamefont
  {Flambaum}}, \bibinfo {author} {\bibfnamefont {N.~R.}\ \bibnamefont
  {Hutzler}},\ and\ \bibinfo {author} {\bibfnamefont {L.~V.}\ \bibnamefont
  {Skripnikov}},\ }\bibfield  {title} {\bibinfo {title} {Electronic structure
  of the ytterbium monohydroxide molecule to search for axionlike particles},\
  }\href {https://doi.org/10.1103/PhysRevA.103.022813} {\bibfield  {journal}
  {\bibinfo  {journal} {Phys. Rev. A}\ }\textbf {\bibinfo {volume} {103}},\
  \bibinfo {pages} {022813} (\bibinfo {year} {2021}{\natexlab{a}})}\BibitemShut
  {NoStop}%
\bibitem [{\citenamefont {Maison}\ \emph
  {et~al.}(2021{\natexlab{b}})\citenamefont {Maison}, \citenamefont
  {Skripnikov}, \citenamefont {Oleynichenko},\ and\ \citenamefont
  {Zaitsevskii}}]{maison2021axion}%
  \BibitemOpen
  \bibfield  {author} {\bibinfo {author} {\bibfnamefont {D.~E.}\ \bibnamefont
  {Maison}}, \bibinfo {author} {\bibfnamefont {L.~V.}\ \bibnamefont
  {Skripnikov}}, \bibinfo {author} {\bibfnamefont {A.~V.}\ \bibnamefont
  {Oleynichenko}},\ and\ \bibinfo {author} {\bibfnamefont {A.~V.}\ \bibnamefont
  {Zaitsevskii}},\ }\bibfield  {title} {\bibinfo {title} {Axion-mediated
  electron–electron interaction in ytterbium monohydroxide molecule},\ }\href
  {https://doi.org/10.1063/5.0051590} {\bibfield  {journal} {\bibinfo
  {journal} {J. Chem. Phys.}\ }\textbf {\bibinfo {volume} {154}},\ \bibinfo
  {pages} {224303} (\bibinfo {year} {2021}{\natexlab{b}})}\BibitemShut
  {NoStop}%
\bibitem [{\citenamefont {Maison}\ and\ \citenamefont
  {Skripnikov}(2022)}]{Maison:2022}%
  \BibitemOpen
  \bibfield  {author} {\bibinfo {author} {\bibfnamefont {D.~E.}\ \bibnamefont
  {Maison}}\ and\ \bibinfo {author} {\bibfnamefont {L.~V.}\ \bibnamefont
  {Skripnikov}},\ }\bibfield  {title} {\bibinfo {title} {Static electric dipole
  moment of the francium atom induced by axionlike particle exchange},\ }\href
  {https://doi.org/10.1103/PhysRevA.105.032813} {\bibfield  {journal} {\bibinfo
   {journal} {Phys. Rev. A}\ }\textbf {\bibinfo {volume} {105}},\ \bibinfo
  {pages} {032813} (\bibinfo {year} {2022})}\BibitemShut {NoStop}%
\bibitem [{\citenamefont {Moody}\ and\ \citenamefont
  {Wilczek}(1984)}]{moody1984new}%
  \BibitemOpen
  \bibfield  {author} {\bibinfo {author} {\bibfnamefont {J.}~\bibnamefont
  {Moody}}\ and\ \bibinfo {author} {\bibfnamefont {F.}~\bibnamefont
  {Wilczek}},\ }\bibfield  {title} {\bibinfo {title} {New macroscopic
  forces?},\ }\href {https://doi.org/10.1103/PhysRevD.30.130} {\bibfield
  {journal} {\bibinfo  {journal} {Phys.\ Rev.\ D}\ }\textbf {\bibinfo {volume}
  {30}},\ \bibinfo {pages} {130} (\bibinfo {year} {1984})}\BibitemShut
  {NoStop}%
\bibitem [{\citenamefont {Khriplovich}(1991)}]{Khriplovich:91}%
  \BibitemOpen
  \bibfield  {author} {\bibinfo {author} {\bibfnamefont {I.~B.}\ \bibnamefont
  {Khriplovich}},\ }\href@noop {} {\emph {\bibinfo {title} {Parity
  non-conservation in atomic phenomena}}}\ (\bibinfo  {publisher} {Gordon and
  Breach},\ \bibinfo {address} {New York},\ \bibinfo {year} {1991})\BibitemShut
  {NoStop}%
\bibitem [{\citenamefont {Dmitriev}\ \emph {et~al.}(1992)\citenamefont
  {Dmitriev}, \citenamefont {Khait}, \citenamefont {Kozlov}, \citenamefont
  {Labzovsky}, \citenamefont {Mitrushenkov}, \citenamefont {Shtoff},\ and\
  \citenamefont {Titov}}]{Dmitriev:92}%
  \BibitemOpen
  \bibfield  {author} {\bibinfo {author} {\bibfnamefont {Y.~Y.}\ \bibnamefont
  {Dmitriev}}, \bibinfo {author} {\bibfnamefont {Y.~G.}\ \bibnamefont {Khait}},
  \bibinfo {author} {\bibfnamefont {M.~G.}\ \bibnamefont {Kozlov}}, \bibinfo
  {author} {\bibfnamefont {L.~N.}\ \bibnamefont {Labzovsky}}, \bibinfo {author}
  {\bibfnamefont {A.~O.}\ \bibnamefont {Mitrushenkov}}, \bibinfo {author}
  {\bibfnamefont {A.~V.}\ \bibnamefont {Shtoff}},\ and\ \bibinfo {author}
  {\bibfnamefont {A.~V.}\ \bibnamefont {Titov}},\ }\bibfield  {title} {\bibinfo
  {title} {Calculation of the spin-ro\-ta\-ti\-onal hamiltonian including
  $\mathcal{P}$- and $\mathcal{P,T}$-odd weak interaction terms for the {HgF}
  and {PbF} molecules},\ }\href@noop {} {\bibfield  {journal} {\bibinfo
  {journal} {Phys.\ Lett.\ A}\ }\textbf {\bibinfo {volume} {167}},\ \bibinfo
  {pages} {280} (\bibinfo {year} {1992})}\BibitemShut {NoStop}%
\bibitem [{\citenamefont {Visscher}\ \emph {et~al.}(1996)\citenamefont
  {Visscher}, \citenamefont {Lee},\ and\ \citenamefont {Dyall}}]{Visscher:96a}%
  \BibitemOpen
  \bibfield  {author} {\bibinfo {author} {\bibfnamefont {L.}~\bibnamefont
  {Visscher}}, \bibinfo {author} {\bibfnamefont {T.~J.}\ \bibnamefont {Lee}},\
  and\ \bibinfo {author} {\bibfnamefont {K.~G.}\ \bibnamefont {Dyall}},\
  }\bibfield  {title} {\bibinfo {title} {Formulation and implementation of a
  relativistic unrestricted coupled-cluster method including noniterative
  connected triples},\ }\href {https://doi.org/10.1063/1.472655} {\bibfield
  {journal} {\bibinfo  {journal} {J.\ Chem.\ Phys.}\ }\textbf {\bibinfo
  {volume} {105}},\ \bibinfo {pages} {8769} (\bibinfo {year}
  {1996})}\BibitemShut {NoStop}%
\bibitem [{\citenamefont {Bartlett}\ and\ \citenamefont
  {Musia{\l}}(2007)}]{Bartlett:2007}%
  \BibitemOpen
  \bibfield  {author} {\bibinfo {author} {\bibfnamefont {R.~J.}\ \bibnamefont
  {Bartlett}}\ and\ \bibinfo {author} {\bibfnamefont {M.}~\bibnamefont
  {Musia{\l}}},\ }\bibfield  {title} {\bibinfo {title} {Coupled-cluster theory
  in quantum chemistry},\ }\href {https://doi.org/10.1103/RevModPhys.79.291}
  {\bibfield  {journal} {\bibinfo  {journal} {Rev. Mod. Phys.}\ }\textbf
  {\bibinfo {volume} {79}},\ \bibinfo {pages} {291} (\bibinfo {year}
  {2007})}\BibitemShut {NoStop}%
\bibitem [{\citenamefont {Skripnikov}\ and\ \citenamefont
  {Titov}(2015)}]{Skripnikov:15a}%
  \BibitemOpen
  \bibfield  {author} {\bibinfo {author} {\bibfnamefont {L.~V.}\ \bibnamefont
  {Skripnikov}}\ and\ \bibinfo {author} {\bibfnamefont {A.~V.}\ \bibnamefont
  {Titov}},\ }\bibfield  {title} {\bibinfo {title} {Theoretical study of
  thorium monoxide for the electron electric dipole moment search: Electronic
  properties of $h^3\delta_1$ in {ThO}},\ }\href
  {https://doi.org/10.1063/1.4904877} {\bibfield  {journal} {\bibinfo
  {journal} {J.\ Chem.\ Phys.}\ }\textbf {\bibinfo {volume} {142}},\ \bibinfo
  {eid} {024301} (\bibinfo {year} {2015})}\BibitemShut {NoStop}%
\bibitem [{\citenamefont {Skripnikov}(2016)}]{Skripnikov:16b}%
  \BibitemOpen
  \bibfield  {author} {\bibinfo {author} {\bibfnamefont {L.~V.}\ \bibnamefont
  {Skripnikov}},\ }\bibfield  {title} {\bibinfo {title} {Combined 4-component
  and relativistic pseudopotential study of {ThO} for the electron electric
  dipole moment search},\ }\href {https://doi.org/10.1063/1.4968229} {\bibfield
   {journal} {\bibinfo  {journal} {J.\ Chem.\ Phys.}\ }\textbf {\bibinfo
  {volume} {145}},\ \bibinfo {pages} {214301} (\bibinfo {year}
  {2016})}\BibitemShut {NoStop}%
\bibitem [{\citenamefont {Skripnikov}\ \emph {et~al.}(2017)\citenamefont
  {Skripnikov}, \citenamefont {Maison},\ and\ \citenamefont
  {Mosyagin}}]{Skripnikov:17a}%
  \BibitemOpen
  \bibfield  {author} {\bibinfo {author} {\bibfnamefont {L.~V.}\ \bibnamefont
  {Skripnikov}}, \bibinfo {author} {\bibfnamefont {D.~E.}\ \bibnamefont
  {Maison}},\ and\ \bibinfo {author} {\bibfnamefont {N.~S.}\ \bibnamefont
  {Mosyagin}},\ }\bibfield  {title} {\bibinfo {title} {Scalar-pseudoscalar
  interaction in the francium atom},\ }\href
  {https://doi.org/10.1103/PhysRevA.95.022507} {\bibfield  {journal} {\bibinfo
  {journal} {Phys.\ Rev.\ A}\ }\textbf {\bibinfo {volume} {95}},\ \bibinfo
  {pages} {022507} (\bibinfo {year} {2017})}\BibitemShut {NoStop}%
\bibitem [{\citenamefont {Boys}(1950)}]{boys1950electronic}%
  \BibitemOpen
  \bibfield  {author} {\bibinfo {author} {\bibfnamefont {S.~F.}\ \bibnamefont
  {Boys}},\ }\bibfield  {title} {\bibinfo {title} {Electronic wave
  functions-{I}. {A} general method of calculation for the stationary states of
  any molecular system},\ }\href {https://doi.org/10.1098/rspa.1950.0036}
  {\bibfield  {journal} {\bibinfo  {journal} {Proc. Math. Phys. Eng. Sci.}\
  }\textbf {\bibinfo {volume} {200}},\ \bibinfo {pages} {542} (\bibinfo {year}
  {1950})}\BibitemShut {NoStop}%
\bibitem [{\citenamefont {Helgaker}\ \emph {et~al.}(2013)\citenamefont
  {Helgaker}, \citenamefont {Jorgensen},\ and\ \citenamefont
  {Olsen}}]{helgaker2013molecular}%
  \BibitemOpen
  \bibfield  {author} {\bibinfo {author} {\bibfnamefont {T.}~\bibnamefont
  {Helgaker}}, \bibinfo {author} {\bibfnamefont {P.}~\bibnamefont
  {Jorgensen}},\ and\ \bibinfo {author} {\bibfnamefont {J.}~\bibnamefont
  {Olsen}},\ }\href@noop {} {\emph {\bibinfo {title} {Molecular
  electronic-structure theory}}}\ (\bibinfo  {publisher} {John Wiley \& Sons},\
  \bibinfo {year} {2013})\BibitemShut {NoStop}%
\bibitem [{\citenamefont {Ten-no}(2004)}]{Tenno04}%
  \BibitemOpen
  \bibfield  {author} {\bibinfo {author} {\bibfnamefont {S.}~\bibnamefont
  {Ten-no}},\ }\bibfield  {title} {\bibinfo {title} {Initiation of explicitly
  correlated slater-type geminal theory},\ }\href
  {https://doi.org/https://doi.org/10.1016/j.cplett.2004.09.041} {\bibfield
  {journal} {\bibinfo  {journal} {Chem. Phys. Lett.}\ }\textbf {\bibinfo
  {volume} {398}},\ \bibinfo {pages} {56} (\bibinfo {year} {2004})}\BibitemShut
  {NoStop}%
\bibitem [{\citenamefont {McMurchie}\ and\ \citenamefont
  {Davidson}(1978)}]{McMurchie1978}%
  \BibitemOpen
  \bibfield  {author} {\bibinfo {author} {\bibfnamefont {L.~E.}\ \bibnamefont
  {McMurchie}}\ and\ \bibinfo {author} {\bibfnamefont {E.~R.}\ \bibnamefont
  {Davidson}},\ }\bibfield  {title} {\bibinfo {title} {One- and two-electron
  integrals over cartesian gaussian functions},\ }\href
  {https://doi.org/https://doi.org/10.1016/0021-9991(78)90092-X} {\bibfield
  {journal} {\bibinfo  {journal} {J. Comput. Phys.}\ }\textbf {\bibinfo
  {volume} {26}},\ \bibinfo {pages} {218} (\bibinfo {year} {1978})}\BibitemShut
  {NoStop}%
\bibitem [{\citenamefont {Obara}\ and\ \citenamefont
  {Saika}(1986)}]{Obara_Saika_1986}%
  \BibitemOpen
  \bibfield  {author} {\bibinfo {author} {\bibfnamefont {S.}~\bibnamefont
  {Obara}}\ and\ \bibinfo {author} {\bibfnamefont {A.}~\bibnamefont {Saika}},\
  }\bibfield  {title} {\bibinfo {title} {Efficient recursive computation of
  molecular integrals over cartesian gaussian functions},\ }\href
  {https://doi.org/10.1063/1.450106} {\bibfield  {journal} {\bibinfo  {journal}
  {J. Chem. Phys.}\ }\textbf {\bibinfo {volume} {84}},\ \bibinfo {pages} {3963}
  (\bibinfo {year} {1986})}\BibitemShut {NoStop}%
\bibitem [{\citenamefont {Gill}\ \emph {et~al.}(1991)\citenamefont {Gill},
  \citenamefont {Johnson},\ and\ \citenamefont {Pople}}]{Gill1991}%
  \BibitemOpen
  \bibfield  {author} {\bibinfo {author} {\bibfnamefont {P.~M.~W.}\
  \bibnamefont {Gill}}, \bibinfo {author} {\bibfnamefont {B.~G.}\ \bibnamefont
  {Johnson}},\ and\ \bibinfo {author} {\bibfnamefont {J.~A.}\ \bibnamefont
  {Pople}},\ }\bibfield  {title} {\bibinfo {title} {Two-electron repulsion
  integrals over gaussian s functions},\ }\href
  {https://doi.org/https://doi.org/10.1002/qua.560400604} {\bibfield  {journal}
  {\bibinfo  {journal} {Int. J. Quantum Chem.}\ }\textbf {\bibinfo {volume}
  {40}},\ \bibinfo {pages} {745} (\bibinfo {year} {1991})}\BibitemShut
  {NoStop}%
\bibitem [{\citenamefont {Pople}\ and\ \citenamefont
  {Hehre}(1978)}]{Pople1978}%
  \BibitemOpen
  \bibfield  {author} {\bibinfo {author} {\bibfnamefont {J.~A.}\ \bibnamefont
  {Pople}}\ and\ \bibinfo {author} {\bibfnamefont {W.~J.}\ \bibnamefont
  {Hehre}},\ }\bibfield  {title} {\bibinfo {title} {Computation of electron
  repulsion integrals involving contracted gaussian basis functions},\ }\href
  {https://doi.org/https://doi.org/10.1016/0021-9991(78)90001-3} {\bibfield
  {journal} {\bibinfo  {journal} {J. Comput. Phys.}\ }\textbf {\bibinfo
  {volume} {27}},\ \bibinfo {pages} {161} (\bibinfo {year} {1978})}\BibitemShut
  {NoStop}%
\bibitem [{\citenamefont {Head‐Gordon}\ and\ \citenamefont
  {Pople}(1988)}]{Head_Gordon_Pople_1988}%
  \BibitemOpen
  \bibfield  {author} {\bibinfo {author} {\bibfnamefont {M.}~\bibnamefont
  {Head‐Gordon}}\ and\ \bibinfo {author} {\bibfnamefont {J.~A.}\ \bibnamefont
  {Pople}},\ }\bibfield  {title} {\bibinfo {title} {A method for two‐electron
  gaussian integral and integral derivative evaluation using recurrence
  relations},\ }\href {https://doi.org/10.1063/1.455553} {\bibfield  {journal}
  {\bibinfo  {journal} {J. Chem. Phys.}\ }\textbf {\bibinfo {volume} {89}},\
  \bibinfo {pages} {5777} (\bibinfo {year} {1988})}\BibitemShut {NoStop}%
\bibitem [{\citenamefont {Hamilton}\ and\ \citenamefont
  {Schaefer}(1991)}]{Hamilton1991}%
  \BibitemOpen
  \bibfield  {author} {\bibinfo {author} {\bibfnamefont {T.~P.}\ \bibnamefont
  {Hamilton}}\ and\ \bibinfo {author} {\bibfnamefont {H.~F.}\ \bibnamefont
  {Schaefer}},\ }\bibfield  {title} {\bibinfo {title} {New variations in
  two-electron integral evaluation in the context of direct scf procedures},\
  }\href {https://doi.org/https://doi.org/10.1016/0301-0104(91)80126-3}
  {\bibfield  {journal} {\bibinfo  {journal} {Chem. Phys.}\ }\textbf {\bibinfo
  {volume} {150}},\ \bibinfo {pages} {163} (\bibinfo {year}
  {1991})}\BibitemShut {NoStop}%
\bibitem [{\citenamefont {Ten-no}(1993)}]{Tenno1993}%
  \BibitemOpen
  \bibfield  {author} {\bibinfo {author} {\bibfnamefont {S.}~\bibnamefont
  {Ten-no}},\ }\bibfield  {title} {\bibinfo {title} {An efficient algorithm for
  electron repulsion integrals over contracted gaussian-type functions},\
  }\href {https://doi.org/https://doi.org/10.1016/0009-2614(93)85195-T}
  {\bibfield  {journal} {\bibinfo  {journal} {Chem. Phys. Lett.}\ }\textbf
  {\bibinfo {volume} {211}},\ \bibinfo {pages} {259} (\bibinfo {year}
  {1993})}\BibitemShut {NoStop}%
\bibitem [{\citenamefont {Yanai}\ \emph {et~al.}(2000)\citenamefont {Yanai},
  \citenamefont {Ishida}, \citenamefont {Nakano},\ and\ \citenamefont
  {Hirao}}]{Yanai2000}%
  \BibitemOpen
  \bibfield  {author} {\bibinfo {author} {\bibfnamefont {T.}~\bibnamefont
  {Yanai}}, \bibinfo {author} {\bibfnamefont {K.}~\bibnamefont {Ishida}},
  \bibinfo {author} {\bibfnamefont {H.}~\bibnamefont {Nakano}},\ and\ \bibinfo
  {author} {\bibfnamefont {K.}~\bibnamefont {Hirao}},\ }\bibfield  {title}
  {\bibinfo {title} {New algorithm for electron repulsion integrals oriented to
  the general contraction scheme},\ }\href
  {https://doi.org/https://doi.org/10.1002/(SICI)1097-461X(2000)76:3<396::AID-QUA8>3.0.CO;2-A}
  {\bibfield  {journal} {\bibinfo  {journal} {Int. J. Quantum Chem.}\ }\textbf
  {\bibinfo {volume} {76}},\ \bibinfo {pages} {396} (\bibinfo {year}
  {2000})}\BibitemShut {NoStop}%
\bibitem [{\citenamefont {Nakai}\ and\ \citenamefont
  {Kobayashi}(2004)}]{Nakai2004}%
  \BibitemOpen
  \bibfield  {author} {\bibinfo {author} {\bibfnamefont {H.}~\bibnamefont
  {Nakai}}\ and\ \bibinfo {author} {\bibfnamefont {M.}~\bibnamefont
  {Kobayashi}},\ }\bibfield  {title} {\bibinfo {title} {New algorithm for the
  rapid evaluation of electron repulsion integrals: elementary basis
  algorithm},\ }\href
  {https://doi.org/https://doi.org/10.1016/j.cplett.2004.02.070} {\bibfield
  {journal} {\bibinfo  {journal} {Chem. Phys. Lett.}\ }\textbf {\bibinfo
  {volume} {388}},\ \bibinfo {pages} {50} (\bibinfo {year} {2004})}\BibitemShut
  {NoStop}%
\bibitem [{\citenamefont {Valeev}(2022)}]{valeev2020libint}%
  \BibitemOpen
  \bibfield  {author} {\bibinfo {author} {\bibfnamefont {E.~F.}\ \bibnamefont
  {Valeev}},\ }\href@noop {} {\bibinfo {title} {Libint: A library for the
  evaluation of molecular integrals of many-body operators over gaussian
  functions}},\ \bibinfo {howpublished} {http://libint.valeyev.net/} (\bibinfo
  {year} {2022}),\ \bibinfo {note} {version 2.8.0}\BibitemShut {NoStop}%
\bibitem [{\citenamefont {Dupuis}\ \emph {et~al.}(1976)\citenamefont {Dupuis},
  \citenamefont {Rys},\ and\ \citenamefont {King}}]{Dupuis1976}%
  \BibitemOpen
  \bibfield  {author} {\bibinfo {author} {\bibfnamefont {M.}~\bibnamefont
  {Dupuis}}, \bibinfo {author} {\bibfnamefont {J.}~\bibnamefont {Rys}},\ and\
  \bibinfo {author} {\bibfnamefont {H.~F.}\ \bibnamefont {King}},\ }\bibfield
  {title} {\bibinfo {title} {Evaluation of molecular integrals over gaussian
  basis functions},\ }\href {https://doi.org/10.1063/1.432807} {\bibfield
  {journal} {\bibinfo  {journal} {J. Chem. Phys.}\ }\textbf {\bibinfo {volume}
  {65}},\ \bibinfo {pages} {111} (\bibinfo {year} {1976})}\BibitemShut
  {NoStop}%
\bibitem [{\citenamefont {King}\ and\ \citenamefont {Dupuis}(1976)}]{King1976}%
  \BibitemOpen
  \bibfield  {author} {\bibinfo {author} {\bibfnamefont {H.~F.}\ \bibnamefont
  {King}}\ and\ \bibinfo {author} {\bibfnamefont {M.}~\bibnamefont {Dupuis}},\
  }\bibfield  {title} {\bibinfo {title} {Numerical integration using rys
  polynomials},\ }\href
  {https://doi.org/https://doi.org/10.1016/0021-9991(76)90008-5} {\bibfield
  {journal} {\bibinfo  {journal} {J. Comput. Phys.}\ }\textbf {\bibinfo
  {volume} {21}},\ \bibinfo {pages} {144} (\bibinfo {year} {1976})}\BibitemShut
  {NoStop}%
\bibitem [{\citenamefont {Rys}\ \emph {et~al.}(1983)\citenamefont {Rys},
  \citenamefont {Dupuis},\ and\ \citenamefont {King}}]{Rys1983}%
  \BibitemOpen
  \bibfield  {author} {\bibinfo {author} {\bibfnamefont {J.}~\bibnamefont
  {Rys}}, \bibinfo {author} {\bibfnamefont {M.}~\bibnamefont {Dupuis}},\ and\
  \bibinfo {author} {\bibfnamefont {H.~F.}\ \bibnamefont {King}},\ }\bibfield
  {title} {\bibinfo {title} {Computation of electron repulsion integrals using
  the rys quadrature method},\ }\href
  {https://doi.org/https://doi.org/10.1002/jcc.540040206} {\bibfield  {journal}
  {\bibinfo  {journal} {J. Comput. Chem.}\ }\textbf {\bibinfo {volume} {4}},\
  \bibinfo {pages} {154} (\bibinfo {year} {1983})}\BibitemShut {NoStop}%
\bibitem [{\citenamefont {Lindh}\ \emph {et~al.}(1991)\citenamefont {Lindh},
  \citenamefont {Ryu},\ and\ \citenamefont {Liu}}]{Lindh1991}%
  \BibitemOpen
  \bibfield  {author} {\bibinfo {author} {\bibfnamefont {R.}~\bibnamefont
  {Lindh}}, \bibinfo {author} {\bibfnamefont {U.}~\bibnamefont {Ryu}},\ and\
  \bibinfo {author} {\bibfnamefont {B.}~\bibnamefont {Liu}},\ }\bibfield
  {title} {\bibinfo {title} {The reduced multiplication scheme of the {Rys}
  quadrature and new recurrence relations for auxiliary function based
  two‐electron integral evaluation},\ }\href
  {https://doi.org/10.1063/1.461610} {\bibfield  {journal} {\bibinfo  {journal}
  {J. Chem. Phys.}\ }\textbf {\bibinfo {volume} {95}},\ \bibinfo {pages} {5889}
  (\bibinfo {year} {1991})}\BibitemShut {NoStop}%
\bibitem [{\citenamefont {Ten-no}(2007)}]{ten2007new}%
  \BibitemOpen
  \bibfield  {author} {\bibinfo {author} {\bibfnamefont {S.}~\bibnamefont
  {Ten-no}},\ }\bibfield  {title} {\bibinfo {title} {New implementation of
  second-order {M}{\o}ller-{P}lesset perturbation theory with an analytic
  {S}later-type geminal},\ }\href {https://doi.org/10.1063/1.2403853}
  {\bibfield  {journal} {\bibinfo  {journal} {J. Chem. Phys.}\ }\textbf
  {\bibinfo {volume} {126}},\ \bibinfo {pages} {014108} (\bibinfo {year}
  {2007})}\BibitemShut {NoStop}%
\bibitem [{\citenamefont {Shiozaki}(2009)}]{shiozaki2009evaluation}%
  \BibitemOpen
  \bibfield  {author} {\bibinfo {author} {\bibfnamefont {T.}~\bibnamefont
  {Shiozaki}},\ }\bibfield  {title} {\bibinfo {title} {Evaluation of
  slater-type geminal integrals using tailored gaussian quadrature},\ }\href
  {https://doi.org/10.1016/j.cplett.2009.07.108} {\bibfield  {journal}
  {\bibinfo  {journal} {Chem. Phys. Lett.}\ }\textbf {\bibinfo {volume}
  {479}},\ \bibinfo {pages} {160} (\bibinfo {year} {2009})}\BibitemShut
  {NoStop}%
\bibitem [{\citenamefont {Kumar}\ \emph {et~al.}(2020)\citenamefont {Kumar},
  \citenamefont {Neese},\ and\ \citenamefont {Valeev}}]{Valeev2020}%
  \BibitemOpen
  \bibfield  {author} {\bibinfo {author} {\bibfnamefont {A.}~\bibnamefont
  {Kumar}}, \bibinfo {author} {\bibfnamefont {F.}~\bibnamefont {Neese}},\ and\
  \bibinfo {author} {\bibfnamefont {E.~F.}\ \bibnamefont {Valeev}},\ }\bibfield
   {title} {\bibinfo {title} {Explicitly correlated coupled cluster method for
  accurate treatment of open-shell molecules with hundreds of atoms},\ }\href
  {https://doi.org/10.1063/5.0012753} {\bibfield  {journal} {\bibinfo
  {journal} {J. Chem. Phys.}\ }\textbf {\bibinfo {volume} {153}},\ \bibinfo
  {pages} {094105} (\bibinfo {year} {2020})}\BibitemShut {NoStop}%
\bibitem [{\citenamefont {Kállay}\ \emph {et~al.}(2021)\citenamefont
  {Kállay}, \citenamefont {Horváth}, \citenamefont {Gyevi-Nagy},\ and\
  \citenamefont {Nagy}}]{Kallay2021}%
  \BibitemOpen
  \bibfield  {author} {\bibinfo {author} {\bibfnamefont {M.}~\bibnamefont
  {Kállay}}, \bibinfo {author} {\bibfnamefont {R.~A.}\ \bibnamefont
  {Horváth}}, \bibinfo {author} {\bibfnamefont {L.}~\bibnamefont
  {Gyevi-Nagy}},\ and\ \bibinfo {author} {\bibfnamefont {P.~R.}\ \bibnamefont
  {Nagy}},\ }\bibfield  {title} {\bibinfo {title} {Size-consistent explicitly
  correlated triple excitation correction},\ }\href
  {https://doi.org/10.1063/5.0057426} {\bibfield  {journal} {\bibinfo
  {journal} {J. Chem. Phys.}\ }\textbf {\bibinfo {volume} {155}},\ \bibinfo
  {pages} {034107} (\bibinfo {year} {2021})}\BibitemShut {NoStop}%
\bibitem [{\citenamefont {K\'{a}llay}\ \emph {et~al.}(2023)\citenamefont
  {K\'{a}llay}, \citenamefont {Horv\'{a}th}, \citenamefont {Gyevi-Nagy},\ and\
  \citenamefont {Nagy}}]{Kallay:2023}%
  \BibitemOpen
  \bibfield  {author} {\bibinfo {author} {\bibfnamefont {M.}~\bibnamefont
  {K\'{a}llay}}, \bibinfo {author} {\bibfnamefont {R.~A.}\ \bibnamefont
  {Horv\'{a}th}}, \bibinfo {author} {\bibfnamefont {L.}~\bibnamefont
  {Gyevi-Nagy}},\ and\ \bibinfo {author} {\bibfnamefont {P.~R.}\ \bibnamefont
  {Nagy}},\ }\bibfield  {title} {\bibinfo {title} {Basis set limit {CCSD(T)}
  energies for extended molecules via a reduced-cost explicitly correlated
  approach},\ }\href {https://doi.org/10.1021/acs.jctc.2c01031} {\bibfield
  {journal} {\bibinfo  {journal} {J. Chem Phys.}\ }\textbf {\bibinfo {volume}
  {19}},\ \bibinfo {pages} {174} (\bibinfo {year} {2023})}\BibitemShut
  {NoStop}%
\bibitem [{\citenamefont {Peccei}(2008)}]{Peccei2008}%
  \BibitemOpen
  \bibfield  {author} {\bibinfo {author} {\bibfnamefont {R.~D.}\ \bibnamefont
  {Peccei}},\ }\bibinfo {title} {The strong cp problem and axions},\ in\ \href
  {https://doi.org/10.1007/978-3-540-73518-2_1} {\emph {\bibinfo {booktitle}
  {Axions: Theory, Cosmology, and Experimental Searches}}},\ \bibinfo {editor}
  {edited by\ \bibinfo {editor} {\bibfnamefont {M.}~\bibnamefont {Kuster}},
  \bibinfo {editor} {\bibfnamefont {G.}~\bibnamefont {Raffelt}},\ and\ \bibinfo
  {editor} {\bibfnamefont {B.}~\bibnamefont {Beltr{\'a}n}}}\ (\bibinfo
  {publisher} {Springer Berlin Heidelberg},\ \bibinfo {address} {Berlin,
  Heidelberg},\ \bibinfo {year} {2008})\ pp.\ \bibinfo {pages}
  {3--17}\BibitemShut {NoStop}%
\bibitem [{\citenamefont {Kelly}\ \emph {et~al.}(2021)\citenamefont {Kelly},
  \citenamefont {Kumar},\ and\ \citenamefont {Liu}}]{HeavyAxion:2021}%
  \BibitemOpen
  \bibfield  {author} {\bibinfo {author} {\bibfnamefont {K.~J.}\ \bibnamefont
  {Kelly}}, \bibinfo {author} {\bibfnamefont {S.}~\bibnamefont {Kumar}},\ and\
  \bibinfo {author} {\bibfnamefont {Z.}~\bibnamefont {Liu}},\ }\bibfield
  {title} {\bibinfo {title} {Heavy axion opportunities at the dune near
  detector},\ }\href {https://doi.org/10.1103/PhysRevD.103.095002} {\bibfield
  {journal} {\bibinfo  {journal} {Phys. Rev. D}\ }\textbf {\bibinfo {volume}
  {103}},\ \bibinfo {pages} {095002} (\bibinfo {year} {2021})}\BibitemShut
  {NoStop}%
\bibitem [{\citenamefont {Giannotti}\ \emph {et~al.}(2010)\citenamefont
  {Giannotti}, \citenamefont {Nita},\ and\ \citenamefont
  {Welch}}]{Giannotti:2010}%
  \BibitemOpen
  \bibfield  {author} {\bibinfo {author} {\bibfnamefont {M.}~\bibnamefont
  {Giannotti}}, \bibinfo {author} {\bibfnamefont {R.}~\bibnamefont {Nita}},\
  and\ \bibinfo {author} {\bibfnamefont {E.}~\bibnamefont {Welch}},\ }\bibfield
   {title} {\bibinfo {title} {Phenomenological implications of heavy axion
  models},\ }\href {https://doi.org/10.1063/1.3489553} {\bibfield  {journal}
  {\bibinfo  {journal} {AIP Conf Proc}\ }\textbf {\bibinfo {volume} {1274}},\
  \bibinfo {pages} {20} (\bibinfo {year} {2010})}\BibitemShut {NoStop}%
\bibitem [{\citenamefont {Abe}\ \emph {et~al.}(2004)\citenamefont {Abe},
  \citenamefont {Yanai}, \citenamefont {Nakajima},\ and\ \citenamefont
  {Hirao}}]{abe2004four}%
  \BibitemOpen
  \bibfield  {author} {\bibinfo {author} {\bibfnamefont {M.}~\bibnamefont
  {Abe}}, \bibinfo {author} {\bibfnamefont {T.}~\bibnamefont {Yanai}}, \bibinfo
  {author} {\bibfnamefont {T.}~\bibnamefont {Nakajima}},\ and\ \bibinfo
  {author} {\bibfnamefont {K.}~\bibnamefont {Hirao}},\ }\bibfield  {title}
  {\bibinfo {title} {A four-index transformation in {D}irac's four-component
  relativistic theory},\ }\href {https://doi.org/10.1016/j.cplett.2004.02.030}
  {\bibfield  {journal} {\bibinfo  {journal} {Chem.\ Phys.\ Lett.}\ }\textbf
  {\bibinfo {volume} {388}},\ \bibinfo {pages} {68} (\bibinfo {year}
  {2004})}\BibitemShut {NoStop}%
\bibitem [{\citenamefont {Dyall}(2004)}]{Dyall:04}%
  \BibitemOpen
  \bibfield  {author} {\bibinfo {author} {\bibfnamefont {K.~G.}\ \bibnamefont
  {Dyall}},\ }\bibfield  {title} {\bibinfo {title} {Relativistic double-zeta,
  triple-zeta, and quadruple-zeta basis sets for the 5d elements {Hf--Hg}},\
  }\href {https://doi.org/10.1007/s00214-004-0607-y} {\bibfield  {journal}
  {\bibinfo  {journal} {Theor. Chem. Acc.}\ }\textbf {\bibinfo {volume}
  {112}},\ \bibinfo {pages} {403} (\bibinfo {year} {2004})}\BibitemShut
  {NoStop}%
\bibitem [{\citenamefont {Dyall}(2012)}]{Dyall:12}%
  \BibitemOpen
  \bibfield  {author} {\bibinfo {author} {\bibfnamefont {K.~G.}\ \bibnamefont
  {Dyall}},\ }\bibfield  {title} {\bibinfo {title} {Core correlating basis
  functions for elements 31--118},\ }\href
  {https://doi.org/10.1007/s00214-012-1217-8} {\bibfield  {journal} {\bibinfo
  {journal} {Theor. Chem. Acc.}\ }\textbf {\bibinfo {volume} {131}},\ \bibinfo
  {pages} {1217} (\bibinfo {year} {2012})}\BibitemShut {NoStop}%
\bibitem [{\citenamefont {Dyall}\ and\ \citenamefont
  {Gomes}(2010)}]{Dyall:2010}%
  \BibitemOpen
  \bibfield  {author} {\bibinfo {author} {\bibfnamefont {K.~G.}\ \bibnamefont
  {Dyall}}\ and\ \bibinfo {author} {\bibfnamefont {A.~S.}\ \bibnamefont
  {Gomes}},\ }\bibfield  {title} {\bibinfo {title} {Revised relativistic basis
  sets for the 5d elements {Hf--Hg}},\ }\href
  {https://doi.org/10.1007/s00214-009-0717-7} {\bibfield  {journal} {\bibinfo
  {journal} {Theor Chem Acc.}\ }\textbf {\bibinfo {volume} {125}},\ \bibinfo
  {pages} {97} (\bibinfo {year} {2010})}\BibitemShut {NoStop}%
\bibitem [{\citenamefont {Dyall}(2016)}]{Dyall:2016}%
  \BibitemOpen
  \bibfield  {author} {\bibinfo {author} {\bibfnamefont {K.~G.}\ \bibnamefont
  {Dyall}},\ }\bibfield  {title} {\bibinfo {title} {Relativistic double-zeta,
  triple-zeta, and quadruple-zeta basis sets for the light elements {H--Ar}},\
  }\href {https://doi.org/10.1007/s00214-016-1884-y} {\bibfield  {journal}
  {\bibinfo  {journal} {Theor. Chem. Acc.}\ }\textbf {\bibinfo {volume}
  {135}},\ \bibinfo {pages} {128} (\bibinfo {year} {2016})}\BibitemShut
  {NoStop}%
\bibitem [{\citenamefont {Oleynichenko}\ \emph {et~al.}(2020)\citenamefont
  {Oleynichenko}, \citenamefont {Zaitsevskii}, \citenamefont {Skripnikov},\
  and\ \citenamefont {Eliav}}]{Oleynichenko:20}%
  \BibitemOpen
  \bibfield  {author} {\bibinfo {author} {\bibfnamefont {A.~V.}\ \bibnamefont
  {Oleynichenko}}, \bibinfo {author} {\bibfnamefont {A.}~\bibnamefont
  {Zaitsevskii}}, \bibinfo {author} {\bibfnamefont {L.~V.}\ \bibnamefont
  {Skripnikov}},\ and\ \bibinfo {author} {\bibfnamefont {E.}~\bibnamefont
  {Eliav}},\ }\bibfield  {title} {\bibinfo {title} {Relativistic {F}ock space
  coupled cluster method for many-electron systems: non-perturbative account
  for connected triple excitations},\ }\href
  {https://doi.org/10.3390/sym12071101} {\bibfield  {journal} {\bibinfo
  {journal} {{S}ymmetry}\ }\textbf {\bibinfo {volume} {12}},\ \bibinfo {pages}
  {1101} (\bibinfo {year} {2020})}\BibitemShut {NoStop}%
\bibitem [{\citenamefont {Cossel}\ \emph {et~al.}(2012)\citenamefont {Cossel},
  \citenamefont {Gresh}, \citenamefont {Sinclair}, \citenamefont {Coffey},
  \citenamefont {Skripnikov}, \citenamefont {Petrov}, \citenamefont {Mosyagin},
  \citenamefont {Titov}, \citenamefont {Field}, \citenamefont {Meyer},
  \citenamefont {Cornell},\ and\ \citenamefont {Ye}}]{Cossel:12}%
  \BibitemOpen
  \bibfield  {author} {\bibinfo {author} {\bibfnamefont {K.~C.}\ \bibnamefont
  {Cossel}}, \bibinfo {author} {\bibfnamefont {D.~N.}\ \bibnamefont {Gresh}},
  \bibinfo {author} {\bibfnamefont {L.~C.}\ \bibnamefont {Sinclair}}, \bibinfo
  {author} {\bibfnamefont {T.}~\bibnamefont {Coffey}}, \bibinfo {author}
  {\bibfnamefont {L.~V.}\ \bibnamefont {Skripnikov}}, \bibinfo {author}
  {\bibfnamefont {A.~N.}\ \bibnamefont {Petrov}}, \bibinfo {author}
  {\bibfnamefont {N.~S.}\ \bibnamefont {Mosyagin}}, \bibinfo {author}
  {\bibfnamefont {A.~V.}\ \bibnamefont {Titov}}, \bibinfo {author}
  {\bibfnamefont {R.~W.}\ \bibnamefont {Field}}, \bibinfo {author}
  {\bibfnamefont {E.~R.}\ \bibnamefont {Meyer}}, \bibinfo {author}
  {\bibfnamefont {E.~A.}\ \bibnamefont {Cornell}},\ and\ \bibinfo {author}
  {\bibfnamefont {J.}~\bibnamefont {Ye}},\ }\bibfield  {title} {\bibinfo
  {title} {Broadband velocity modulation spectroscopy of {H}f{F}+: Towards a
  measurement of the electron electric dipole moment},\ }\href
  {https://doi.org/10.1016/j.cplett.2012.06.037} {\bibfield  {journal}
  {\bibinfo  {journal} {Chem.\ Phys.\ Lett.}\ }\textbf {\bibinfo {volume}
  {546}},\ \bibinfo {pages} {1 } (\bibinfo {year} {2012})}\BibitemShut
  {NoStop}%
\bibitem [{\citenamefont {Visscher}\ and\ \citenamefont
  {Dyall}(1997)}]{Visscher:1997}%
  \BibitemOpen
  \bibfield  {author} {\bibinfo {author} {\bibfnamefont {L.}~\bibnamefont
  {Visscher}}\ and\ \bibinfo {author} {\bibfnamefont {K.~G.}\ \bibnamefont
  {Dyall}},\ }\bibfield  {title} {\bibinfo {title} {Dirac--fock atomic
  electronic structure calculations using different nuclear charge
  distributions},\ }\href {https://doi.org/10.1006/adnd.1997.0751} {\bibfield
  {journal} {\bibinfo  {journal} {At. Data Nucl. Data Tables}\ }\textbf
  {\bibinfo {volume} {67}},\ \bibinfo {pages} {207} (\bibinfo {year}
  {1997})}\BibitemShut {NoStop}%
\bibitem [{\citenamefont {Sahoo}(2017)}]{sahoo2017improved}%
  \BibitemOpen
  \bibfield  {author} {\bibinfo {author} {\bibfnamefont {B.~K.}\ \bibnamefont
  {Sahoo}},\ }\bibfield  {title} {\bibinfo {title} {Improved limits on the
  hadronic and semihadronic {$CP$} violating parameters and role of a dark
  force carrier in the electric dipole moment of $^{199}\mathrm{Hg}$},\ }\href
  {https://doi.org/10.1103/PhysRevD.95.013002} {\bibfield  {journal} {\bibinfo
  {journal} {Phys. Rev. D}\ }\textbf {\bibinfo {volume} {95}},\ \bibinfo
  {pages} {013002} (\bibinfo {year} {2017})}\BibitemShut {NoStop}%
\bibitem [{\citenamefont {Gharibnejad}\ and\ \citenamefont
  {Derevianko}(2015)}]{gharibnejad2015dark}%
  \BibitemOpen
  \bibfield  {author} {\bibinfo {author} {\bibfnamefont {H.}~\bibnamefont
  {Gharibnejad}}\ and\ \bibinfo {author} {\bibfnamefont {A.}~\bibnamefont
  {Derevianko}},\ }\bibfield  {title} {\bibinfo {title} {Dark forces and atomic
  electric dipole moments},\ }\href
  {https://doi.org/10.1103/PhysRevD.91.035007} {\bibfield  {journal} {\bibinfo
  {journal} {Phys. Rev. D}\ }\textbf {\bibinfo {volume} {91}},\ \bibinfo
  {pages} {035007} (\bibinfo {year} {2015})}\BibitemShut {NoStop}%
\bibitem [{DIR()}]{DIRAC19}%
  \BibitemOpen
  \href@noop {} {}\bibinfo {note} {DIRAC, a relativistic ab initio electronic
  structure program, Release DIRAC19 (2019), written by A. S. P. Gomes, T.
  Saue, L. Visscher, H. J. Aa. Jensen, and R. Bast, with contributions from I.
  A. Aucar, V. Bakken, K. G. Dyall, S. Dubillard, U. Ekstroem, E. Eliav, T.
  Enevoldsen, E. Fasshauer, T. Fleig, O. Fossgaard, L. Halbert, E. D.
  Hedegaard, T. Helgaker, J. Henriksson, M. Ilias, Ch. R. Jacob, S. Knecht, S.
  Komorovsky, O. Kullie, J. K. Laerdahl, C. V. Larsen, Y. S. Lee, H. S.
  Nataraj, M. K. Nayak, P. Norman, M. Olejniczak, J. Olsen, J. M. H. Olsen, Y.
  C. Park, J. K. Pedersen, M. Pernpointner, R. Di Remigio, K. Ruud, P. Salek,
  B. Schimmelpfennig, B. Senjean, A. Shee, J. Sikkema, A. J. Thorvaldsen, J.
  Thyssen, J. van Stralen, M. L. Vidal, S. Villaume, O. Visser, T. Winther, and
  S. Yamamoto (see http://diracprogram.org).}\BibitemShut {Stop}%
\bibitem [{\citenamefont {Saue}\ \emph {et~al.}(2020)\citenamefont {Saue},
  \citenamefont {Bast}, \citenamefont {Gomes}, \citenamefont {Jensen},
  \citenamefont {Visscher}, \citenamefont {Aucar}, \citenamefont {Di~Remigio},
  \citenamefont {Dyall}, \citenamefont {Eliav}, \citenamefont {Fasshauer},
  \citenamefont {Fleig}, \citenamefont {Halbert}, \citenamefont {Hedegård},
  \citenamefont {Helmich-Paris}, \citenamefont {Iliaš}, \citenamefont {Jacob},
  \citenamefont {Knecht}, \citenamefont {Laerdahl}, \citenamefont {Vidal},
  \citenamefont {Nayak}, \citenamefont {Olejniczak}, \citenamefont {Olsen},
  \citenamefont {Pernpointner}, \citenamefont {Senjean}, \citenamefont {Shee},
  \citenamefont {Sunaga},\ and\ \citenamefont {van Stralen}}]{Saue:2020}%
  \BibitemOpen
  \bibfield  {author} {\bibinfo {author} {\bibfnamefont {T.}~\bibnamefont
  {Saue}}, \bibinfo {author} {\bibfnamefont {R.}~\bibnamefont {Bast}}, \bibinfo
  {author} {\bibfnamefont {A.~S.~P.}\ \bibnamefont {Gomes}}, \bibinfo {author}
  {\bibfnamefont {H.~J.~A.}\ \bibnamefont {Jensen}}, \bibinfo {author}
  {\bibfnamefont {L.}~\bibnamefont {Visscher}}, \bibinfo {author}
  {\bibfnamefont {I.~A.}\ \bibnamefont {Aucar}}, \bibinfo {author}
  {\bibfnamefont {R.}~\bibnamefont {Di~Remigio}}, \bibinfo {author}
  {\bibfnamefont {K.~G.}\ \bibnamefont {Dyall}}, \bibinfo {author}
  {\bibfnamefont {E.}~\bibnamefont {Eliav}}, \bibinfo {author} {\bibfnamefont
  {E.}~\bibnamefont {Fasshauer}}, \bibinfo {author} {\bibfnamefont
  {T.}~\bibnamefont {Fleig}}, \bibinfo {author} {\bibfnamefont
  {L.}~\bibnamefont {Halbert}}, \bibinfo {author} {\bibfnamefont {E.~D.}\
  \bibnamefont {Hedegård}}, \bibinfo {author} {\bibfnamefont {B.}~\bibnamefont
  {Helmich-Paris}}, \bibinfo {author} {\bibfnamefont {M.}~\bibnamefont
  {Iliaš}}, \bibinfo {author} {\bibfnamefont {C.~R.}\ \bibnamefont {Jacob}},
  \bibinfo {author} {\bibfnamefont {S.}~\bibnamefont {Knecht}}, \bibinfo
  {author} {\bibfnamefont {J.~K.}\ \bibnamefont {Laerdahl}}, \bibinfo {author}
  {\bibfnamefont {M.~L.}\ \bibnamefont {Vidal}}, \bibinfo {author}
  {\bibfnamefont {M.~K.}\ \bibnamefont {Nayak}}, \bibinfo {author}
  {\bibfnamefont {M.}~\bibnamefont {Olejniczak}}, \bibinfo {author}
  {\bibfnamefont {J.~M.~H.}\ \bibnamefont {Olsen}}, \bibinfo {author}
  {\bibfnamefont {M.}~\bibnamefont {Pernpointner}}, \bibinfo {author}
  {\bibfnamefont {B.}~\bibnamefont {Senjean}}, \bibinfo {author} {\bibfnamefont
  {A.}~\bibnamefont {Shee}}, \bibinfo {author} {\bibfnamefont {A.}~\bibnamefont
  {Sunaga}},\ and\ \bibinfo {author} {\bibfnamefont {J.~N.~P.}\ \bibnamefont
  {van Stralen}},\ }\bibfield  {title} {\bibinfo {title} {The dirac code for
  relativistic molecular calculations},\ }\href
  {https://doi.org/10.1063/5.0004844} {\bibfield  {journal} {\bibinfo
  {journal} {J.\ Chem.\ Phys.}\ }\textbf {\bibinfo {volume} {152}},\ \bibinfo
  {pages} {204104} (\bibinfo {year} {2020})}\BibitemShut {NoStop}%
\bibitem [{MRC()}]{MRCC2020}%
  \BibitemOpen
  \bibfield  {title} {\bibinfo {title} {{{\sc mrcc}}},\ }\bibinfo {note} {m.
  K\'{a}llay, P. R. Nagy, D. Mester, Z. Rolik, G. Samu, J. Csontos, J.
  Cs\'{o}ka, P. B. Szab\'{o}, L. Gyevi-Nagy, B. H\'{e}gely, I. Ladj\'{a}nszki,
  L. Szegedy, B. Lad\'{o}czki, K. Petrov, M. Farkas, P. D. Mezei, and \'{a}.
  Ganyecz: The {\sc mrcc} program system: Accurate quantum chemistry from water
  to proteins, J. Chem. Phys. 152, 074107 (2020); {\sc mrcc}, a quantum
  chemical program suite written by M. K\'{a}llay, P. R. Nagy, D. Mester, Z.
  Rolik, G. Samu, J. Csontos, J. Cs\'{o}ka, P. B. Szab\'{o}, L. Gyevi-Nagy, B.
  H\'{e}gely, I. Ladj\'{a}nszki, L. Szegedy, B. Lad\'{o}czki, K. Petrov, M.
  Farkas, P. D. Mezei, and \'{a}. Ganyecz. See www.mrcc.hu.}\BibitemShut
  {Stop}%
\bibitem [{\citenamefont {K\'{a}llay}\ and\ \citenamefont
  {Surj\'{a}n}(2001)}]{Kallay:1}%
  \BibitemOpen
  \bibfield  {author} {\bibinfo {author} {\bibfnamefont {M.}~\bibnamefont
  {K\'{a}llay}}\ and\ \bibinfo {author} {\bibfnamefont {P.~R.}\ \bibnamefont
  {Surj\'{a}n}},\ }\bibfield  {title} {\bibinfo {title} {Higher excitations in
  coupled-cluster theory},\ }\href {https://doi.org/10.1063/1.1383290}
  {\bibfield  {journal} {\bibinfo  {journal} {J.\ Chem.\ Phys.}\ }\textbf
  {\bibinfo {volume} {115}},\ \bibinfo {pages} {2945} (\bibinfo {year}
  {2001})}\BibitemShut {NoStop}%
\bibitem [{\citenamefont {K\'{a}llay}\ \emph {et~al.}(2002)\citenamefont
  {K\'{a}llay}, \citenamefont {Szalay},\ and\ \citenamefont
  {Surj\'{a}n}}]{Kallay:2}%
  \BibitemOpen
  \bibfield  {author} {\bibinfo {author} {\bibfnamefont {M.}~\bibnamefont
  {K\'{a}llay}}, \bibinfo {author} {\bibfnamefont {P.~G.}\ \bibnamefont
  {Szalay}},\ and\ \bibinfo {author} {\bibfnamefont {P.~R.}\ \bibnamefont
  {Surj\'{a}n}},\ }\bibfield  {title} {\bibinfo {title} {A general
  state-selective multireference coupled-cluster algorithm},\ }\href
  {https://doi.org/10.1063/1.1483856} {\bibfield  {journal} {\bibinfo
  {journal} {J.\ Chem.\ Phys.}\ }\textbf {\bibinfo {volume} {117}},\ \bibinfo
  {pages} {980} (\bibinfo {year} {2002})}\BibitemShut {NoStop}%
\bibitem [{\citenamefont {Maison}\ \emph {et~al.}(2019)\citenamefont {Maison},
  \citenamefont {Skripnikov},\ and\ \citenamefont {Glazov}}]{Maison:2019}%
  \BibitemOpen
  \bibfield  {author} {\bibinfo {author} {\bibfnamefont {D.~E.}\ \bibnamefont
  {Maison}}, \bibinfo {author} {\bibfnamefont {L.~V.}\ \bibnamefont
  {Skripnikov}},\ and\ \bibinfo {author} {\bibfnamefont {D.~A.}\ \bibnamefont
  {Glazov}},\ }\bibfield  {title} {\bibinfo {title} {Many-body study of the $g$
  factor in boronlike argon},\ }\href
  {https://doi.org/10.1103/PhysRevA.99.042506} {\bibfield  {journal} {\bibinfo
  {journal} {Phys. Rev. A}\ }\textbf {\bibinfo {volume} {99}},\ \bibinfo
  {pages} {042506} (\bibinfo {year} {2019})}\BibitemShut {NoStop}%
\bibitem [{\citenamefont {Maison}\ \emph {et~al.}(2020)\citenamefont {Maison},
  \citenamefont {Skripnikov}, \citenamefont {Flambaum},\ and\ \citenamefont
  {Grau}}]{Maison:20a}%
  \BibitemOpen
  \bibfield  {author} {\bibinfo {author} {\bibfnamefont {D.~E.}\ \bibnamefont
  {Maison}}, \bibinfo {author} {\bibfnamefont {L.~V.}\ \bibnamefont
  {Skripnikov}}, \bibinfo {author} {\bibfnamefont {V.~V.}\ \bibnamefont
  {Flambaum}},\ and\ \bibinfo {author} {\bibfnamefont {M.}~\bibnamefont
  {Grau}},\ }\bibfield  {title} {\bibinfo {title} {Search for
  $\mathcal{CP}$-violating nuclear magnetic quadrupole moment using the
  {LuOH$^+$} cation},\ }\href {https://doi.org/10.1063/5.0028983} {\bibfield
  {journal} {\bibinfo  {journal} {J.\ Chem.\ Phys.}\ }\textbf {\bibinfo
  {volume} {153}},\ \bibinfo {pages} {224302} (\bibinfo {year}
  {2020})}\BibitemShut {NoStop}%
\bibitem [{\citenamefont {Skripnikov}(2017)}]{Skripnikov:17c}%
  \BibitemOpen
  \bibfield  {author} {\bibinfo {author} {\bibfnamefont {L.~V.}\ \bibnamefont
  {Skripnikov}},\ }\bibfield  {title} {\bibinfo {title} {Communication:
  Theoretical study of {HfF$^+$} cation to search for the
  $\mathcal{T},\mathcal{P}$-odd interactions},\ }\href
  {https://doi.org/10.1063/1.4993622} {\bibfield  {journal} {\bibinfo
  {journal} {J.\ Chem.\ Phys.}\ }\textbf {\bibinfo {volume} {147}},\ \bibinfo
  {pages} {021101} (\bibinfo {year} {2017})}\BibitemShut {NoStop}%
\bibitem [{\citenamefont {Fleig}(2017)}]{Fleig:17}%
  \BibitemOpen
  \bibfield  {author} {\bibinfo {author} {\bibfnamefont {T.}~\bibnamefont
  {Fleig}},\ }\bibfield  {title} {\bibinfo {title}
  {$\mathcal{P},\mathcal{T}$-odd and magnetic hyperfine-interaction constants
  and excited-state lifetime for {HfF$^+$}},\ }\href
  {https://doi.org/10.1103/PhysRevA.96.040502} {\bibfield  {journal} {\bibinfo
  {journal} {Phys.\ Rev.\ A}\ }\textbf {\bibinfo {volume} {96}},\ \bibinfo
  {pages} {040502(R)} (\bibinfo {year} {2017})}\BibitemShut {NoStop}%
\bibitem [{\citenamefont {Petrov}\ \emph {et~al.}(2007)\citenamefont {Petrov},
  \citenamefont {Mosyagin}, \citenamefont {Isaev},\ and\ \citenamefont
  {Titov}}]{Petrov:07a}%
  \BibitemOpen
  \bibfield  {author} {\bibinfo {author} {\bibfnamefont {A.~N.}\ \bibnamefont
  {Petrov}}, \bibinfo {author} {\bibfnamefont {N.~S.}\ \bibnamefont
  {Mosyagin}}, \bibinfo {author} {\bibfnamefont {T.~A.}\ \bibnamefont
  {Isaev}},\ and\ \bibinfo {author} {\bibfnamefont {A.~V.}\ \bibnamefont
  {Titov}},\ }\bibfield  {title} {\bibinfo {title} {Theoretical study of
  {HfF$^+$} in search of the electron electric dipole moment},\ }\href
  {https://doi.org/10.1103/PhysRevA.76.030501} {\bibfield  {journal} {\bibinfo
  {journal} {Phys.\ Rev.\ A}\ }\textbf {\bibinfo {volume} {76}},\ \bibinfo
  {pages} {030501(R)} (\bibinfo {year} {2007})}\BibitemShut {NoStop}%
\bibitem [{\citenamefont {Petrov}\ \emph {et~al.}(2009)\citenamefont {Petrov},
  \citenamefont {Mosyagin},\ and\ \citenamefont {Titov}}]{Petrov:09b}%
  \BibitemOpen
  \bibfield  {author} {\bibinfo {author} {\bibfnamefont {A.~N.}\ \bibnamefont
  {Petrov}}, \bibinfo {author} {\bibfnamefont {N.~S.}\ \bibnamefont
  {Mosyagin}},\ and\ \bibinfo {author} {\bibfnamefont {A.~V.}\ \bibnamefont
  {Titov}},\ }\bibfield  {title} {\bibinfo {title} {Theoretical study of
  low-lying electronic terms and transition moments for {HfF$^+$} for the
  electron {EDM} search},\ }\href {https://doi.org/10.1103/PhysRevA.79.012505}
  {\bibfield  {journal} {\bibinfo  {journal} {Phys.\ Rev.\ A}\ }\textbf
  {\bibinfo {volume} {79}},\ \bibinfo {pages} {012505} (\bibinfo {year}
  {2009})}\BibitemShut {NoStop}%
\end{thebibliography}

%

\end{document}